\input psfig.sty
\def\topfig#1#2#3{
    \topinsert

\psfig{figure=#1,width=\hsize,angle=90,bbllx=54bp,%
bblly=54bp,bburx=558bp,bbury=756bp}
    {\tenpoint \baselineskip=18 pt \noindent {\bf #2:} {\rm #3}}
    \smallskip
    \endinsert}
\pssilent
\font\twelverm   = cmr12
\font\ninerm     = cmr9
\font\twelvebf   = cmbx12
\font\ninebf     = cmbx9
\font\twelveit   = cmti12

\font\twelvei    = cmmi12
\font\ninei      = cmmi9
\font\twelvett   = cmtt10 at 12truept

\font\twelvesl   = cmsl12
\font\twelvesy   = cmsy10 at 12truept
\font\ninesy     = cmsy9
\font\twelvex    = cmex10 at 12truept
\font\fourtebf   = cmbx10 scaled 1440
\font\tabrm=cmr10 at 10 truept
\def\twelvepoint{ \def\rm{\fam0 \twelverm }
    \textfont0=\twelverm \scriptfont0=\ninerm \scriptscriptfont0=\sevenrm
    \textfont1=\twelvei  \scriptfont1=\ninei \scriptscriptfont1=\seveni
    \textfont2=\twelvesy \scriptfont2=\ninesy \scriptscriptfont2=\sevensy
    \textfont3=\twelvex  \scriptfont3=\twelvex \scriptscriptfont3=\twelvex
    \textfont\itfam=\twelveit      \def\it{\fam\itfam \twelveit}
    \textfont\slfam=\twelvesl      \def\sl{\fam\slfam \twelvesl}
    \textfont\ttfam=\twelvett      \def\tt{\fam\ttfam \twelvett}
    \textfont\bffam=\twelvebf    \scriptfont\bffam=\ninebf
    \scriptscriptfont\bffam=\sevenbf \def\bf{\fam\bffam \twelvebf}
    \normalbaselineskip=14pt
    \normalbaselines \rm      }
\def\tenpoint{ \def\rm{ \fam0 \tenrm }
    \textfont0=\tenrm \scriptfont0=\sevenrm \scriptscriptfont0=\fiverm
    \textfont1=\teni  \scriptfont1=\seveni  \scriptscriptfont1=\fivei
    \textfont2=\tensy \scriptfont2=\sevensy \scriptscriptfont2=\fivesy
    \textfont3=\tenex \scriptfont3=\tenex   \scriptscriptfont3=\tenex
    \textfont\itfam=\tenit      \def\it{\fam\itfam \tenit}
    \textfont\slfam=\tensl      \def\sl{\fam\slfam \tensl}
    \textfont\ttfam=\tentt      \def\tt{\fam\ttfam \tentt}
    \textfont\bffam=\tenbf      \scriptfont\bffam=\sevenbf
    \scriptscriptfont\bffam=\fivebf \def\bf{\fam\bffam \tenbf}
    \normalbaselineskip=12pt
    \let\sc=\eightrm  \let\big=\tenbig
    \normalbaselines \rm      }
\twelvepoint
\hoffset= 0 true in
\vsize 9.0 true in
\hsize 6.5 true in
\parindent 20 truept
\parskip=\smallskipamount

\def\nuc#1#2{${\rm ^{#2}#1}$}

\def\sect#1{ \centerline {\bf #1} \nobreak}
\def\subsect#1{ \centerline {\sl #1} \nobreak}
\def\title#1{ \centerline {\fourtebf #1} \nobreak}
\def\subtitle#1{ \centerline {\bf #1} \nobreak}
\def\author#1{\centerline{\bf #1}}
\def\authaddr#1{\centerline{\sl #1}}
\def\etal{{\rm et al}.}

\def\t9{\rm T_9 \thinspace}
\def\gcc{\rm\thinspace g \thinspace cm^{-3}}
\def\kel{\rm \thinspace K}

\def\rqe{$r_{QSE}(^AZ)$}

\def\thinrule{\hrule height0.2pt}

\topinsert
\vskip 2in
\endinsert

\title{Silicon Burning I:}
\subtitle{Neutronization and the Physics of Quasi-Equilibrium}

\vskip 1 in

\author{W. Raphael Hix}
\authaddr{Harvard-Smithsonian Center for Astrophysics, }
\authaddr{60 Garden Street, Cambridge, MA 02138.}
\authaddr{e-mail: raph@cfa.harvard.edu}
\medskip
\centerline{and}
\medskip
\author{Friedrich-Karl Thielemann}
\authaddr{Institut f\"ur theoretische Physik, Universit\"at Basel,}
\authaddr{Klingelbergstrasse 82, CH-4056 Basel Switzerland}
\authaddr{e-mail: fkt@quasar.physik.unibas.ch}

\vfill\eject

\sect{Abstract}

\bigskip

As the ultimate stage of stellar nucleosynthesis, and the source of the iron
peak nuclei, silicon burning is important to our understanding of the
evolution of massive stars and supernovae.  Our reexamination of silicon
burning, using results gleaned from simulation work done with a large nuclear
network (299 nuclei and more than 3000 reactions) and from independent
calculations of equilibrium abundance distributions, offers new insights
into the quasi-equilibrium mechanism and the approach to nuclear
statistical equilibrium. We find that
the degree to which the matter has been neutronized is of great importance,
not only to the final products but also to the rate of energy generation
and the membership of the quasi-equilibrium groups.  A small increase in
the global neutronization results in much larger free neutron fluences,
increasing the abundances of more neutron-rich nuclei.  As a result,
incomplete silicon burning results in neutron richness among the isotopes
of the iron peak much larger than the global neutronization would
indicate.  Finally, we briefly discuss the limitations and pitfalls of
models for silicon burning currently employed within hydrodynamic
models.  In a forthcoming paper we will present a new approximation
to the full nuclear network which preserves the most important features
of the large nuclear network calculations at a significant improvement in
computational speed.  Such improved methods are ideally suited for
hydrodynamic calculations which involve the production of iron peak nuclei,
where the larger network calculation proves unmanageable.

\bigskip

\vfill\eject

\sect{1. Introduction}

\medskip

With the exhaustion of hydrogen in the core of a massive star, an
inexorable contraction begins, heating and compressing the core, delayed
for a time as each succeeding nuclear fuel ignites and is transformed.
Beginning with the helium ash of hydrogen burning, most of these burning
stages consist of fusion reactions among the nuclei of the ash.  The first
exception is neon burning.  Following carbon burning, which leaves behind
ash composed of O, Ne, and Mg, the temperature and density continue to rise.
Before the temperature is adequate to allow fusion reactions among O nuclei
to overcome the Coulomb repulsion, the photon field becomes sufficiently
energetic to photodissociate Ne.  In general, a photodisintegration channel
becomes important when the Q-value of the reaction, that is the energy
difference between fuel and products, is smaller than approximately $30 k_B
T$.  For such temperatures, the high energy tail of the Planck distribution
provides enough photons of sufficient energy that the photodisintegration
reaction represents a comparable flow.  With a Q-value of 4.7 MeV, $\rm
^{20}Ne (\gamma,\alpha) ^{16}O$ becomes important for temperatures above
$1.5 \times 10^9 \kel$.  Thus the next stage is neon burning, typified by
the photodisintegration of neon and the subsequent capture of the ejected
$\alpha$-particles by the remaining heavy ions.  With the exhaustion of
this energy source, the collapse continues until the O nuclei are
sufficiently energetic to fuse.  Once more nuclear burning offers a respite
until the oxygen fuel is exhausted and the collapse continues, further
raising the temperature and density.  With typical Q-values for reactions
among stable nuclei above silicon being 8-12 MeV, photodisintegration begins
to play an important role for all nuclei once the temperature exceeds $3
\times 10^9 \kel$.  Nuclei with the smallest binding energies are destroyed
in favor of their more tightly bound neighbors.  Very quickly this leaves
the core composed of silicon isotopes and their $\alpha$-nuclei neighbors,
S, Ar, and Ca, the most bound of the light nuclei.  Continued contraction
further increases temperatures, favoring even more tightly bound nuclei,
the iron peak.  What follows is a complex series of photodissociation and
capture reactions, converting silicon into iron peak elements.  Unlike
previous burning stages, where a few reactions (either heavy ion reactions
among the principal constituents or conversion among neighbors initiated
by photodisintegration) dominated, here the fuel nuclei are linked to the
product nuclei by a multitude of reaction chains and cycles.  Since these
chains of reactions wind their way through many nuclei intermediate between
silicon and the iron peak nuclei,  it is necessary to keep track of many
more nuclei than was the case for prior burning stages.  With four potential
particles in the incoming channel, p, n, $\alpha$, or $\gamma$, and these
same four choices for the outgoing channels, discounting elastic scattering
reactions, there are, in principle, at least 12 reactions per nucleus.
Thus silicon burning is a complicated web of reactions, making it necessary
to keep track of the abundances of a large number of nuclei.  This is a
major contribution to the complexity of silicon burning.

Silicon burning is further complicated by the closeness to equilibrium of
many of these pairs of photodissociations and capture reactions.  Net
fluxes are often orders of magnitude smaller than either the forward or
reverse reaction rates would indicate.  Considering that the end state of
silicon burning is an equilibrium, nuclear statistical equilibrium (NSE),
it is hardly surprising that equilibria among nuclei arise during silicon
burning.  From a purely numerical point of view, the net reaction flux being
the difference of two large numbers raises the potential danger of round-off
errors.  However, equilibria are attractive for the simplification they
provide.  As we will see, equilibria are the key to understanding silicon
burning and simplifying our modeling thereof.

The work of Bodansky, Clayton \& Fowler (1968; hereafter BCF) examined the
process by which this equilibrium distribution is formed.  Burbidge,
Burbidge, Fowler \& Hoyle (1957; hereafter $\rm B^2FH$) had postulated that
the intermediate nuclei, principally the
$\alpha$-particle nuclei from \nuc{Mg}{24} to \nuc{Ca}{40}, were formed
by the $\alpha$-process, a series of $\alpha$ particle captures, which was
separate from the e-process which formed the iron peak.  Taking their cue
from Hayashi \etal\ (1959) and the nuclear network calculations of Truran,
Cameron, \& Gilbert (1965), BCF showed that the intermediate nuclei are
formed via a partial equilibrium in which groups of nuclei exist which are
internally in equilibrium under the exchange of photons, protons, neutrons
and $\alpha$-particles, i.e. in equilibrium with respect to strong and
electromagnetic reactions.  These quasi-equilibrium (QSE) groups remain
out of equilibrium with respect to other groups until the final complete
equilibrium is achieved.  BCF concluded that the abundances of the
intermediate nuclei could be explained by a single QSE group which reached
from \nuc{Si}{28} through the iron peak, but which failed to reach complete
equilibrium, or to completely exhaust silicon.  Thus the $\alpha$- and
e processes
of $\rm B^2FH$ were unified, coinciding with silicon burning, which had
been determined to be a fundamental late stage of nuclear burning in
massive stars.  Further, BCF examined the influence of weak reactions on
silicon burning and concluded that, while these reactions could have
important effects on the abundance distribution, the ratio of the total
number of protons to the total number of neutrons did not vary far from
unity.  In contrast to Fowler \& Hoyle (1964), BCF concluded that the
dominance of \nuc{Fe}{56} was due to the production of \nuc{Ni}{56} which
subsequently decayed to \nuc{Co}{56} and then to \nuc{Fe}{56}.  This scenario
agrees well with observations of supernovae.  The lack of an equilibrium
among weak reactions requires the monitoring of an additional degree of
freedom which affects the equilibrium solution, the degree to which the
material is neutronized.  In the literature this is parameterized in two
ways, as $Y_e$, the electron (molar) abundance, or as $\eta$, the neutron
excess parameter:
$$
\eqalignno{
Y_e&=\sum_i Z_i Y_i = \sum_i \left( Z_i \over A_i \right) X_i &({\rm 1a})\cr
\eta&=\sum_i \left[N_i - Z_i \right] Y_i = \sum_i \left[{N_i - Z_i} \over A_i
 \right] X_i = \sum_i \left[{A_i -2 Z_i} \over A_i \right] X_i &({\rm 1b})
 \cr \eta&= 1-2 Y_e \ , &(\rm{1c}) \cr
}
$$
where $Z_i$, $N_i$, $A_i$, $X_i$, $Y_i$ are the atomic number, neutron
number, mass number, mass fraction, and abundance of nucleus $i$.
Physically, $Y_e$ is the ratio of protons to nucleons (identical to the ratio
of electrons to nucleons, hence the subscript $e$) or the total proton
fraction, and $\eta$ is the fraction of excess neutrons per nucleon.  BCF
showed that the abundances which result from silicon burning are strongly
dependent on the degree of neutronization, a view reinforced by Hartmann,
Woosley, \& El Eid (1985).  In this paper we will show that not only the
nuclear products but the entire mechanism is strongly affected by the degree
of neutronization.

Woosley, Arnett \& Clayton (1973; hereafter WAC), in the context of
parameterized explosive silicon burning, showed that rather than a single
QSE group between silicon and the iron peak, there are initially two groups,
roughly divided by $A \simeq 45$.  WAC further demonstrated that considering
only the reactions which link the groups yields a good approximation of the
time necessary for the two groups to merge and form a single QSE group.
Indeed, WAC contended that this linkage was dominated by a single reaction,
$\rm ^{45}Sc(p,\gamma)^{46}Ti$, with perhaps a quarter of the flow going
through less important reactions like $\rm ^{42}Ca(\alpha,\gamma) ^{46}Ti$
and $\rm ^{45}Ti(n,\gamma)^{46}Ti$.  In later sections, we will show that
for material which has been more highly neutronized, these two QSE groups
form further from mutual equilibrium and that this division persists for a
much longer time.  In addition there is some ambiguity in defining the boundary
between the QSE groups and the most important reactions linking the groups
as $Y_e$ is varied.  As we will also show, the dependence of the behavior
of the QSE groups on the degree of neutronization has important consequences
for the energy generation and other physical manifestations of silicon
burning, in addition to determining the nuclear products.

Thielemann \& Arnett (1985; hereafter TA), examined silicon burning in
the context of hydrostatic models of massive stars and noticed behavior
largely in keeping with that described by WAC for the explosive case.
For conditions characteristic of the cores of more massive stars, high
temperature, low density, and consequently larger $Y_e$, TA found that the
bottleneck between the QSE groups, coinciding roughly with $Z=21$, was
bridged on the proton-rich side of stability.  However, for conditions more
characteristic of lower mass stars, lower temperature, higher density, and
smaller $Y_e$, the bridge was found to be proton capture on neutron-rich
isotopes of Ca.  This more complex behavior differed with the assertions of
WAC and of Weaver, Woosley, \& Fuller (1985), that the single reaction $\rm
^{45}Sc (p,\gamma)^{46}Ti$ is responsible for the lion's share of the flow
between the QSE groups.  As we will show in later sections,
our efforts allow us to explain the differences between the results of TA
and WAC.

\bigskip

\sect{2. Nuclear Reaction Network Calculation}

\medskip

Silicon burning in the cores of massive stars takes place under a range of
temperatures and densities and with a range of electron fractions.  We have
examined silicon burning in a cube of this parameter space with temperatures
in the range 3.5 to 5 $\times 10^9 \kel$, densities between $10^7$ and $10^{10}
\gcc$, and $Y_e$ between .498 and .46, using a nuclear network of 299 nuclei,
shown in Table 1. This region of parameter space roughly spans the region
that previous investigations of nucleosynthesis in massive stars (Thielemann
\& Arnett 1985, Nomoto \& Hashimoto 1988) have exhibited during core silicon
burning.  Further, the lower density portions of this parameter space
overlap, to a large extent, the parameter space that prior investigations
(WAC, Woosley, Pinto, \& Weaver 1988, Thielemann, Hashimoto, \& Nomoto 1990,
Aufderheide, Baron, \& Thielemann 1991) have determined appropriate
for explosive silicon burning.  We have included the effects of Coulomb
screening on the equilibria for this same region of parameter space, as
discussed in Hix, Thielemann, Fushiki, \& Truran (1996). The results of
nuclear network calculations for this region of parameter space and the
comparison of these network calculations to equilibrium calculations are
discussed in the present paper.  In a subsequent paper (Hix \& Thielemann 1996,
hereafter Paper II), we will discuss the applicability of quasi-equilibrium
to explosive burning.  The final goal is an improvement to the network,
using what we have learned about quasi-equilibrium, which speeds the
calculation of the energy generation and nuclear abundance changes due to
silicon burning, preserving much of the accuracy of the network calculation
while greatly reducing the computational overhead.

\topinsert
\centerline{Table 1: Nuclear set Included in Calculations}
\nobreak
\medskip
\moveright  .85in
\hbox { \hsize= .75in
\valign{\thinrule \smallskip \thinrule \medskip
\centerline{#}\strut \vskip -12pt \thinrule \smallskip &#\strut&#\strut&
#\strut&#\strut&#\strut&#\strut&#\strut&#\strut&#\strut&#\strut& #\strut&
#\strut&#\strut&#\strut&#\strut&#\strut&#\strut  \smallskip \thinrule \cr
Element&n&H&He\rlap*&Li&Be\rlap*&B\rlap*&C&N&O&F&Ne&Na&Mg&Al&Si&P&S\cr
$A_{min}$&1&1&3&6&7&8&10&12&14&17&18&20&21&23&25&27&29 \cr
$A_{max}$&1&3&6&8&11&12&15&17&20&21&25&26&28&30&33&35&38 \cr
\noalign{\vrule}
Element&Cl&Ar&K&Ca&Sc&Ti&V&Cr&Mn&Fe&Co&Ni&Cu&Zn&Ga&Ge&  \cr
$A_{min}$&31&33&35&37&40&42&44&46&48&50&52&54&57&59&61&63&  \cr
$A_{max}$&40&44&46&49&50&52&54&56&58&62&63&67&69&72&74&78& \cr }}
\vskip -5pt
\hskip .5in
{* excepting ${\rm ^5He}$, ${\rm ^8Be}$, and ${\rm ^9B}$ \hfil}
\endinsert

For our study of silicon burning, we consider 299 nuclei, listed in
Table 1, linked by more than 3000 reactions.  This nuclear set stretches
from protons and neutrons to germanium.  With 11 or more isotopes per
element around and above iron and 7 or more in the region around silicon,
this nuclear set is more complete than that used in earlier work
discussed in \S 1.  This improvement is most important for
material which has undergone significant prior electron capture and hence
has a $Y_e$ significantly less than .498.  Rather than complicate our task
by including hydrodynamics, we have instead done a parameter study of
silicon burning as a function of three variables, temperature, density and
$Y_e$.  In the course of a hydrodynamic model calculation, the star's core
would wind its way through our cube of  parameter space.  In particular,
electron capture will cause $Y_e$ to decrease with time.  However, the
timescales for changes in $Y_e$ via weak reactions, even at densities
sufficient for significant electron capture, are much longer than the
timescales for the strong and electromagnetic reactions.  Thus it is
possible to treat them separately. Therefore we have neglected weak
reactions, allowing us to treat $Y_e$ as a constant parameter rather
than merely considering the $Y_e$ at one point in time.  The reaction
rates for light nuclei are based on experimental information from Caughlan
\& Fowler (1988), Bao \& K\"appeler (1987), Wiescher \etal\ (1986, 1987,
1989), Wiescher, G\"orres, \& Thielemann (1988), and also
from Wagoner (1969) and Wagoner, Fowler, \& Hoyle (1967) if otherwise
unavailable.  For intermediate and heavy nuclei, where higher level
densities permit application of statistical model calculations, rates by
Thielemann, Arnould, \& Truran (1987) are used (see also Cowan, Thielemann,
\& Truran 1991).  We utilized a reaction
network at constant temperature, density, and $Y_e$, with $Y_e$ set by an
initial distribution of silicon isotopes.  There are of course several
such distributions with the same $Y_e$ but, with the exception of a brief
initial adjustment phase which does not enter into our analysis, such
variations of the initial distribution showed the same results.  The basics
of a nuclear reaction network are summarized in Woosley (1986) or Thielemann,
Nomoto, \& Hashimoto (1994).  It is particularly important to note that the
weak and intermediate screening prescriptions of Graboske \etal\ (1973) were
used, as well as the strong screening prescription of Itoh \etal\ (1990).
Transition among these prescriptions is performed by utilizing the smallest
predicted screening enhancement.

\bigskip

\sect{3. The Physics of Quasi-Equilibrium}

\medskip

Previous authors, notably BCF and WAC, have shown that quasi-equilibrium
(QSE) is a most important aid to understanding the process of silicon
burning.  This is fortunate as it is much easier to follow the evolution
of equilibrium groups rather than become lost in the welter of individual
abundances and reactions.  Since \nuc{Si}{28} is the principal
fuel, we will begin our discussion of QSE here.  If we take \nuc{Si}{28}
to be the focus of our quasi-equilibrium group, then the abundance of a
nucleus $^AZ$ which is in equilibrium with \nuc{Si}{28} with respect to
the exchange of free nucleons, $\alpha$-particles, and photons is
$$
Y_{QSE}(^AZ) = \left(C(^AZ) \over C({\rm ^{28}Si})
\right) Y({\rm ^{28}Si}) {Y_n}^{N-14} {Y_p}^{Z-14} \ ,
\eqno(2)
$$
where $Y_n$, $Y_p$, and $Y({\rm ^{28}Si})$ are the abundances of free
neutrons, free protons, and \nuc{Si}{28}, and we have defined
$$
C(^AZ)= {G(^AZ) \over 2^A} {\left(\rho N_A \over \theta \right)}^{A-1}
A^{3 \over 2} \exp {\left( B(^AZ) \over {k_B T} \right)}
\eqno(3)
$$
for later convenience. $G(^AZ)$ and $B(^AZ)$ are the partition function and
binding energy of the nucleus $^AZ$, $N_A$ is Avogadro's number, $k_B$ is
Boltzmann's constant, and $\rho$ and $T$ are the density and temperature
of the plasma.  Thus, the abundance of a nucleus in quasi-equilibrium with
\nuc{Si}{28} is a function of three other abundances, those of free protons,
free neutrons, and \nuc{Si}{28}, and the thermodynamic conditions and
properties of the nucleus, subsumed here within $C(^AZ)$.  It is the
evolution of these three abundances which determine the behavior of the
entire QSE group.  We reserve $Y_{QSE}(^AZ)$ to represent the abundance of
a nuclear species in quasi-equilibrium with \nuc{Si}{28}.  Equation (2) is
identical to the expressions for quasi-equilibrium introduced by BCF and
WAC, provided the $\alpha$-particles, protons and neutrons are internally
in equilibrium.  We find this provision to be justified by the time
quasi-equilibrium is established.

This derivation ignores screening, an omission which Hix \etal\ (1996)
found can be significant for equilibria under the conditions we
are discussing.  A rederivation including the screening corrections to
the reaction balance produces
\def\hfact#1#2{\prod_{#1}^{#2} \left[ \exp (H_{p,\gamma}(Z')) \over
\exp (H_{\gamma,p}(Z')) \right] }
$$
Y_{QSE}(^AZ) = \left(C(^AZ) \over C({\rm ^{28}Si}) \right) Y({\rm ^{28}Si})
{Y_n}^{N-14} {Y_p}^{Z-14} \hfact{Z'=14}{Z-1} \ ,
\eqno(4)
$$
where $\exp(H_{ij}(Z'))$  is the screening correction for the reaction
$^{A'}Z'(i,j)^{A''}Z''$, and we utilize the convention that, for indices
less than the bottom index ($Z<14$), the product is instead a division.
The form of this equation is perhaps not surprising.  We can rewrite Equation
(4) in the suggestive form
$$
Y_{QSE}(^AZ) = \left( C(^AZ) {Y_n}^N {Y_p}^Z \hfact{Z'=1}{Z-1} \over
C ({\rm ^{28}Si}) {Y_n}^{14} {Y_p}^{14} \hfact{Z'=1}{13} \right)
Y({\rm ^{28}Si}) \ .
\eqno(5)
$$
As $Y({\rm ^{28}Si})$, $Y_p$, and $Y_n$ approach their NSE values,
$$
Y({\rm ^{28}Si}) \Rightarrow Y_{NSE}({\rm ^{28}Si}) \equiv C({\rm ^{28}Si})
{Y_n}^{14} {Y_p}^{14} \hfact{Z'=1}{13} \ ,
\eqno(6)
$$
forcing
$$
Y_{QSE}(^AZ) \Rightarrow C(^AZ) {Y_n}^N {Y_p}^Z \hfact{Z'=1}{Z-1} \equiv
Y_{NSE}(^AZ) \ .
\eqno(7)
$$
Thus the quasi-equilibrium group blends simply into the NSE distribution
as \nuc{Si}{28} and the free nucleons approach equilibrium.

To compare the network results to quasi-equilibrium, we define
$$
r_{QSE}(^AZ) \equiv \log \left[ Y_{QSE}(^AZ) \over Y(^AZ) \right] \ ,
\eqno(8)
$$
with $Y(^AZ)$ being the network abundance; $Y_{QSE}(^AZ)$ is calculated
from the network abundances of free nucleons and \nuc{Si}{28}.  For
comparison, this definition is compatible with $r_{qe}$ as defined by WAC.
Figures 1a-g
show the results of this comparison for varying degrees of silicon exhaustion,
at different temperatures, densities, and values of $Y_e$.  Before we consider
the results, it is instructive to determine the signature that multiple
quasi-equilibrium groups leave on \rqe.

If a nucleus $^AZ$ is in quasi-equilibrium with, for example, \nuc{Ni}{56},
then there is an expression analogous to Equation (4) or Equation (5) for this
abundance.  Taking Equation (5) as the template, we can define a second
quasi-equilibrium abundance with respect to \nuc{Ni}{56},
$$
Y_{QSE2}(^AZ) \equiv \left(C(^AZ) {Y_n}^N {Y_p}^Z \hfact{Z'=1}{Z-1}
\over C({\rm ^{56}Ni}) {Y_n}^{28} {Y_p}^{28} \hfact{Z'=1}{27} \right)
Y({\rm ^{56}Ni}) \ .
\eqno(9)
$$
For species which are part of this second QSE group, i.e., $Y(^AZ)=Y_{QSE2}
(^AZ)$, using Equation (9) and Equation (5) we can evaluate
$$
{Y_{QSE}(^AZ) \over Y(^AZ)} = {Y_{QSE}(^AZ) \over Y_{QSE2}(^AZ)}= {{\left(
C(^AZ) {Y_n}^N {Y_p}^Z \hfact{Z'=1}{Z-1} \over C({\rm ^{28}Si}) {Y_n}^{14}
{Y_p}^{14} \hfact{Z'=1}{13} \right) Y({\rm ^{28}Si}) } \over {\left(C
(^AZ) {Y_n}^N {Y_p}^Z \hfact{Z'=1}{Z-1} \over C({\rm ^{56}Ni}) {Y_n}^{28}
{Y_p}^{28} \hfact{Z'=1}{27} \right) Y({\rm ^{56}Ni}) }} \ .
\eqno(10)
$$
Since the numerators of these fractions are identical, cancellation leaves
$$
{Y_{QSE}(^AZ) \over Y(^AZ)}=\left(C({\rm ^{56}Ni}) {Y_n}^{28} {Y_p}^{28}
\hfact{Z'=1}{27} \over C({\rm ^{28}Si}) {Y_n}^{14} {Y_p}^{14} \hfact{Z'=1}{13}
\right)  {Y({\rm ^{28}Si}) \over Y({\rm ^{56}Ni})} = {Y_{QSE}
(^{56}{\rm Ni}) \over Y(^{56}{\rm Ni})}
\eqno(11) \ .
$$
Thus every member of a quasi-equilibrium group has the same value of \rqe.
For the silicon group, $r_{QSE}(^AZ)=0$, but other groups will also be apparent
but offset from zero.  Although free nucleons do not come into equilibrium
with \nuc{Si}{28} until NSE is established, it is also instructive to examine
$r_{QSE}(p)$ or $r_{QSE}(n)$.  From Equations (3) and (5) we have
$$
Y_{QSE}(p)={{Y_p Y({\rm ^{28}Si})} \over  {C({\rm ^{28}Si}) {Y_n}^{14}
{Y_p}^{14} \hfact{Z'=1}{13}}} \ .
\eqno(12)
$$
Dividing by $Y_p$ gives
$$
{Y_{QSE}(p) \over Y_p} = {{Y({\rm ^{28}Si})} \over  {C({\rm ^{28}Si})
{Y_n}^{14} {Y_p}^{14} \hfact{Z'=1}{13} }}= {Y_{QSE}(n) \over Y_n} \ .
\eqno(13)
$$
The denominator of Equation (13) is identical to the expression for the NSE
abundance of \nuc{Si}{28}, reflecting the abundance of \nuc{Si}{28} if it
was in equilibrium with the free nucleons.  Thus changes in $r_{QSE}(p)
(=r_{QSE}(n))$ reflect the approach of the silicon group to NSE.

\bigskip

\sect{4. Comparison of Network results to Quasi-Equilibrium}

\medskip

\topfig{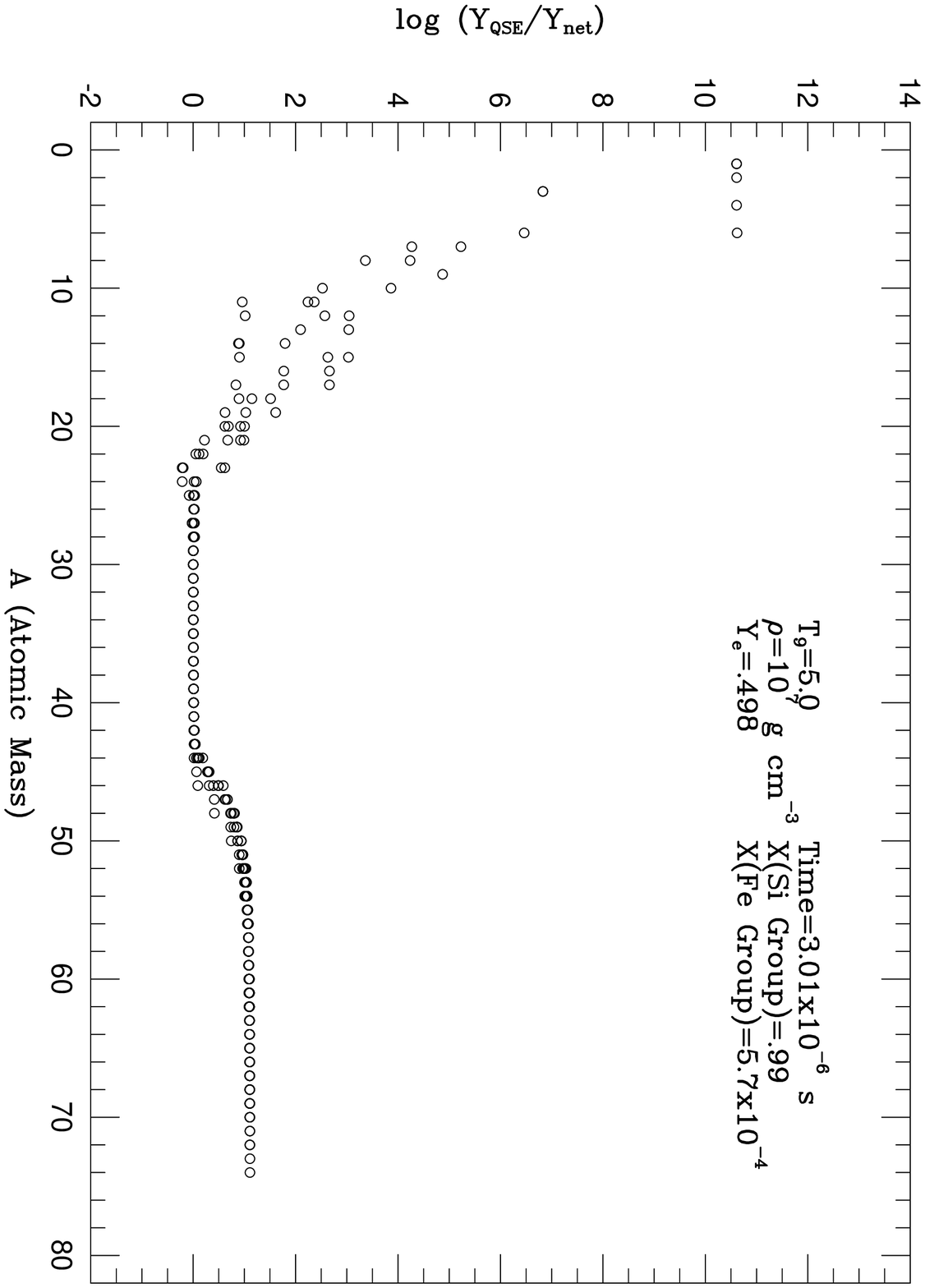}{Figure 1a}{Comparison of the silicon quasi-equilibrium
abundance to the network abundance, as a function of A for $\t9=5.0$,
$\rho=10^7 \gcc$, and $Y_e=.498$, with X(Si group)=.99.}

\topfig{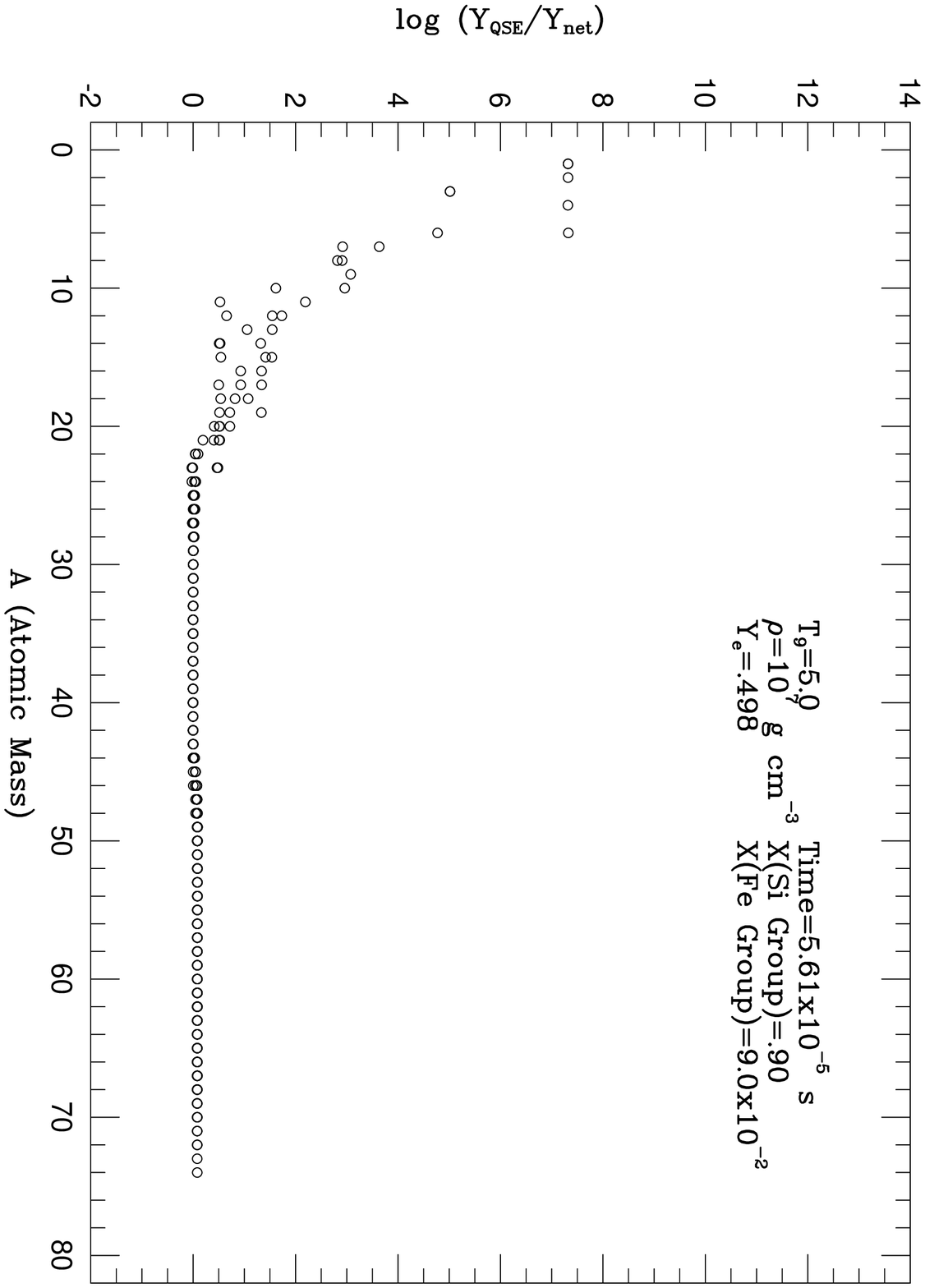}{Figure 1b}{Comparison of the silicon quasi-equilibrium
abundance to the network abundance, as a function of A for $\t9=5.0$,
$\rho=10^7 \gcc$, and $Y_e=.498$, with X(Si group)=.90.}

\topfig{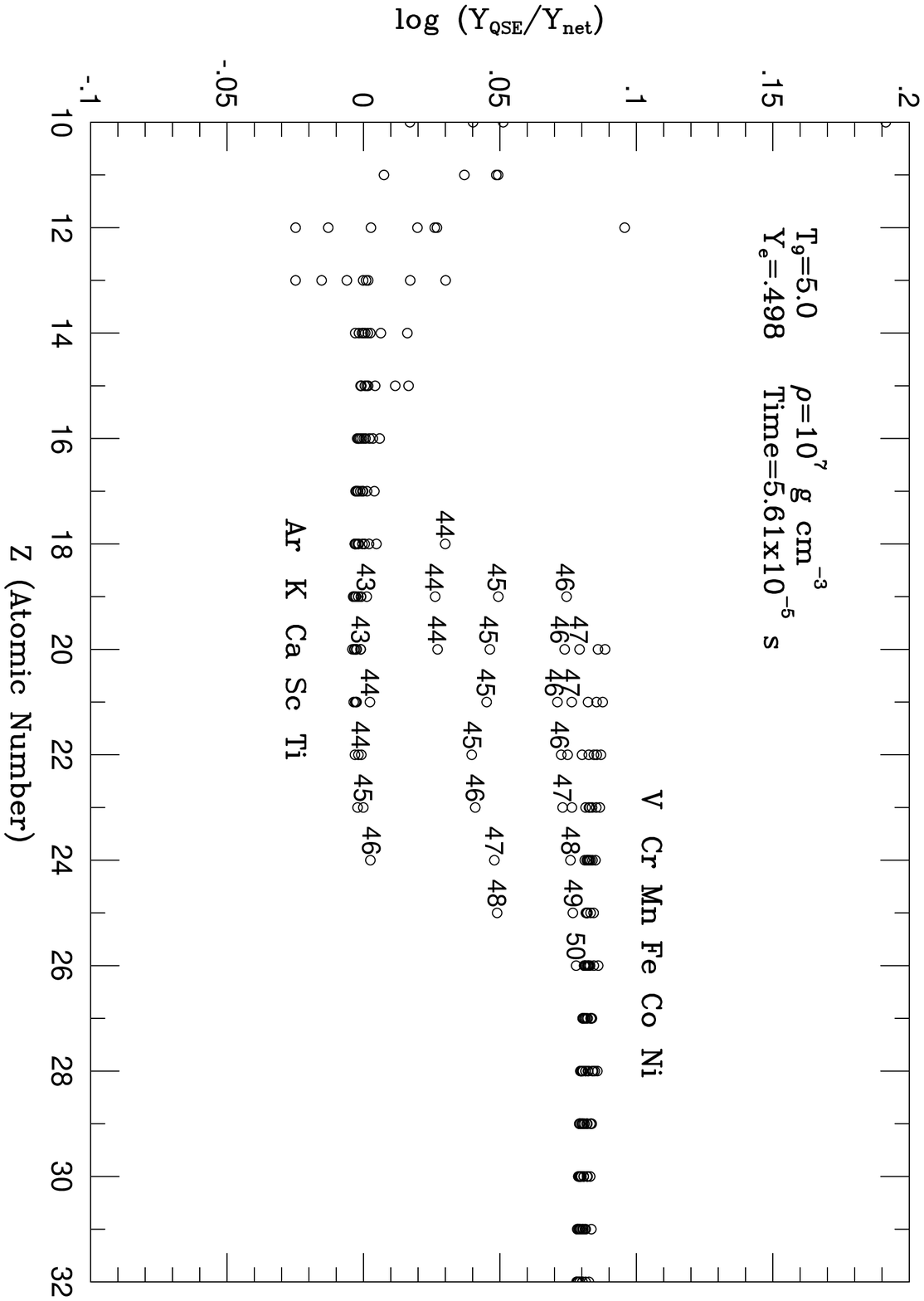}{Figure 1c}{Detail of the comparison of the silicon
quasi-equilibrium abundance to the network abundance, as a function of Z
for $\t9=5.0$, $\rho=10^7 \gcc$, and $Y_e=.498$, with X(Si group)=.90.}

In Figures 1a-g we compare the results of the full network with abundance
predictions calculated from the network abundances of n, p, and \nuc{Si}{28}
assuming quasi-equilibrium.  Elements in quasi-equilibrium will show
similar values of \rqe\ in each of Figures 1a-g.  Figure 1a shows what is a
representative pattern when \rqe\ is plotted against A.  From roughly A=24
to A=45, there is a large cluster of nuclei in quasi-equilibrium with
\nuc{Si}{28}, with a second quasi-equilibrium group stretching from roughly
A=50 to the top of the network.  Finally, there is a third small group
composed of protons, neutrons, $\alpha$-particles and some other light
nuclei.  Species in between this lightest group and the silicon QSE group
do not form large QSE groups, although they are closer to quasi-equilibrium
with silicon than the light element group.  With a temperature of $5 \times
10^9 \kel$, a density of $10^7 \gcc$
and $Y_e$ of .498, this case is among the fastest we will consider.  After
an elapsed time of $3.0 \times 10^{-6}$ s, with 86\% of the mass contained
in \nuc{Si}{28} (X(\nuc{Si}{28})=.86), and 99\% of the mass in the silicon
QSE group (X(Si group)=.99), quasi-equilibrium is clearly well established.
Comparison with Figure 1b shows that at a later time (elapsed time $= 5.6
\times 10^{-5}$ s, X(Si group)=.90, X(\nuc{Si}{28})=.65), this pattern
is maintained although all nuclei are closer to equilibrium.  In fact all of
the nuclei above Na are within 25\% of their silicon QSE abundance.
However examination of Figure 1c, a detail of Figure 1b, plotted now as a
function of Z, shows that even within this narrow margin it is still
possible to resolve the 2 QSE groups and a scattering of nuclei between
them.  Since \rqe\ is plotted against $Z$ in this figure, with important
nuclei labeled by their respective atomic masses, it is also immediately
apparent that the boundary between the groups is not as simple as $Z=21$,
since neutron-rich isotopes of K and Ca are clearly members of the upper
QSE group while proton-rich isotopes of V and Cr are in quasi-equilibrium
with Si.  The description of the boundary as $A=45$ is more successful,
giving way to N=23 from Ti onward.  Examination of Table 2 allows a more
quantitative comparison of the values of \rqe, a point we will return to
in a moment.  The convergence toward equilibrium continues as time elapses.
By an elapsed time of $2\times10^{-3}$ s (X(Si group)=.5), the elements
that formed the upper group are within 5\% of their silicon QSE abundance,
even nuclei with mass fractions as small as $10^{-22}$.

\topfig{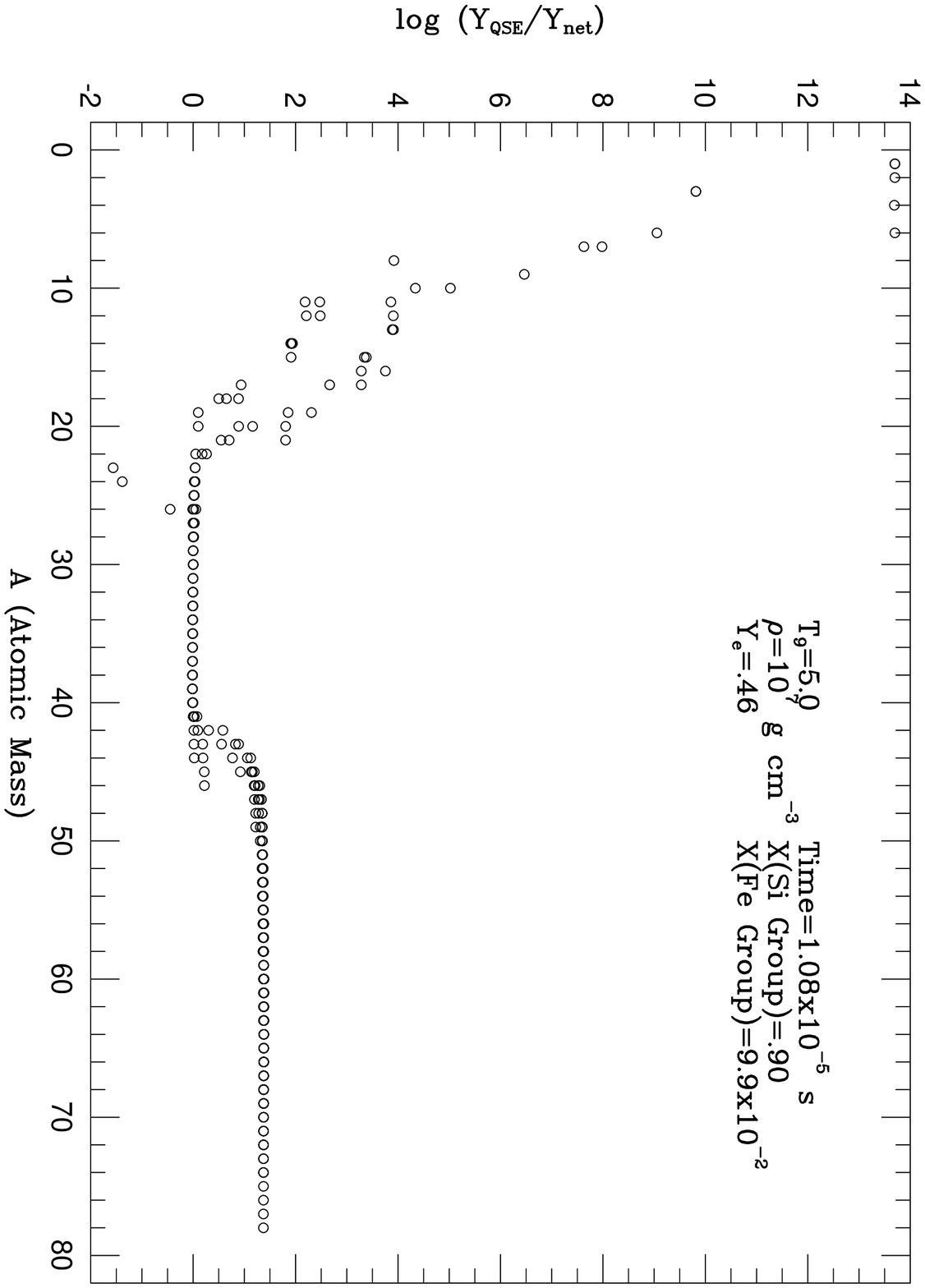}{Figure 1d}{Comparison of the silicon quasi-equilibrium
abundance to the network abundance, as a function of A for $\t9=5.0$,
 $\rho=10^7 \gcc$, and $Y_e=.46$, with X(Si group)=.90.}

\topfig{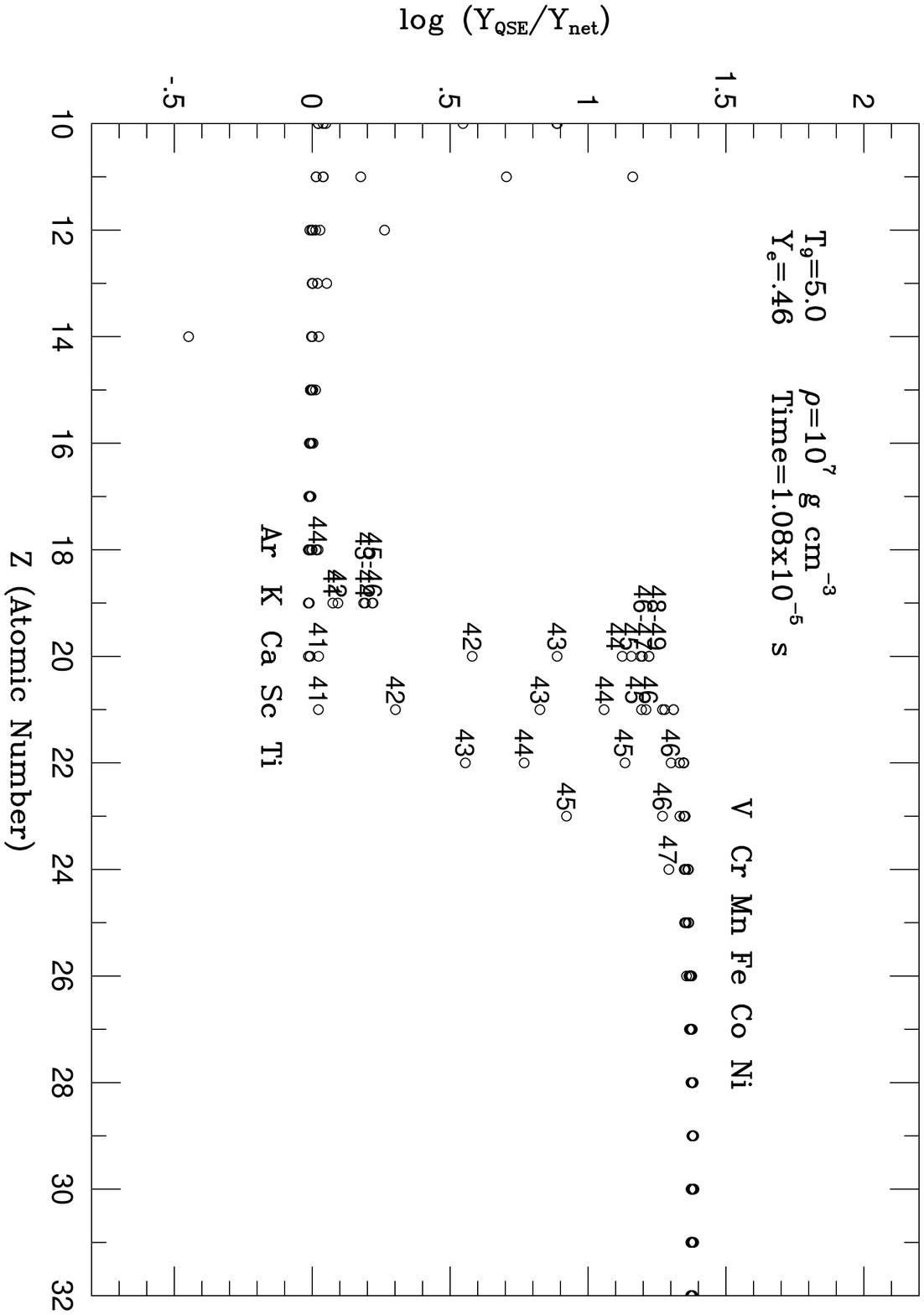}{Figure 1e}{Detail of the comparison of the silicon
quasi-equilibrium abundance to the network abundance, as a function of Z
for $\t9=5.0$, $\rho=10^7 \gcc$, and $Y_e=.46$, with X(Si group)=.90.}

\topinsert
\centerline{Table 2 Quasi-Equilibrium for Selected Nuclei}
\nobreak
\moveright 1.3in
\vbox{ \tabrm \offinterlineskip
\tabskip 1 pc
 \halign{\strut#&\hfil${\rm#}$\hfil&\vrule#&\hfil$#$\hfil&\hfil$#$\hfil
 &\vrule#&\hfil$#$\hfil&\hfil$#$\hfil \cr
\noalign{\smallskip \thinrule \smallskip \thinrule}
&\ &\multispan 5 {\rm \hfil Conditions \hfil}&\ \cr
\noalign{ \thinrule }
&T_9,\ \rho&&\multispan 2 {\hfil $5,\ 10^7$ \hfil}&&
\multispan 2 ${ \hfil 5,\ 10^7 \hfil}$ \cr
&Y_e&&.498&.46&&.498&46  \cr
&X_{Si}&&.900&.899&&.485&.521 \cr
\noalign{ \hrule height 0.6pt}
&Species&&\multispan 5 ${\rm \hfil r_{QSE}(^AZ)\hfil}$ \cr
\noalign{ \thinrule }
&^{24}Mg&&0.026&0.028&&0.030&0.028 \cr
&^{32}S&&-.002&-.005&&-.003&-.006  \cr
&^{36}Ar&&-.003&-.011&&-.006&-.013  \cr
&^{43}K&&0.001&0.187&&-.006&0.079  \cr
&^{45}K&&0.049&0.217&&-.003&0.105  \cr
&^{40}Ca&&-.004&-.007&&-.008&-.015 \cr
&^{42}Ca&&-.002&0.580&&-.009&0.096 \cr
&^{44}Ca&&0.027&1.124&&-.007&0.166  \cr
&^{46}Ca&&0.074&1.191&&-.005&0.188  \cr
&^{48}Ca&&0.086&1.221&&-.003&0.217  \cr
&^{42}Sc&&-.004&0.302&&-.009&0.033  \cr
&^{43}Sc&&-.003&0.826&&-.009&0.133  \cr
&^{44}Sc&&0.002&1.058&&-.009&0.162  \cr
&^{45}Sc&&0.045&1.194&&-.008&0.186  \cr
&^{46}Sc&&0.071&1.210&&-.007&0.193  \cr
&^{47}Sc&&0.076&1.270&&-.007&0.222  \cr
&^{42}Ti&&-.002&\bullet&&-.008&0.019  \cr
&^{44}Ti&&-.001&0.768&&-.010&0.142  \cr
&^{45}Ti&&0.040&1.134&&-.009&0.183  \cr
&^{46}Ti&&0.072&1.301&&-.009&0.226  \cr
&^{44}V&&-.002&\bullet&&-.008&0.109  \cr
&^{45}V&&0.000&0.922&&-.009&0.153  \cr
&^{46}V&&0.040&1.270&&-.010&0.215  \cr
&^{47}V&&0.073&1.333&&-.009&0.235  \cr
&^{46}Cr&&0.003&\bullet&&-.009&0.203   \cr
&^{47}Cr&&0.048&1.293&&-.009&0.232  \cr
&^{48}Cr&&0.076&1.349&&-.009&0.239   \cr
&^{48}Mn&&0.049&\bullet&&-.009&\bullet  \cr
&^{49}Mn&&0.078&1.350&&-.009&0.241  \cr
&^{50}Fe&&0.078&\bullet&&-.008&0.243  \cr
&^{56}Fe&&0.081&1.376&&-.012&0.237  \cr
&^{58}Fe&&0.081&1.375&&-.011&0.236  \cr
\noalign{ \thinrule }
}}
\endinsert

Previous authors have discussed the progress of silicon
burning in terms of the degree of exhaustion of \nuc{Si}{28}.  For this
investigation, where $Y_e$ is a parameter, this description is suspect.
The abundance of \nuc{Si}{28} in a composition purely of silicon decreases
as $Y_e$ decreases.  In place of X(\nuc{Si}{28}), we have chosen to use
the mass fraction within the entire silicon QSE group, X(Si group).  This
choice has several advantages.  Since the mass fraction is dominated by
the two QSE groups, the mass fraction within the silicon group does reflect
the degree to which material remains to be passed into the upper group
(which in NSE will compose 95\% or more of the mass).  Also, this choice is
useful for all $Y_e$ in a way not
possible for a single nucleus.  This choice is, however, limited in the sense
that individual reactions depend on the abundances of individual nuclei.
Knowledge of the mass fraction in the silicon group is not sufficient to
determine the photodisintegration rate of \nuc{Si}{28}, for example.
This is an issue which we must keep in mind as we proceed.

As a further demonstration of this difficulty, consider the case where $\t9
=5.0$, $\rho=10^7 \gcc$, but $Y_e=.46$.  Figure 1d is comparable to Figure
1b, in the sense that X(Si group) is approximately the same (.91 for Figure
1d with $Y_e=.46$ as compared to .90).  But the fraction of \nuc{Si}{28}
differs greatly, with X(\nuc{Si}{28})=.008 for $Y_e=.46$.  Furthermore, for
$Y_e=.46$, X(\nuc{Si}{28}) is actually rising slightly at X(Si group)
$\sim .9$.  It should be noted, however, that for $Y_e=.46$, X(all isotopes
of Si) is .64, thus this extremely small abundance of \nuc{Si}{28} is
specific to this nucleus, and is due to the preference for more neutron-rich
nuclei at lower $Y_e$.  There is also considerable difference in the elapsed
time, $1.1 \times 10^{-5}$ s for $Y_e=.46$ compared to $5.6 \times
10^{-5}$ s for $Y_e=.498$.  Although these elapsed times are small, they
are significant, representing the time required to consume 10\% of the
silicon fuel.  Thus such differences in elapsed time are important, implying
faster transfer of mass to the iron peak
group and consequently, as we will discuss in \S 7, a larger energy
generation rate.  For now however, the concern is quasi-equilibrium.
Comparison of Figure 1d with Figure 1b shows that the gap between the silicon
group and the iron peak group is much larger for $Y_e=.46$.  Indeed, the
underabundance of the iron peak group is more in keeping with that shown
in Figure 1a.  Comparison of Figures 1c and 1e reveals that, rather than being
25\% underabundant, the nuclei in the QSE group around iron are underabundant
by a factor of 25 for $Y_e=.46$.  This is in spite of the similar fractions
of the mass which have been transferred to the iron peak group, .09 and .1
for $Y_e=.498$ and $.46$, respectively.  Clearly, the relative underabundance
of the iron peak group at low $Y_e$ reflects a much larger QSE abundance for
these species rather than an actual dearth of mass transferred.  Another
point to note in Figure 1d is the larger value of $r_{QSE}(p)$ for $Y_e=.46$
compared to $Y_e=.498$.  This reflects the smaller NSE mass fraction
of the silicon group (.00005 for $Y_e=.46$, compared to .018 for $Y_e=
.498$), for lower $Y_e$ and hence for the same degree of fuel exhaustion,
the composition is further from NSE.  These arguments indicate pointedly
that the quasi-equilibrium behavior of silicon burning is strongly
influenced by the state of neutronization of the material.

Another illustration of the impact of neutronization, which is not easily
discernible from Figures 1b and d, but is very apparent in Figures 1c and e,
is the changing constituents of the QSE groups with declining $Y_e$.  For
$Y_e=.498$ the distribution is well described as 2 groups, separated by a
narrow fringe of nuclei with $A=45$ up to Ti and then $N=23$.  The
similarity of the value of \rqe\ for those 4 species with $N=23$
(\nuc{Ti}{45}, \nuc{V}{46}, \nuc{Cr}{47} and \nuc{Mn}{48}) implies that the
reactions linking them are close to being balanced.  Comparison of Figure 1c
and 1e reveals a much more complicated pattern.  In addition to the vastly
different scale, there are many more nuclei in the boundary region.  Instead
of only a pair of Ca isotopes (\nuc{Ca}{44} and \nuc{Ca}{45}), the isotopes
from \nuc{Ca}{42} to \nuc{Ca}{49} are all significantly displaced from the
QSE groups.  A similar situation exists for Sc and Ti.  Those isotopes of
Ti, V, Cr, and Mn which were in the silicon group or the boundary for $Y_e
=.498$ are significantly closer to the upper group.  The nuclei with the
largest A which are within a factor of 2 of quasi-equilibrium with the
silicon group are \nuc{Sc}{42} and \nuc{K}{46}.  The $N=23$ nuclei are
much closer to the upper group, with \nuc{Cr}{47}, for example, having an
abundance 20\% larger than its iron peak group QSE value as opposed to
being a factor of 20 underabundant for QSE with the silicon group. From Ca
on, the $N=22$ and $N=21$ elements occupy the boundary.  The vertical
structure of these lines of constant $N$ implies that the reactions linking
them, pairs of $(p,\gamma),\ (\gamma,p)$ reactions, are not in equilibrium.
Similarly the spacing of isotopes of Ca, for example, indicates an
unbalanced flow by neutron capture upward from the silicon group.  We will
examine in detail the reaction flows between the groups in \S 6.  One
more interesting note is that, in contrast to the behavior of other
nuclei, the neutron-rich isotopes of K (\nuc{K}{44}, \nuc{K}{45}, and
\nuc{K}{46}), although actually further from their silicon group QSE
abundances, seem to
be closer to membership in the silicon group.  Table 2 shows quantitatively
the movement of nuclei upward from the silicon group, with those abundances
below the minimum resolution of the network ($10^{-25}$) represented by
$\bullet$.  Comparison of the second and third columns shows clearly the
migration
of the boundary toward lower Z and A.  The abundance of a nucleus like
\nuc{Sc}{44}, which for $Y_e=.498$, is within .5\% of its QSE abundance
and thus clearly a member of the silicon group, is, at $Y_e=.46$,  better
predicted by QSE with the iron peak group.  For $Y_e=.46$ and X(Si group)
$\sim .9$, the silicon QSE abundance of \nuc{Sc}{44} is over 11 times its
network abundance, while its iron peak group QSE abundance is slightly less
than half its network abundance.  And \nuc{Sc}{44} is clearly not an
isolated example.  Thus the group boundaries defined by earlier work were
clearly dependent on the narrow range of parameters examined.  We will
therefore investigate how our broader range in parameter space affects
these boundaries.

\topfig{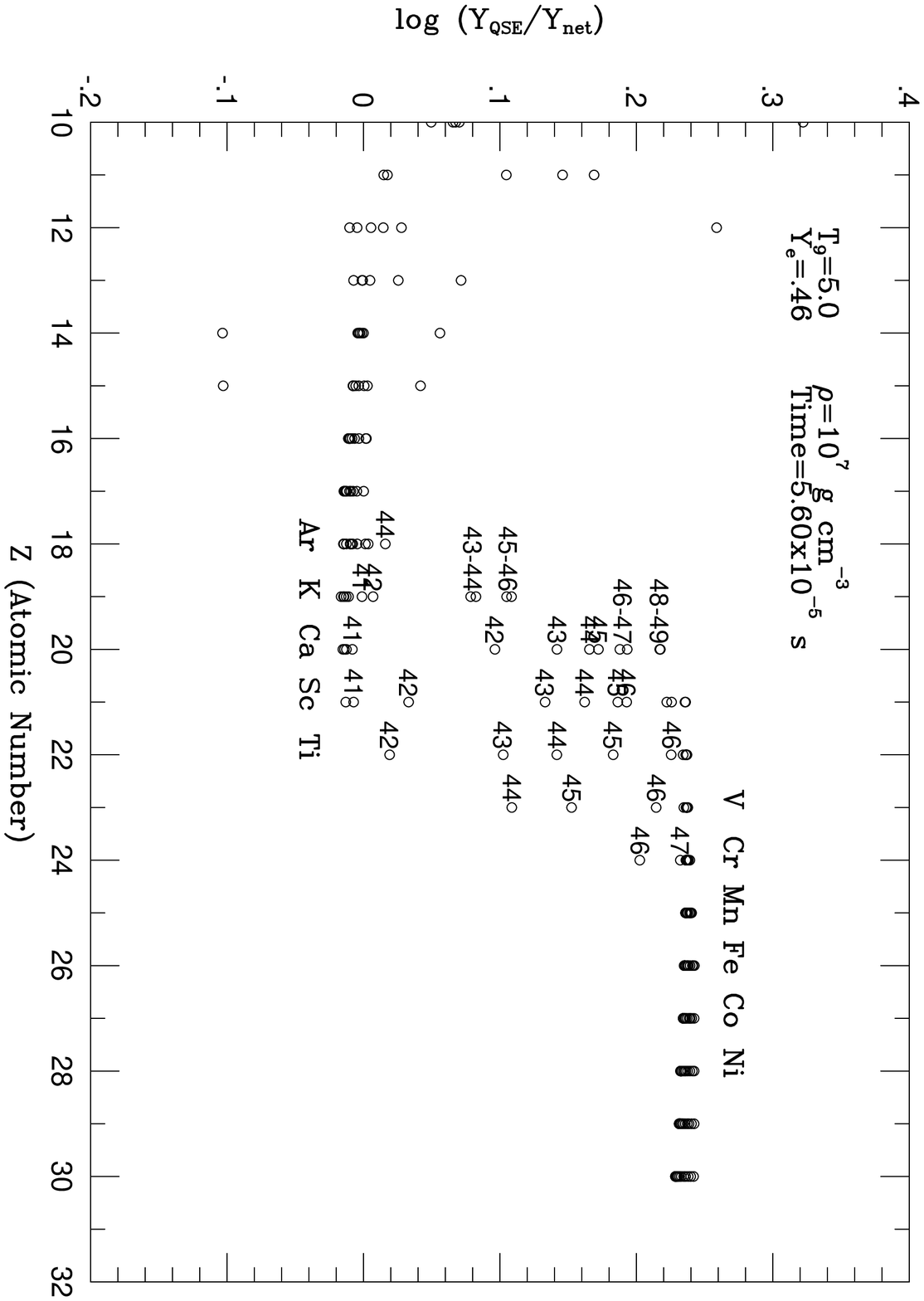}{Figure 1f}{Detail of the comparison of the silicon
quasi-equilibrium abundance to the network abundance, as a function of Z
for $\t9=5.0$, $\rho=10^7 \gcc$, and $Y_e=.46$, with X(Si group)=.52.}

One possibility is that the differences in the boundaries are due to the
fact that the lower $Y_e$ cases are further from equilibrium. This would
result in a temporal variation in the boundary with the boundary discussed
above for lower $Y_e$ approaching that of the $Y_e=.498$ case at
X(Si group) $\sim .9$, as the lower $Y_e$ cases reach a similar degree of
convergence.  We find that this is not the case.  Consider the situations
for X(Si group) $\sim .5$.  This corresponds to elapsed times of $2.1 \times
10^{-3}$ s and $5.6 \times 10^{-5}$ s, for $Y_e=.498$ and $.46$,
respectively, with $\t9=5.0$ and $\rho= 10^7 \gcc$.  As the fourth column of
Table 2 demonstrates, for $Y_e=.498$ the abundances of all of the species
which composed the iron peak QSE group are within 2-3\% of their silicon
quasi-equilibrium abundances.  Figure 1f illustrates the comparison of the
network abundances with those predicted by quasi-equilibrium for $Y_e=.46$
and X(Si group) $\sim .5$.  The members of the iron peak group are within
50\% of their silicon quasi-equilibrium abundances. Comparison of Figures
1c, 1e, and 1f reveals nonetheless that the pattern for $Y_e=.46$, X(Si group)
$\sim .5$ is more like that of $Y_e=.46$, X(Si group) $\sim .9$, even although
the iron peak group has converged much closer to quasi-equilibrium with Si.
Although the increased resolution reveals much more detail among the upper
members of the boundary, and the isotopes of K continue their slow convergence
toward the silicon group, a greater degree of silicon exhaustion leaves
the pattern essentially unchanged.  While the break which leaves \nuc{Sc}{42}
and \nuc{Ti}{42}, newly emerged from below the minimum mass threshold of
the network, close to the silicon group is made more prominent by the
approach of these nuclei to the silicon group, the constituents of the
QSE groups are essentially unchanged.  Furthermore, the dissimilarity of
Figures 1c and 1f, both corresponding to elapsed times of $5.6 \times
10^{-5}$ s, removes any doubt that the differences in the group
boundaries are simply a time dependency.  Thus we can conclude that the
differences in the boundary between the QSE groups is a genuine effect of
the change in $Y_e$ and not due to temporal variation or closeness to
equilibrium.  As we will discuss in \S 7, with the $Y_e=.46$ case
having exhausted 5 times as much fuel as the similar case with $Y_e=.498$
in the same elapsed time, these differences in QSE group structure have
broader implications than simply the abundances of these boundary nuclei,
resulting in large variations in the rate of energy generation.

\topfig{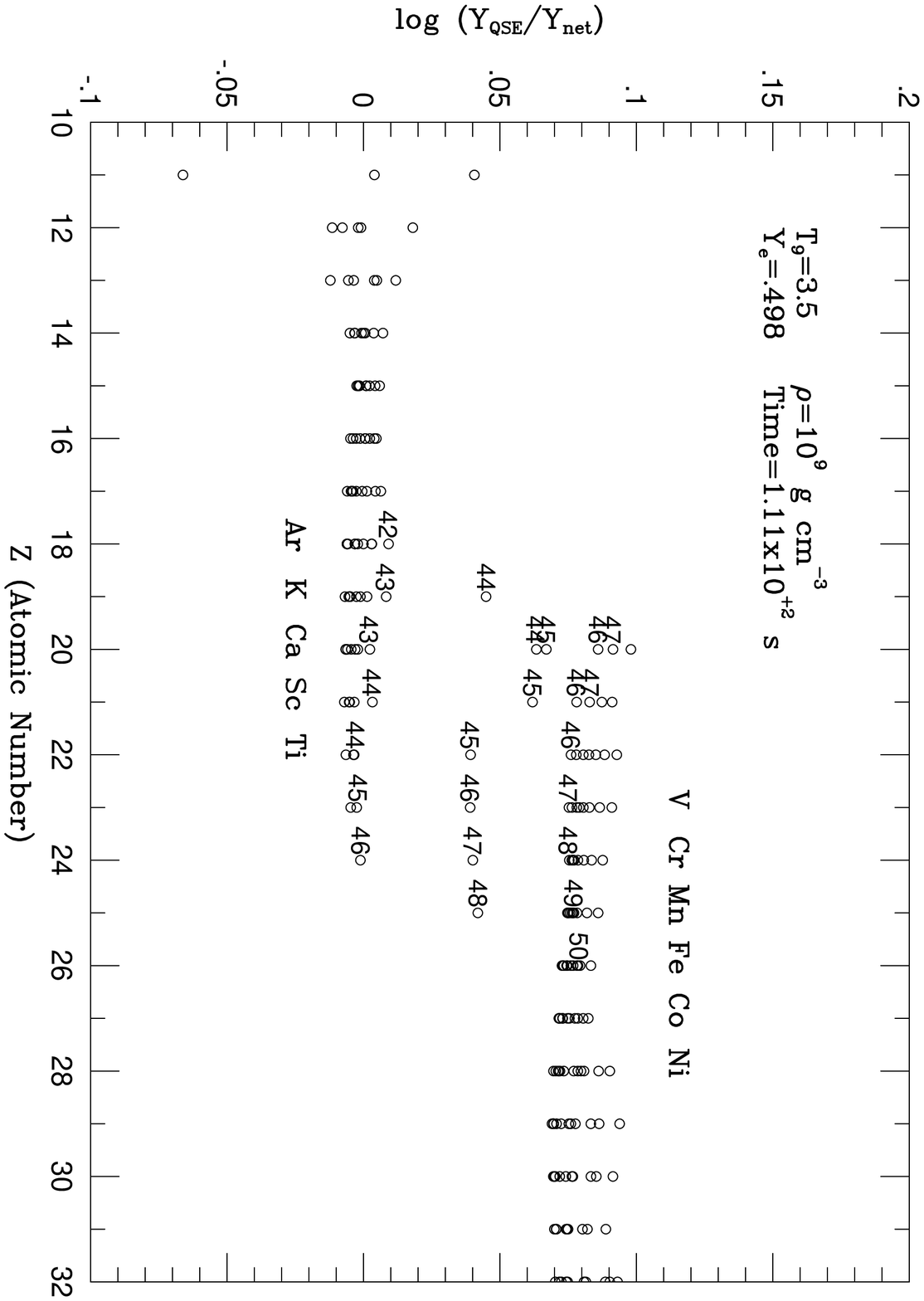}{Figure 1g}{Detail of the comparison of the silicon
quasi-equilibrium abundance to the network abundance, as a function of Z
for $\t9=3.5$, $\rho=10^9 \gcc$, and $Y_e=.498$, with X(Si group)=.90.}

Having shown that the quasi-equilibrium behavior during silicon burning
is dependent on $Y_e$ and the degree of silicon exhaustion, we now turn
to our two remaining parameters.  The conditions considered so far, high
temperature and low density, are the most like those considered by WAC.
They are also the conditions under which the inclusion of screening is the
least critical.  For conditions like those of Figure 1g, $\t9= 3.5$ and
$\rho=10^9 \gcc$, the inclusion of screening in the quasi-equilibrium is
very important.  Under these conditions the screening portion of Equation (10)
can be as large as $10^6$, much larger than the factor of 5 which is the
maximum for $\t9=5.0$ and $\rho =10^7 \gcc$, but much smaller than the
factor of order $10^{13}$ which is possible for $\t9=3.5$ and $\rho =
10^{10} \gcc$.  Comparison of Figure 1c and 1g supports the inclusion of this
screening term.  With $Y_e=.498$ and X(Si group) = .9, the conditions of
Figure 1g differ from the conditions of Figure 1c only in temperature and
density.
Since the screening term is dependent on Z and not on N, its omission would
result in a separate horizontal line for each element, with Ge being
displaced from Si by a factor of almost $10^6$.  The similarity of Figure 1c
and 1g argues strongly that the inclusion of screening is correct.  This
similarity also argues that changes of temperature and density do not
strongly affect the structure of the QSE groups, although the elapsed time,
111 s, is considerably different, reflecting the much slower reaction rates
at low temperatures.  The most striking difference between these cases is
the larger values of \rqe\ among the lighter nuclei.  The much greater value
of $r_{QSE}(p)$ (16.92 compared with 7.32) for this low temperature/high
density case is consistent with the argument that $r_{QSE}(p)$  reflects
how far from NSE the distribution is.  Since the NSE mass fraction of the
silicon group for these conditions is less than $2\times10^{-5}$, the
smallest of the cases shown in this paper, one would expect $r_{QSE}(p)$
for this degree of silicon exhaustion to be the largest in this case.

Clearly changes in temperature and density affect the reaction rates and
hence timescales, but how does this affect adherence to the quasi-equilibrium
distributions?  Comparison of Figure 1g with Figure 1c shows that, for $Y_e=
.498$, the variation of temperature and density has little effect on the
QSE group behavior.  Once again, for X(Si group) = .9, the members of the
iron peak group are within 25\% of their silicon QSE group abundances.  The
spread within the iron peak group is noticeably larger, but still less than
approximately 5\%.  Further investigation reveals that this greater disorder,
i.e. larger variation within the QSE groups, is solely a function of
decreasing temperature.  Most importantly, although there is some difference
in relative placement, the nuclei intermediate between the two QSE groups
are the same.  Examination of cases with lower $Y_e$ supports the finding
that the variation of temperature and density does not affect the grouping
of nuclei, although it can radically alter relative abundances among the
collection intermediate to the QSE groups.

While previous authors have shown the value of quasi-equilibrium as a
key to understanding the process of silicon burning, the work presented in
this section implies that this understanding is incomplete.  As one might
expect, variation of the temperature or density can produce large differences
in the  behavior of silicon burning, most notably in the timescale.  The
important result of this research is the importance of neutronization in the
quasi-equilibrium behavior and timescale.  For similar degrees of silicon
exhaustion, a decrease in $Y_e$ results in a drastic increase in the QSE
abundances of the iron peak group.  As a result, for similar degrees of
silicon exhaustion the iron peak group is much farther from
quasi-equilibrium and the abundances of these nuclei a much smaller fraction
of their silicon QSE abundance.  This delays the merging of the two QSE
groups into a single group spanning the elements from Mg to Ge.  While
high $Y_e$ cases reach this merged group
stage with approximately 90\% of the mass remaining within what was the
silicon group, for lower $Y_e$ this mass fraction is much smaller.
For $Y_e=.498$, for both $\t9=5.0$, $\rho=10^7 \gcc$ and $\t9=3.5$,
$\rho=10^9 \gcc$, a single QSE group (within 10\%) has formed by X(Si group)
= .85.  For $Y_e=.48$ the iron peak nuclei do not reach 90\% of their silicon
QSE group abundance until 50\% or more of the mass in the silicon group is
exhausted.  At $Y_e=.46$ for $\t9=5.0$, $\rho=10^7 \gcc$ and $\t9=3.5$,
$\rho=10^9 \gcc$ the iron peak group reaches 90\% of its silicon QSE
abundance with silicon group mass fractions of .25 and .3, respectively.
Although variations of temperature and density clearly effect great changes
in the timescales of silicon burning, it is the degree of neutronization
which most strongly affects the quasi-equilibrium behavior.  This behavior,
previously only hinted at by TA, is very important to modeling the
mechanism of silicon burning, since low $Y_e$ results in the persistence of
2 separate QSE groups until much later than previously thought. Attempts,
like that of BCF, to model silicon burning by using the quasi-equilibrium
abundances and calculating the loss from a single QSE group are, as a result,
oversimplified.  Our results show that such a method would be inaccurate
unless this separation between the upper and lower QSE groups is included,
as the gap can persist through significant silicon consumption, causing
large miscalculations in the iron peak abundances.

Furthermore, for smaller $Y_e$, the upper boundary of the silicon group
moves downward, while at the same time the nuclei which formerly comprised
the boundary layer have, for the most part, joined the iron peak group.
This movement of the boundary has a number of effects, the most important
being that it mandates there be different reactions which link the QSE
groups as
$Y_e$ varies.  It is hard to see how the reaction $\rm ^{45}Sc(p,\gamma)
^{46}Ti$ can be the link between QSE groups for $Y_e=.46$, since \nuc{Sc}{45}
is virtually a member of the iron peak group under these conditions.  The
similarity in \rqe\ for \nuc{Sc}{45} and \nuc{Ti}{46} for $Y_e=.46$ implies
that the $(p,\gamma),(\gamma,p)$ reaction pair linking them are almost
balanced, making this an unlikely flow path.  In \S 6 we will
investigate the question of how the linking reactions vary as conditions
change, but first we will use the convenience of quasi-equilibrium to
follow the nuclear abundances produced during silicon burning.

\bigskip

\sect{5. Species Formation during Silicon Burning }

\medskip

\topfig{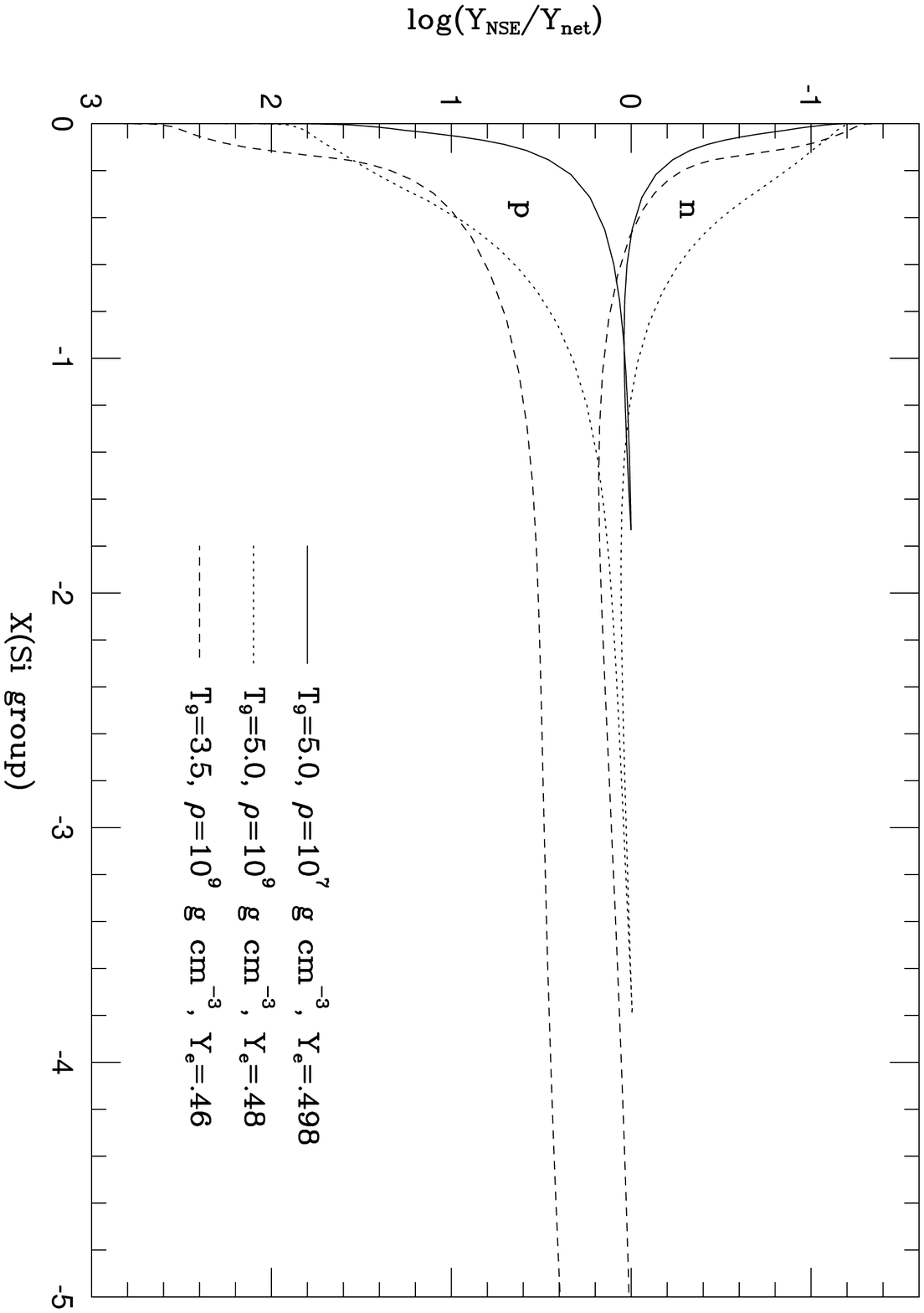}{Figure 2}{Convergence of the free nucleon abundances
to equilibrium as a function of degree of silicon exhaustion.  The cases
portrayed are $\t9=5.0$, $\rm \rho=10^7 \gcc$, and  $Y_e=.498$ (solid
lines), $\t9=5.0$, $\rm \rho=10^9 \gcc$, and $Y_e=.48$ (dotted lines),
and $\t9=3.5$, $\rm \rho=10^9 \gcc $, and $Y_e=.46$ (dashed lines).}

\topfig{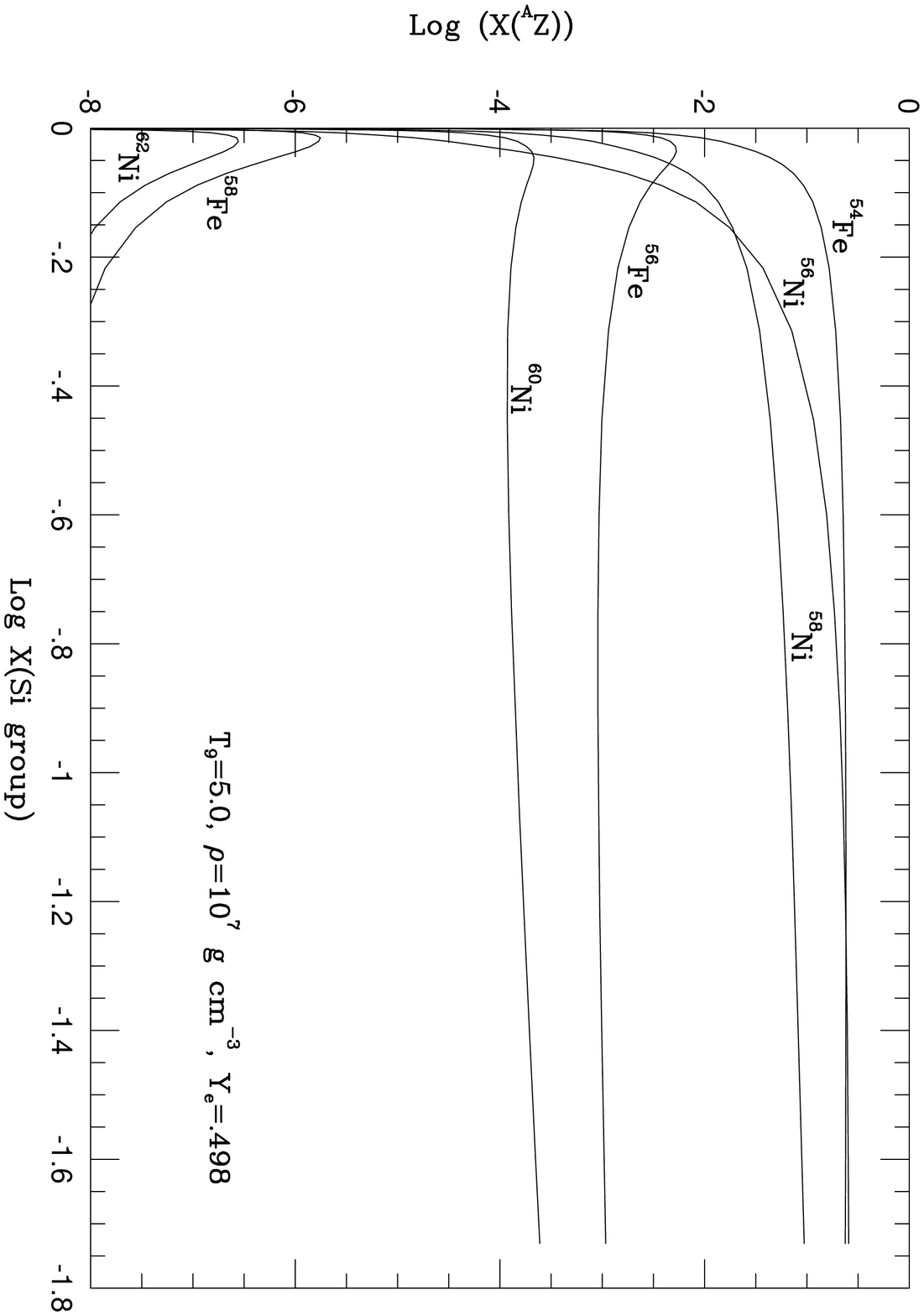}{Figure 3a}{Evolution of the mass fractions of the
dominant members of the iron peak group as a function of the degree of
silicon exhaustion for $\t9=5.0$, $\rho=10^7 \gcc$, and $Y_e=.498$.
Compare with the solid line in Figure 2.}

\topfig{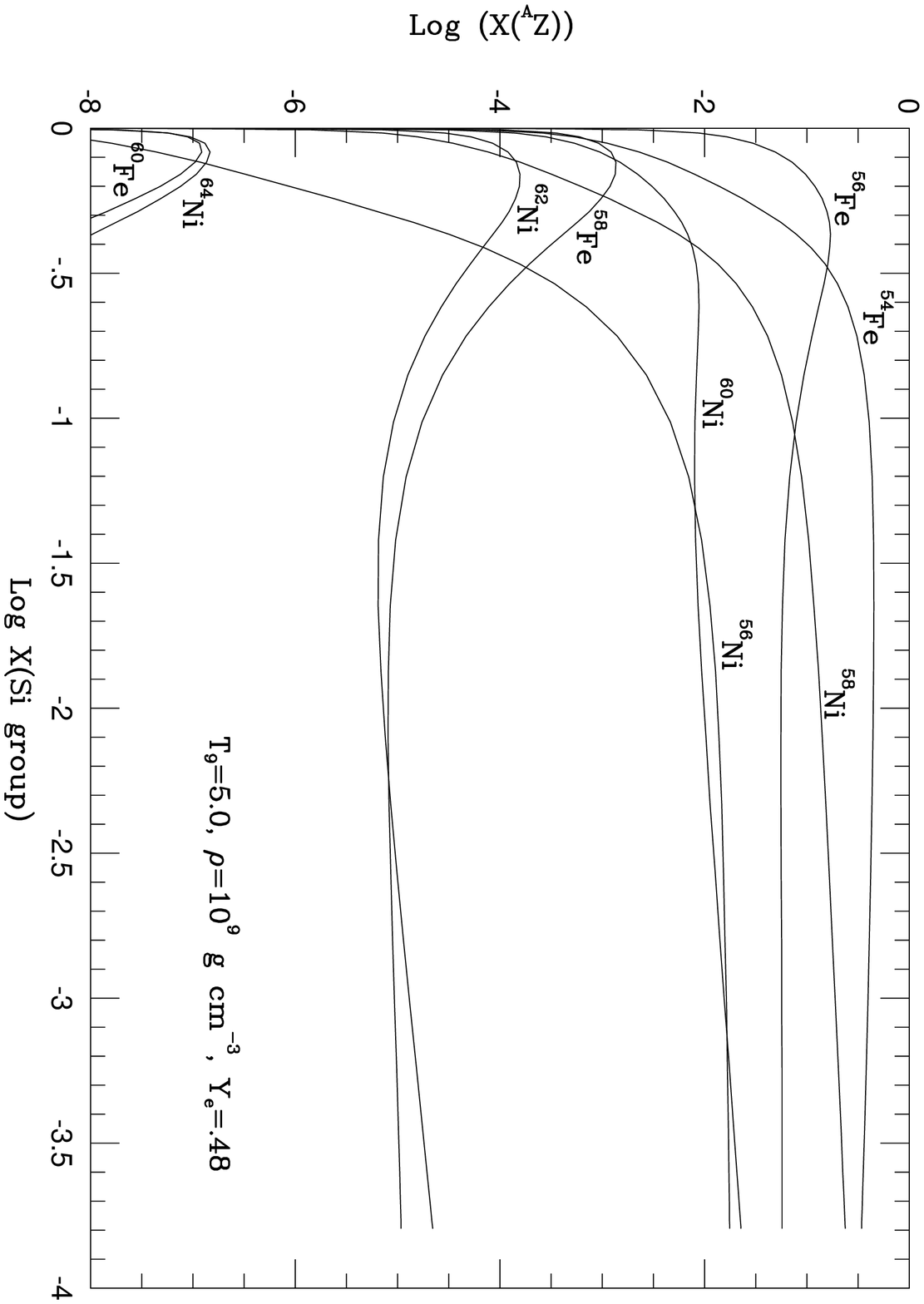}{Figure 3b}{Evolution of the mass fractions of the
dominant members of the iron peak group as a function of the degree of
silicon exhaustion for $\t9=5.0$, $\rho=10^9 \gcc$, and $Y_e=.48$.
Compare with the dotted line in Figure 2.}

\topfig{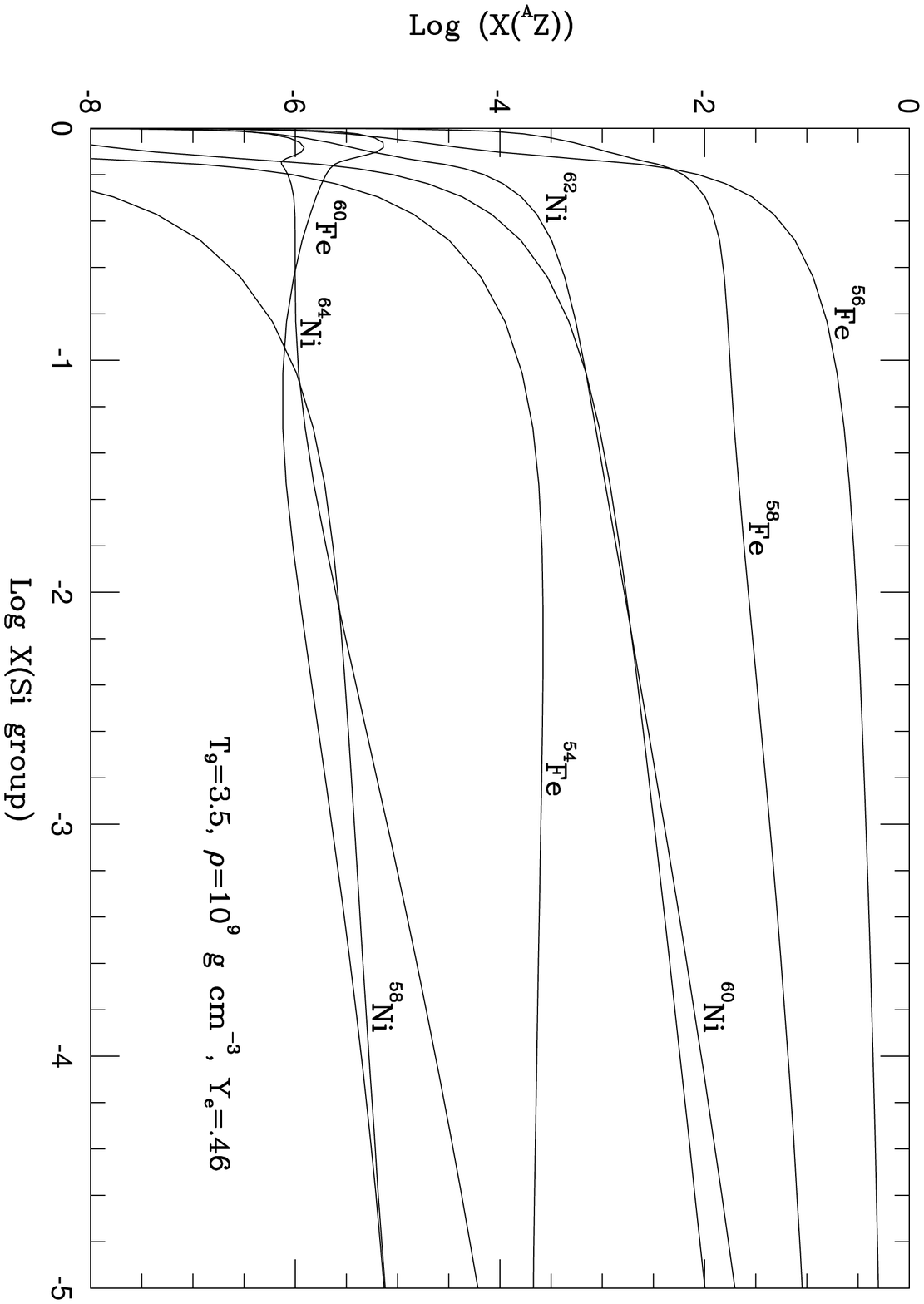}{Figure 3c}{Evolution of the mass fractions of the
dominant members of the iron peak group as a function of the degree of
silicon exhaustion for $\t9=3.5$, $\rho=10^9 \gcc$, and $Y_e=.46$.
Compare with the dashed line in Figure 2.}

\topfig{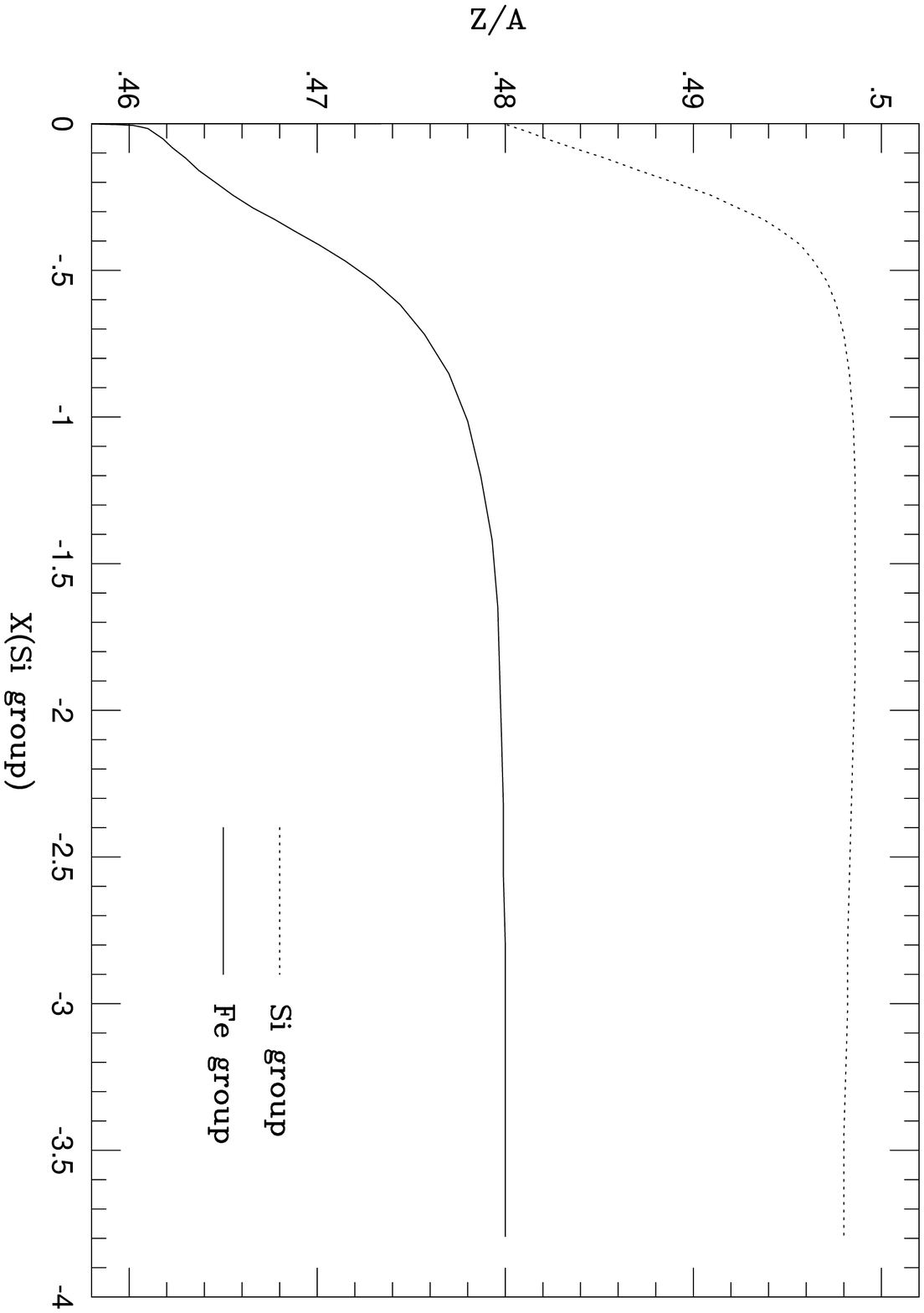}{Figure 4}{Evolution of the average $Z/A$ for the iron
peak group (solid line) and the silicon group (dotted line) with the degree
of silicon exhaustion for $\t9=5.0$, $\rho=10^9 \gcc$, and $Y_e=.48$.
Compare with Figure 3b.}

The applicability of quasi-equilibrium greatly simplifies the study of the
nuclei produced by silicon burning.  Instead of following the intricacies
of the production of 300 nuclei, much can be learned by examining four:
free protons,
free neutrons, and one member of each of the QSE groups.  While the focal
members of the QSE groups map the spacing between the groups, it is the
abundances of the free particles which control the relative abundances within
the QSE groups.  The gap between QSE groups, seen in Figures 1a-g, which
diminishes as silicon approaches exhaustion, reflects the evolution of the
focal elements of the QSE groups.  Figure 2 shows samples of the evolution
of the free proton and free neutron mass fractions as a function of time.
Although the actual abundances vary by many orders of magnitude, comparison
with the appropriate NSE abundances reveals a similarity in pattern.  For
each of the cases portrayed in Figure 2, $\t9=5.0$, $\rho=10^7 \gcc$, and
$Y_e= .498$ (solid lines), $\t9=5.0$, $\rho=10^9 \gcc$, and $Y_e=.48$
(dotted lines), and $\t9=3.5$, $\rho=10^9 \gcc$, and
$Y_e=.46$ (dashed lines), the pattern is similar.  Although the rates of
convergence vary greatly, in each case the network abundance of free
protons is initially more than an order of magnitude smaller than its
equilibrium abundance and converges gradually to equilibrium as the mass
fraction of the silicon group approaches its equilibrium value.  The behavior
of the neutrons is slightly more complicated.  These abundances begin more
than an order of magnitude larger than their equilibrium abundance,
and converge down toward equilibrium, briefly becoming underabundant before
reaching equilibrium.  As with the protons, the details of this behavior; the
rate of convergence, the maximal and minimal abundances, and the
equilibrium mass fractions, vary widely.  But this simple pattern is
maintained.  What effect do such variations have?  The relative abundance
of two isotopes of a given element are related by

$$
{Y_{QSE}(^{A'}Z) \over Y_{QSE}(^{A}Z)} \propto Y_n^{(A'-A)}\ .
\eqno(14)
$$
Thus an overly large neutron abundance tips the the production of an element
toward more neutron-rich isotopes.  Clearly this occurs because the increased
neutron abundance increases the rate of neutron capture.  Similarly the
dearth of free protons will favor the enhancement of the members of a QSE
group with lower atomic number.  Figure 3a shows the evolution of the abundance
for key members of the iron peak group as the degree of silicon exhaustion
increases for $\t9=5.0$, $\rho=10^7 \gcc $, and  $Y_e=.498$.  For comparison,
the corresponding evolution of the free nucleons is shown as the solid line
in Figure 2.  At early time, (X(Si group $\sim .9)$), the iron peak group is
dominated by \nuc{Fe}{54} and \nuc{Fe}{56}.  As X(Si group) decreases, the
abundance of \nuc{Fe}{54} levels out while \nuc{Fe}{56} actually decreases
in abundance as the overabundance of free neutrons drops.  Meanwhile, as
the free proton abundance rises toward equilibrium, \nuc{Ni}{56} and
\nuc{Ni}{58} rise to prominence.  This results in temporal variations in
the relative abundances of the nuclei which dominate the distribution.  It
is not until very late that \nuc{Ni}{56}, which is the most abundant
nucleus in the NSE distribution for these conditions, becomes most abundant
in the network distribution.  The relative abundances of \nuc{Fe}{54},
\nuc{Fe}{56}, and \nuc{Ni}{58} at X(Si group) $\sim .9$ are similar to
those found in the NSE distribution for $Y_e \sim .475$.  At X(Si group)
$\sim .5$, the relative abundances of \nuc{Fe}{54}, \nuc{Ni}{56}, and
\nuc{Ni}{58} are comparable to those found in NSE for $Y_e \sim .49$.  In
another similarity to the NSE distribution, nuclei with similar Z/A react
in concert.  For example the abundances of \nuc{Fe}{56} and \nuc{Ni}{60}
react to a decreasing silicon group mass fraction by rising sharply, peaking
around X(Si group) $\sim .9$, and then falling off.  For \nuc{Fe}{56} this
shallow decline turns around for X(Si group) $\sim .1$, coincident with the
minimum of the free neutron distribution.  As the free neutrons converge
back toward their equilibrium value the abundance of \nuc{Fe}{56} also
increases slightly.  For \nuc{Ni}{60} this decline after maximum is turned
around for X(Si group)$ \sim .4$ as the increase in the free proton fraction
overcomes the decline in the free neutron fraction.  Even after the
abundance of \nuc{Fe}{56} begins to increase again, the rate of increase in
the abundance of \nuc{Ni}{60}, aided by its greater dependence on the free
proton abundance, is greater.  Other such pairings of nuclei observed in
the NSE distributions, like \nuc{Fe}{54} and \nuc{Ni}{58} or \nuc{Fe}{58}
and \nuc{Ni}{62}, also behave in concert.

This excessive neutron-richness of the iron peak group at early times is also
apparent for other thermodynamic conditions.   For $\t9=5.0$, $ \rho=10^9
\gcc$, and $Y_e=.48$, shown in Figure 3b (and as the dotted line in Figure 2),
the relative distribution of \nuc{Fe}{54}, \nuc{Fe}{56}, \nuc{Ni}{58}, and
\nuc{Ni}{60}, at X(Si group) $\sim .5$, is more in keeping with equilibrium
distributions with $Y_e \sim .47$.  Once again, relatively neutron-rich
isotopes, in this case \nuc{Fe}{58}, \nuc{Ni}{60}, and \nuc{Ni}{62}, peak
early, exhibit a local minimum in concert with a similar local minimum in
the abundance of free neutrons, and then rise toward equilibrium.
Here again, the increasing abundance of free protons causes the Ni isotopes
to rise more quickly than their Fe brethren.  Thus the behavior of this
example is consistent with the previous case, though even more neutron-rich,
as one might expect from this lower $Y_e$ case.  For $\t9=3.5$, $\rho=10^9
\gcc$, and $Y_e=.46$, shown in Figure 3c (and as the dashed line in Figure 2),
the early behavior is even more neutron-rich.  For X(Si group) $\sim .5$,
the relative distribution of \nuc{Fe}{56}, \nuc{Fe}{58}, \nuc{Ni}{60}, and
\nuc{Ni}{62} reflects a $Y_e < .45$.  Even for material which is highly
neutronized, the iron peak group is especially neutron-rich at early times.

Over this wide range of conditions, the abundance distributions within
the iron peak group exhibit a number of similarities.  First, the relative
abundances are very reflective of the changes in the free nucleon
abundances.  Second, at early times, particularly when the silicon group still
dominates the mass fraction, the iron peak group exhibits a higher degree of
neutronization than the abundance distribution as a whole.  The solid
line in Figure 4, which plots the average Z/A of the iron peak group as the
degree of silicon exhaustion increases, for $\t9=5.0$,\ $ \rho=10^9 \gcc$,
and $Y_e=.48$ (the same conditions as in Figure 3b), shows that this excessive
neutronization is not an illusion of the most abundant nuclei.  For the
iron peak group to show excessive neutronization there must be a group of
nuclei which are neutron poor relative to the global $Y_e$.  This is
required by the constancy of total neutron number in the absence of weak
interactions.  Since the mass fractions of the light particles are very
small, the likely source of these excess neutrons is the silicon group.
The dotted line in Figure 4, which shows the average Z/A for the silicon
group as it is depleted, vindicates this assertion.  Clearly, in
equilibrium, the neutron-richness of the silicon group is much lower than
the global neutronization indicates.  While in small part this is due to
the absence of the most neutron-rich isotopes of Ca, Sc, and Ti from the
silicon group, the major cause is the relative binding energies within the
silicon group.

\topinsert
\centerline{Table 3 Average Z/A as a function of Element}
\nobreak
\centerline{in NSE at $\t9=5.0$, $\rho=10^9 \gcc$, and
$Y_e=.48$}
\nobreak
\medskip
\moveright .85in
\hbox { \hsize=.75 in
\valign{\thinrule \smallskip \thinrule \vskip 9pt plus 2pt minus 2pt
\centerline{#}\strut \vskip -14pt plus 2pt minus 2pt \thinrule
\smallskip &#\strut&#\strut&#\strut&#\strut&#\strut&#\strut&#\strut
&#\strut&#\strut&#\strut&#\strut \smallskip \thinrule \cr
Element&n&H&He&Li&Be&B&C&N&O&F&Ne\cr
Z/A&0&1&.500&.500&.548&.464&.500&.500&.500&.528&.500 \cr
\noalign{\vrule}
Element&Na&Mg&Al&Si&P&S&Cl&Ar&K&Ca&Sc  \cr
Z/A&.479&.500&.482&.500&.485&.498&.486&.496&.487&.499&.481  \cr
\noalign{\vrule}
Element&Ti&V&Cr&Mn&Fe&Co&Ni&Cu&Zn&Ga&Ge \cr
Z/A&.477&.473&.474&.473&.478&.481&.483&.480&.480&.478&.477  \cr
}}
\endinsert

As Table 3 indicates, under conditions of nuclear statistical equilibrium,
even with $Y_e=.48$, elements which compose the silicon group are dominated
by their $N=Z$ isotopes.  This is particularly true for the dominant elements,
Si, S, Ar, and Ca, where the $N=Z$ isotope is also an $\alpha$-nucleus.
As for nuclei with Z between 2 and 10, the valley of maximal binding energy
is narrow for elements  within the silicon group.  With the high temperatures
of silicon burning, and the resultant photon distribution, the more
neutron-rich isotopes within the silicon group, with binding energies
smaller than their $N=Z$ neighbors, are more quickly photodissociated,
freeing large numbers of free neutrons.  These neutrons are captured by
nuclei in the iron peak group, where the binding energy valley has curved
to the neutron-rich side and widened, and thus the neutron-rich isotopes
are more durable.  Thus it is the tendency toward $N=Z$ nuclei
within the silicon group which fuels the extreme neutron excesses in the
iron peak group during silicon burning.  This provides a mechanism for
producing small quantities of more neutron-rich iron peak group nuclei,
characteristic of a larger $\eta$.  As Figure 4 shows for $\t9=5.0$, $\rho=
10^9 \gcc$, and $Y_e=.48$, if only 10\% of the silicon was exhausted, the
effective $\eta$ within the iron peak group would be .08, twice its
equilibrium value.  Thus, by such a mechanism it is possible to produce
neutron-rich members of the iron peak group from material which has never
experienced the high densities necessary to produce such a large global
neutron excess.  By tying up the bulk of the nucleons in N=Z nuclei, and
thereby amplifying the effective neutronization among the iron peak nuclei,
incomplete silicon burning faintly echoes the emerging understanding
of the r-process (see Woosley \& Hoffman 1992).
This neutron enrichment of the iron peak group is clearly a consideration
when one is modeling objects where explosive burning of silicon, and hence
incomplete silicon burning, are important to the elemental production.

\bigskip

\sect{6. Reaction Flows}

\medskip

With the majority of the mass concentrated within the two QSE groups, the
process of silicon burning is dominated by these groups of nuclei.  The
previous section indicates that, with knowledge of the free nucleon
abundances, it is possible to ignore the reactions within the
quasi-equilibrium groups, since these reactions simply reflect the changes in
abundances required for nuclei to remain in quasi-equilibrium under changes
in the free-proton and free-neutron fractions.  Thus, instead of studying the
thousands of reactions which make up silicon burning, we can concentrate
on those reactions which enter or leave the QSE groups.  Ideally, these few
important reactions would be the same for all physical conditions.  However,
in \S 5 we demonstrated that small changes in neutronization result
in much larger changes in the free-neutron abundance, thus favoring neutron
capture and more neutron-rich nuclei.  This enhanced importance of neutron
captures affects the boundary region and results in changes in the location
of the boundaries of the silicon and iron peak QSE groups, as we discussed
in \S 4.  As a result we expect that the dominant reactions linking these
groups must change as a function of $Y_e$.  However, before we examine the
complicated behavior that joins the silicon and iron peak groups, let us
examine the lower boundary condition for the QSE groups.

\bigskip

\subsect{6.1. Downward Flows from Silicon}

\medskip

While we have heretofore concentrated on the upper edge of the silicon
quasi-equilibrium group, the flow from the lower edge is also important to
the destruction of silicon and the other constituents of the silicon group.
Previous authors have shown that these reactions govern the rate at which
silicon is destroyed.  Examination of Figure 1a, 1b, or 1d shows that the
network abundances of the nuclei below $A \sim 24$ are orders of magnitude
less than their silicon QSE abundances, with the degree of underabundance
increasing as Z decreases.  Further, there is little evidence of
quasi-equilibrium behavior among these nuclei.  Figure 1c and 1e reveal that
the principal isotopes of Mg and Al are close to their silicon
QSE abundances.  Within a margin of $\pm 10 \%$, all of the isotopes
of Mg and Al, except \nuc{Mg}{21} and perhaps \nuc{Mg}{22}, are
members of the silicon group.  This closeness to quasi-equilibrium does not
hold for Na and Ne.  Thus the bottom edge of the silicon group is Mg.  If the
principal isotopes of Mg and Al are in QSE with \nuc{Si}{28}, then the
rate of photodissociation of \nuc{Si}{28}, either by $\rm ^{28}Si
(\gamma,\alpha)^{24}Mg$ or by $\rm ^{28}Si(\gamma,p)^{27}Al$, is
essentially balanced by the reverse capture reactions.  Thus it is the net
flow from this bottom edge which governs the downward flow from silicon.
BCF contended that, with \nuc{Mg}{24} in quasi-equilibrium with
\nuc{Si}{28}, the rate of destruction of \nuc{Si}{28} was governed by the
photodissociation of \nuc{Mg}{24}, with $\rm ^{24}Mg(\gamma,\alpha)
^{20}Ne$ dominating $\rm ^{24}Mg(\gamma,p)^{23}Na$.  This was
supported by WAC, who found that these reactions did indeed dominate
the downward flow.  However, the reaction network used by WAC was
very narrow at this lower boundary of the silicon group with four isotopes
of Mg, and only single isotopes of Al, Na, and Ne.  Further, because of
doubt over the accuracy of the then best available reaction rates, a
number of reactions which those authors described as potentially
important, like $\rm ^{24}Mg(p,\alpha)^{21}Na$ and $\rm ^{24}Mg
(n,\alpha)^{21}Ne$, were excluded from their reaction network.  Thus
we now reexamine the reactions which cross this lower boundary to see
the importance of our improved reaction network in addition to examining
the effects of the range of $Y_e$.

\topfig{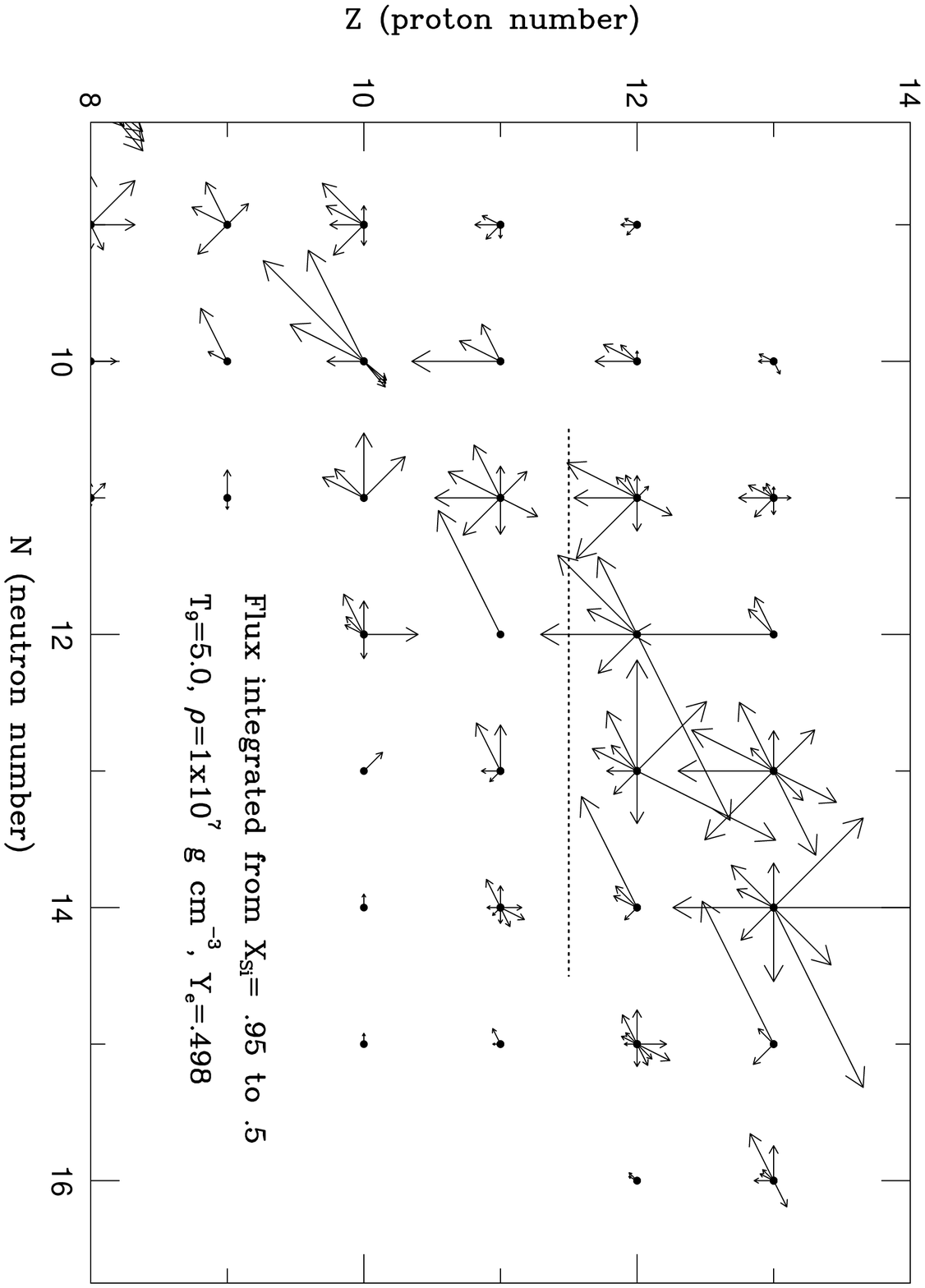}{Figure 5a}{Vector field representing the integrated
reaction fluxes from X(Si group) =.95 to .5, along the lower boundary of
the silicon group, for $\t9=5.0$, $\rho=10^7 \gcc$, and $Y_e=.498$.
The vector magnitudes have a logarithmic dependence on the size of
the integrated flux.}

\topfig{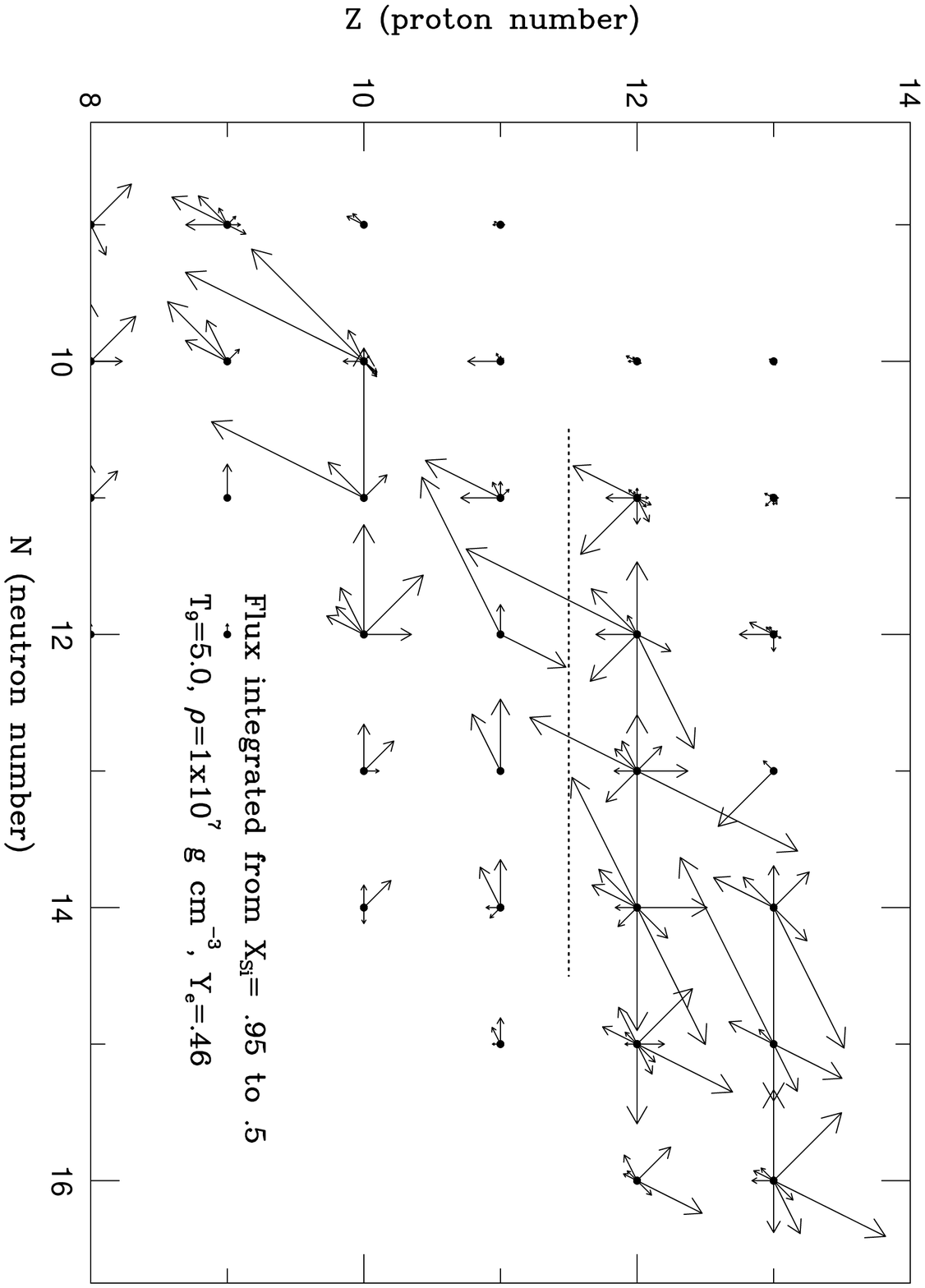}{Figure 5b}{Vector field representing the integrated
reaction fluxes from X(Si group) =.95 to .5, along the lower boundary of
the silicon group, for $\t9=5.0$, $\rho=10^7 \gcc$, and $Y_e=.46$.
The vector magnitudes have a logarithmic dependence on the size of
the integrated flux.}

\topfig{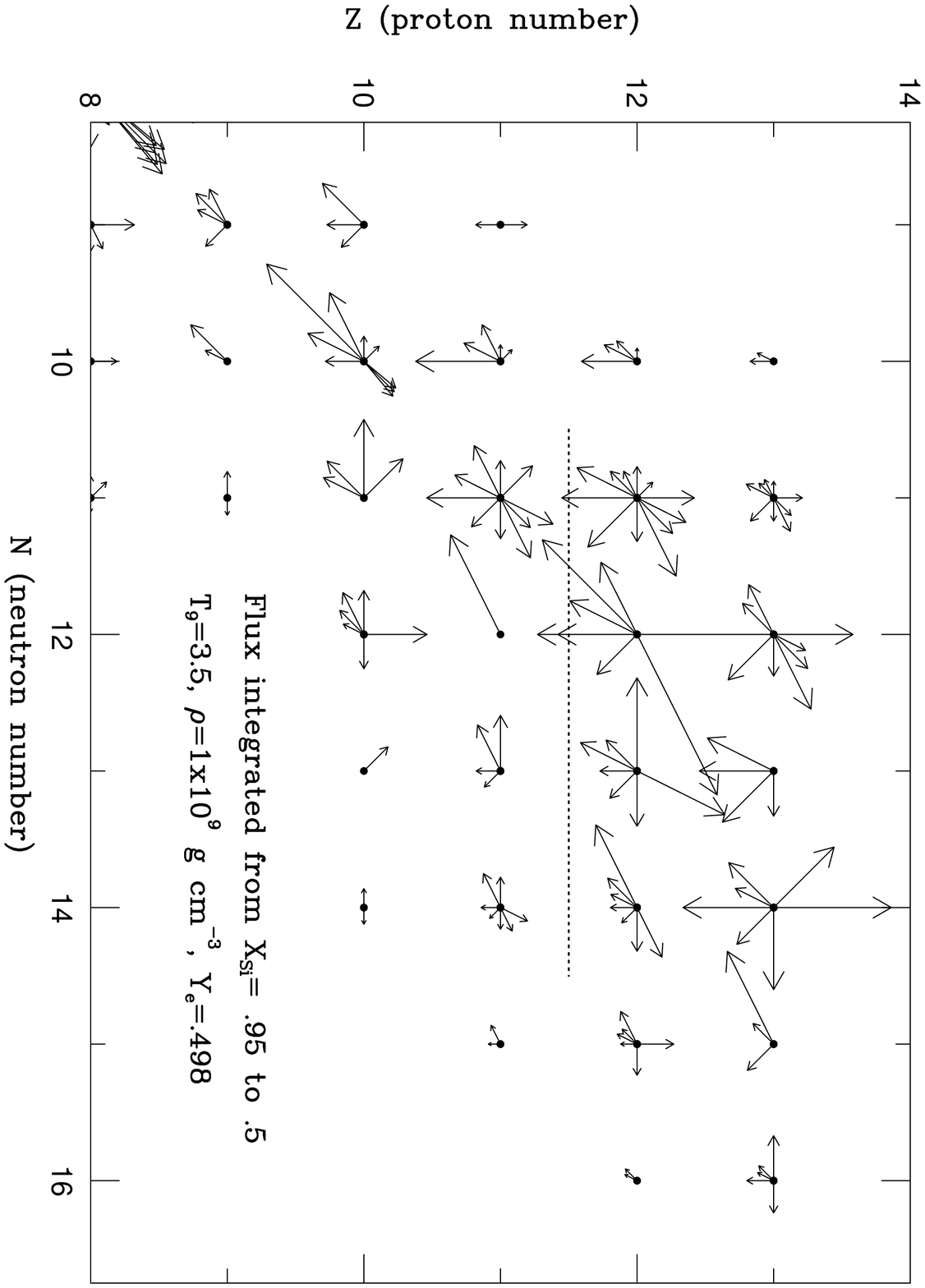}{Figure 5c}{Vector field representing the integrated
reaction fluxes from X(Si group) =.95 to .5, along the lower boundary of
the silicon group, for $\t9=3.5$, $\rho=10^9 \gcc$, and $Y_e=.498$.
The vector magnitudes have a logarithmic dependence on the size of
the integrated flux.}

\topfig{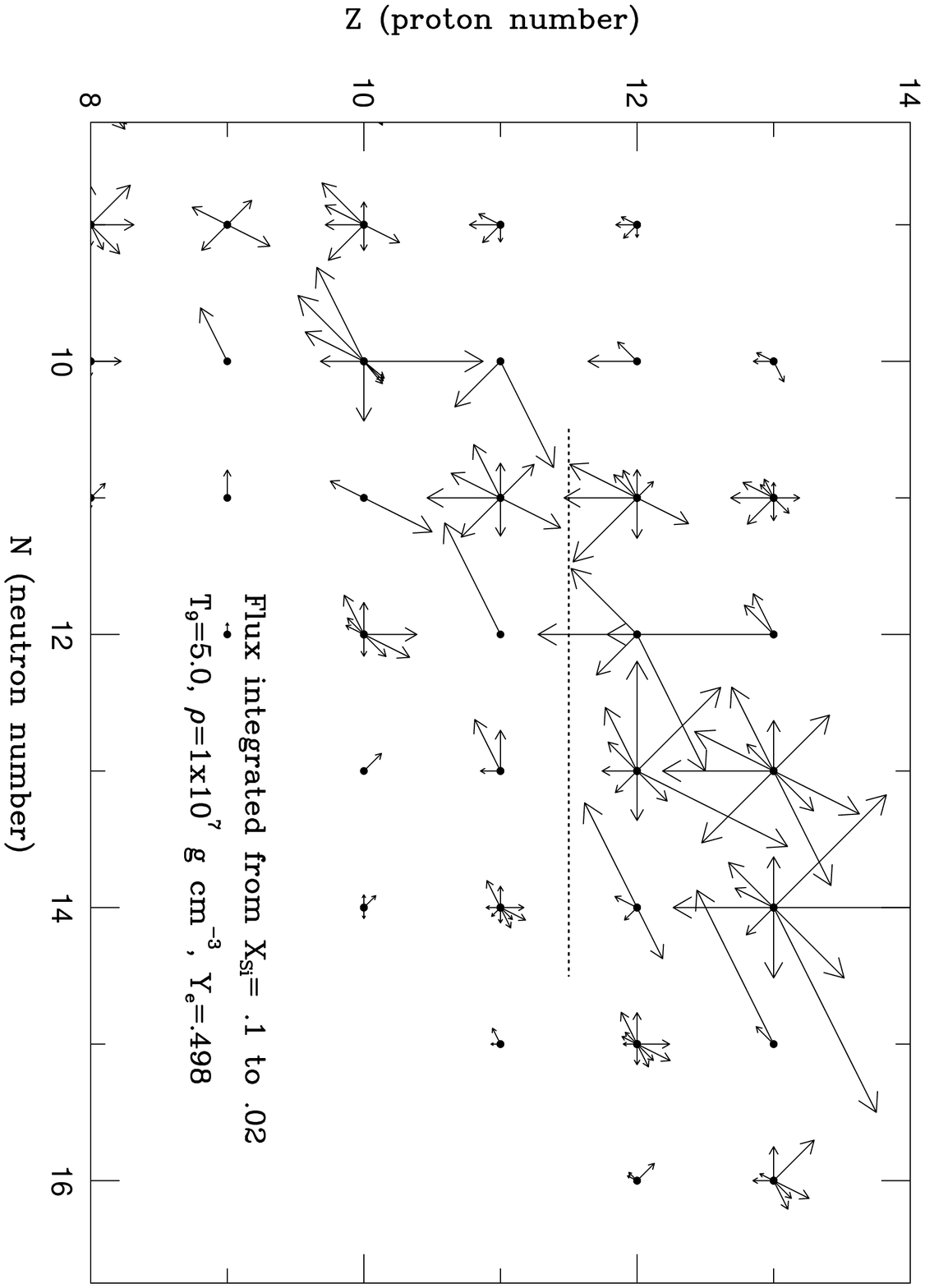}{Figure 5d}{Vector field representing the integrated
reaction fluxes from X(Si group) =.1 to .02, along the lower boundary of
the silicon group, for $\t9=5.0$, $\rho=10^7 \gcc$, and $Y_e=.498$.
The vector magnitudes have a logarithmic dependence on the size of
the integrated flux.}

While the variation of $Y_e$ has considerable influence on the relative
abundances of the Mg isotopes, there is not the pronounced shift in group
membership at this lower boundary that occurs at the upper boundary.
Comparison of Figures 1c, 1f, and 1g, shows changes in the relative spacing
of the Mg isotopes, and a somewhat more pronounced spread in those cases
further from NSE, but little evidence of significant movement of the
lower boundary.  Thus the effects of variation of $Y_e$ on the lower
boundary are simply changes in relative abundances.  Since  changes in
$Y_e$ have no counteracting effect on the reaction cross sections, these
changes in relative abundances significantly alter the reaction balances.
In this and the following subsection, we will show a series of vector
diagrams of the integrated reaction fluxes, forward reaction rate minus its
respective reverse, integrated over time.  The direction of the vector
reflects the dominant reaction of the pair, pointing from target to product.
In order to show best the many reactions which contribute and yet not have
the dominant vectors be visually overwhelming, the magnitudes of these
vectors have a logarithmic dependence on the actual magnitude of the flux.
As an example, in Figure 5a, for $\t9=5.0$, $\rho=10^7 \gcc$ and $Y_e=.498$,
the flux through $\rm ^{24}Mg (\gamma,\alpha) ^{20}Ne$ is 12 times greater
than that through $\rm ^{23}Mg (n,\alpha) ^{20}Ne$, although the vector
representing $\rm ^{24}Mg (\gamma,\alpha) ^{20}Ne$ is only 1.4 times longer
than the vector representing $\rm ^{23}Mg (n,\alpha) ^{20}Ne$.  Figure 5a
shows the reaction fluxes integrated from X(Si group) = .95 to .5, which
is the portion of the burning where the boundary between the silicon group
and elements of lower Z is most important.  Even at X(Si group) = .5, the
abundances of \nuc{Na}{23} and \nuc{Ne}{20} are less than $1\over2$ and
$6\over10$ of their respective silicon quasi-equilibrium abundances.  As
expected, the downward flows from the silicon group strongly dominate mass
transfer across this border.  Only a few small fluxes, most notably $\rm
^{22}Na (n,p) ^{22}Mg$ and $\rm ^{22}Na (\alpha,n) ^{25}{Al}$, point upward
across the boundary.  At first glance, this figure also seems to agree with
the results of WAC, with the flux through $\rm ^{24}Mg (\gamma,\alpha)
^{20}Ne$ being nearly 3 times the flux through $\rm ^{24}Mg (\gamma,p)
^{23}Na$.  But the largest reaction flux downward from the silicon group
is actually $\rm ^{26}Mg (p,\alpha) ^{23}Na$ with a flux 5.3 times larger
than that through $\rm ^{24}Mg (\gamma,p) ^{23}Na$.  In addition there is
a group of lesser flows, including $\rm ^{24}Mg (p,\alpha) ^{21}Na$, $\rm
^{23}Mg (n,p) ^{23}Na$, $\rm ^{23}Mg (n,\alpha) ^{20}Ne$, and $\rm ^{26}Al
(n,\alpha) ^{23}Na$, which in sum rival the importance of $\rm ^{24}Mg
(\gamma,\alpha) ^{20}Ne$.  This is certainly more complicated
 than the simple approach taken by BCF.  Thus our more extensive
network indicates that the apparent simplicity of the results of BCF
and WAC is due to limitations in their calculation.

This is further reinforced by examination of the reaction fluxes for lower
$Y_e$.  In Figure 5b, identical to Figure 5a, except that $Y_e=.46$,  the
reactions  $\rm ^{24}Mg (n,\alpha) ^{21}Ne$, $\rm ^{26}Mg (p,\alpha)
^{23}Na$, $\rm ^{25}Mg (n,\alpha) ^{22}Ne$, $\rm ^{23}Mg (n,p)
^{23}Na$, $\rm ^{23}Mg (n,\alpha) ^{20}Ne$, and $\rm ^{24}Mg (n,p)
^{24}Na$ all carry more flux than  $\rm ^{24}Mg (\gamma,\alpha)
^{20}Ne$.  This tendency at lower $Y_e$ to favor reactions with neutrons
in the incoming channel and neutron-rich nuclei as targets is very
much in keeping with the much larger free-neutron abundances found for
these conditions.  There is also a noteworthy flux upward via $\rm ^{23}Na
(\alpha,n) ^{26}Al$.  This flux is actually part of a cycle, $\rm ^{23}Na
(\alpha,n) ^{26}Al (n,p) ^{26}Mg (p,\alpha) ^{23}Na$, with the flux through
$\rm ^{23}Na (\alpha,n) ^{26}Al$ representing less than 7\% of the flux
through  $\rm ^{26}Mg (p,\alpha) ^{23}Na$.  It is a myriad of cycles like
this within the QSE groups which keep the $\alpha$-particles in equilibrium
with the free nucleons.

Variations of temperature and density do not cause strong changes in the
relative importance of reactions like those due to variations of $Y_e$.
Figure 5c, with $\t9=3.5$, $\rho = 10^9 \gcc$, and $Y_e=.498$, is similar to
Figure 5a.  In this case $\rm ^{24}Mg (\gamma, \alpha) ^{20}Ne$ does actually
carry the largest flux, but the same reactions we found important in Figure
5a are also important here.  Thus the
variation of temperature, which alters the balance of each set of forward
and reverse reactions, and density, which enhances captures, can result in
differences in relative fluxes, but in general does not open up drastically
different paths, since these variations do not result in the large free
neutron fraction found to be important for low $Y_e$.  With the passage of
time and the approach to equilibrium, the reaction flows dwindle.  The
approach to equilibrium also implies that the isotopes of Na and Ne are
not as underabundant.  Thus the dominance of the flows downward from the
silicon group over those flows directed upward lessens.  This is illustrated
in Figure 5d, which presents the integrated flux from X(Si group) = .1 to .02,
for $\t9=5.0$, $\rho=10^7 \gcc$ and $Y_e=.498$.  Over the
corresponding interval of time, the abundance of \nuc{Ne}{20} converges
from 80\% of its QSE abundance to its NSE value.  The sum of the
upward fluxes, principally $\rm ^{21}Na (\alpha,p) ^{24}Mg$, is almost
as large as the sum of the downward fluxes.  Furthermore, the balances
between some pairs of reactions have changed.  Representative of this
trend, the balance of $\rm ^{21}Ne (\alpha,n) ^{24}Mg$ and $\rm ^{24}Mg
(n,\alpha) ^{21}Ne$, which favored $\rm ^{24}Mg (n,\alpha) ^{21}Ne$
at early times, favors $\rm ^{21}Ne (\alpha,n) ^{24}Mg$ as equilibrium
is approached.

Thus the use of our larger and improved network indicates that there are
a number of reactions, in addition to the photodisintegration reactions
of \nuc{Mg}{24}, which are important to understanding the depletion of the
silicon group toward lighter nuclei.  Under some conditions, particularly
low $Y_e$, these alternative reactions dominate the $\rm ^{24}Mg (\gamma,
\alpha) ^{20}Ne$ and $\rm ^{24}Mg (\gamma,p) ^{23}Na$ reactions which
previous authors found to be the most important.  We do agree with previous
authors in finding that, as equilibrium is approached, the decreasing
abundance of the Mg isotopes, mandated by quasi-equilibrium, and the closer
proximity of the Ne and Na isotopes to their silicon quasi-equilibrium
abundances, results in a decline in the net flux of mass downward from the
silicon group.  This equilibration of the lighter nuclei with the silicon
group is the final phase of silicon burning, occurring well after the
abundances are dominated by the iron peak elements and a single
quasi-equilibrium group stretches upward from Mg.  By this point, as we
will show in \S\S 6.2 and 7, the conversion of silicon into iron peak
elements no longer dominates the energy production but serves as
a source of free particles, driving the distribution to NSE.

\bigskip

\subsect{6.2. Bridging the Gap between Silicon and Iron}

\medskip

While the downward flow from the silicon group generates the free nucleons
needed to build silicon into iron peak elements, and hence governs the
timescale for silicon burning, the links between the silicon and iron peak
quasi-equilibrium groups mediate the formation of the iron peak.  For times
prior to the establishment of a single QSE group, a small number of reactions
dominate the flow into the iron peak group.  It is the slowness of these
reactions which allows the persistence of two separate groups.  WAC found that,
with their network, the flow was dominated by a group of reactions ending in
\nuc{Ti}{46}.  They contended that the dominant flow was $\rm ^{45}Sc (p,
\gamma) ^{46}Ti$, aided at high temperature by $\rm ^{42}Ca (\alpha,\gamma)
^{46}Ti$ and $\rm ^{45}Ti (n,\gamma) ^{46}Ti$.  This contradicted earlier
work by BCF and Michaud \& Fowler (1972), which
had selected $\rm ^{44}Ti (\alpha,p) ^{47}V$ as the principal bridge.
WAC argued that this reaction was only important at late times when
a single QSE group was a good approximation.  We will examine both of
these contentions within the context of our larger network, and also
examine the influence of $Y_e$, something not done by previous authors.

\topfig{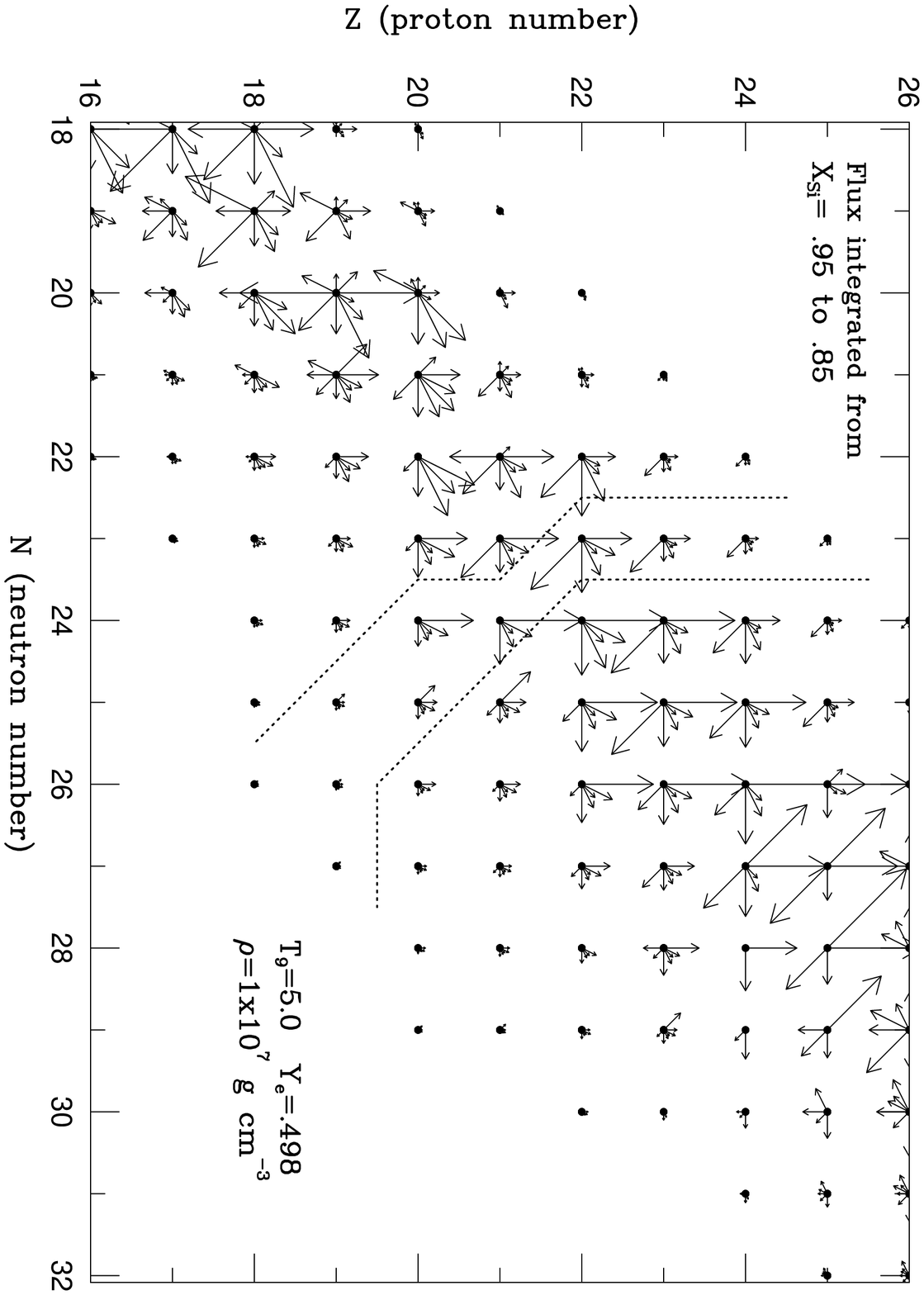}{Figure 6a}{Vector field representing the integrated
reaction fluxes from X(Si group) =.95 to .85, along the boundary between
the silicon group and the iron peak group, for $\t9=5.0$, $\rho=10^7 \gcc$,
and $Y_e=.498$.  The vector magnitudes have a logarithmic dependence
on the size of the integrated flux.}

\topfig{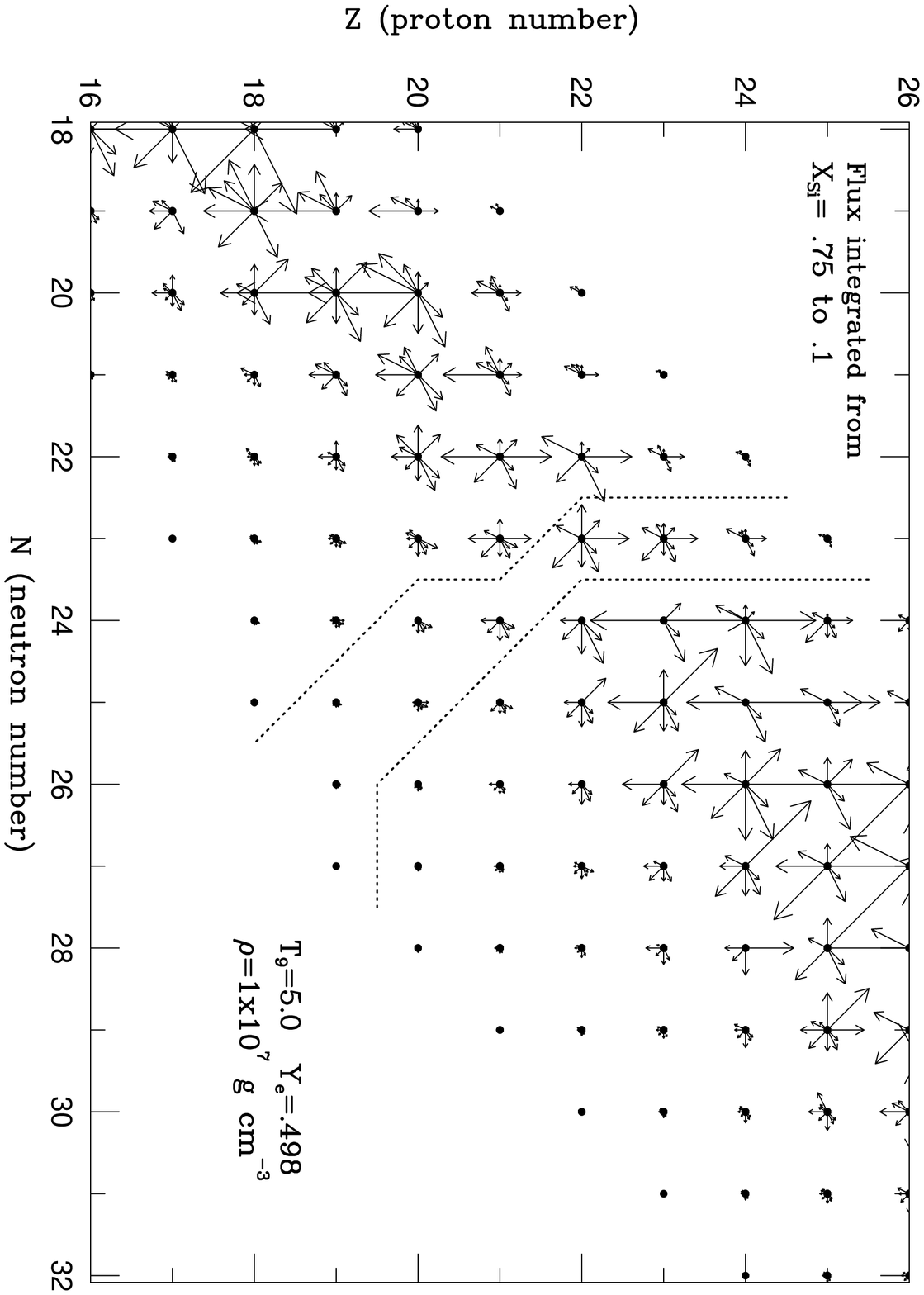}{Figure 6b}{Vector field representing the integrated
reaction fluxes from X(Si group) =.75 to .1, along the boundary between
the silicon group and the iron peak group, for $\t9=5.0$, $\rho=10^7 \gcc$,
and $Y_e=.498$.  The vector magnitudes have a logarithmic dependence
on the size of the integrated flux.}

\topfig{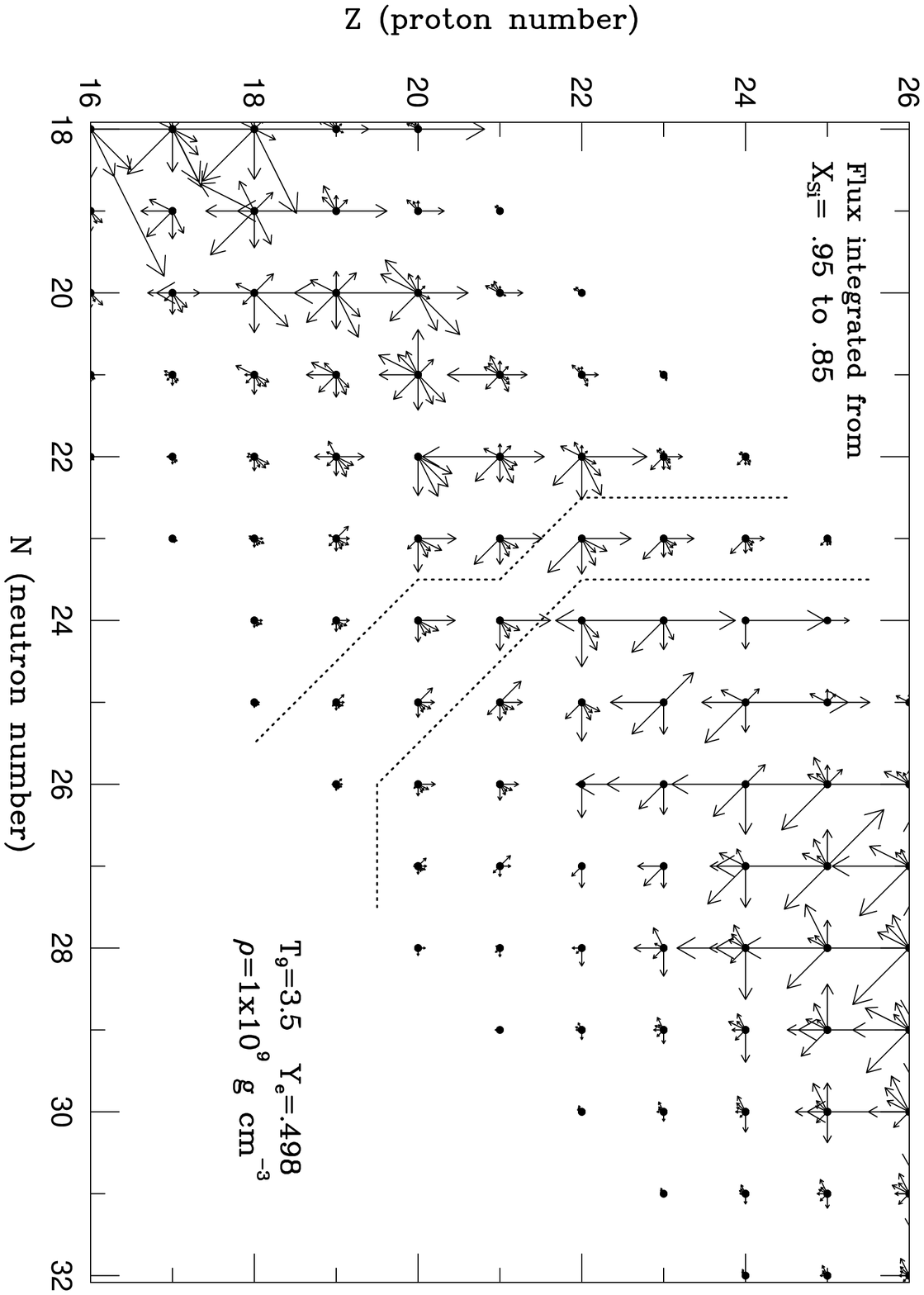}{Figure 6c}{Vector field representing the integrated
reaction fluxes from X(Si group) =.95 to .85, along the boundary between
the silicon group and the iron peak group, for $\t9=3.5$, $\rho=10^9 \gcc$,
and $Y_e=.498$.  The vector magnitudes have a logarithmic dependence
on the size of the integrated flux.}

\topfig{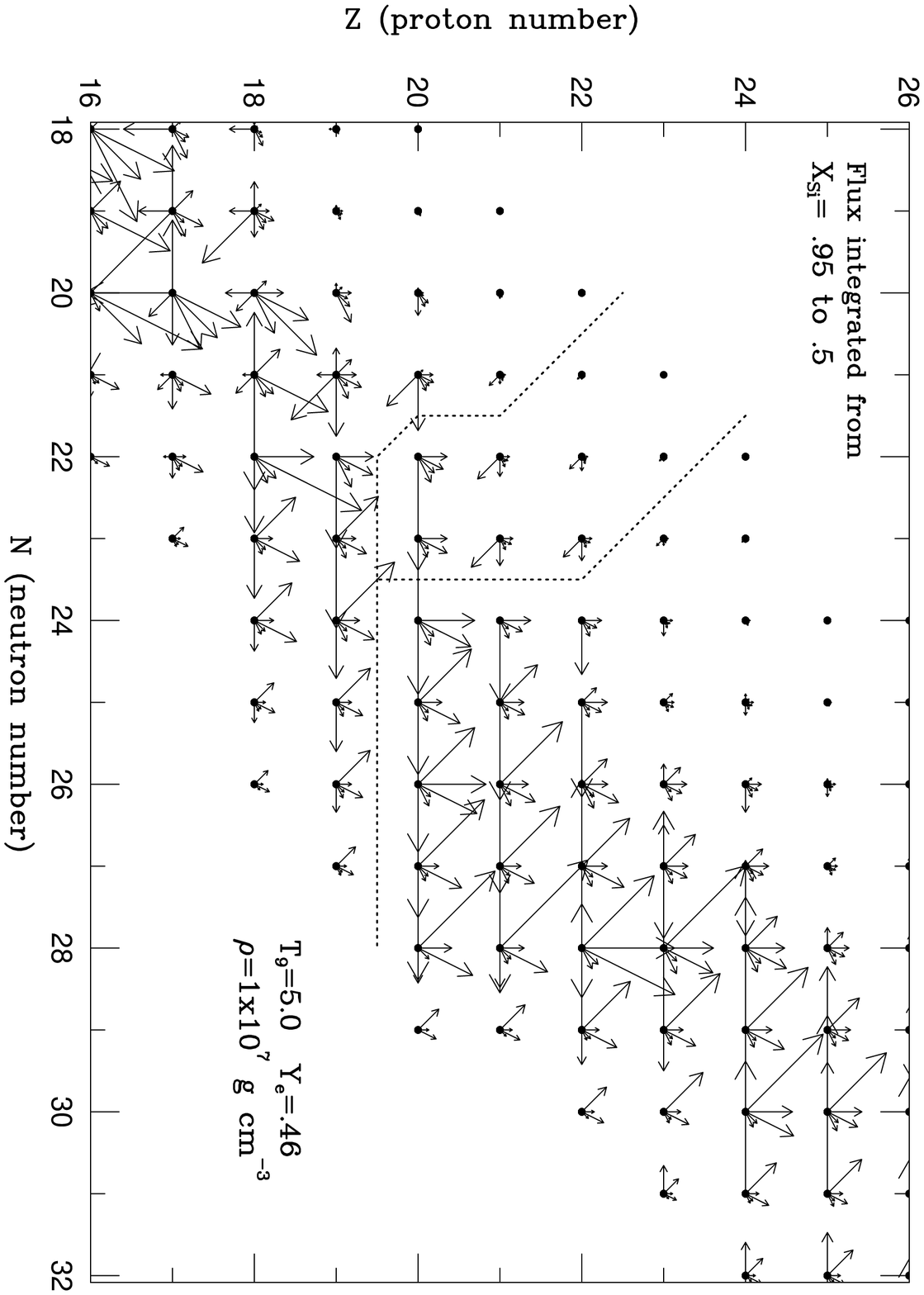}{Figure 6d}{Vector field representing the integrated
reaction fluxes from X(Si group) =.95 to .5, along the boundary between
the silicon group and the iron peak group, for $\t9=5.0$, $\rho=10^7 \gcc$,
and $Y_e=.46$.  The vector magnitudes have a logarithmic dependence
on the size of the integrated flux.}

\topfig{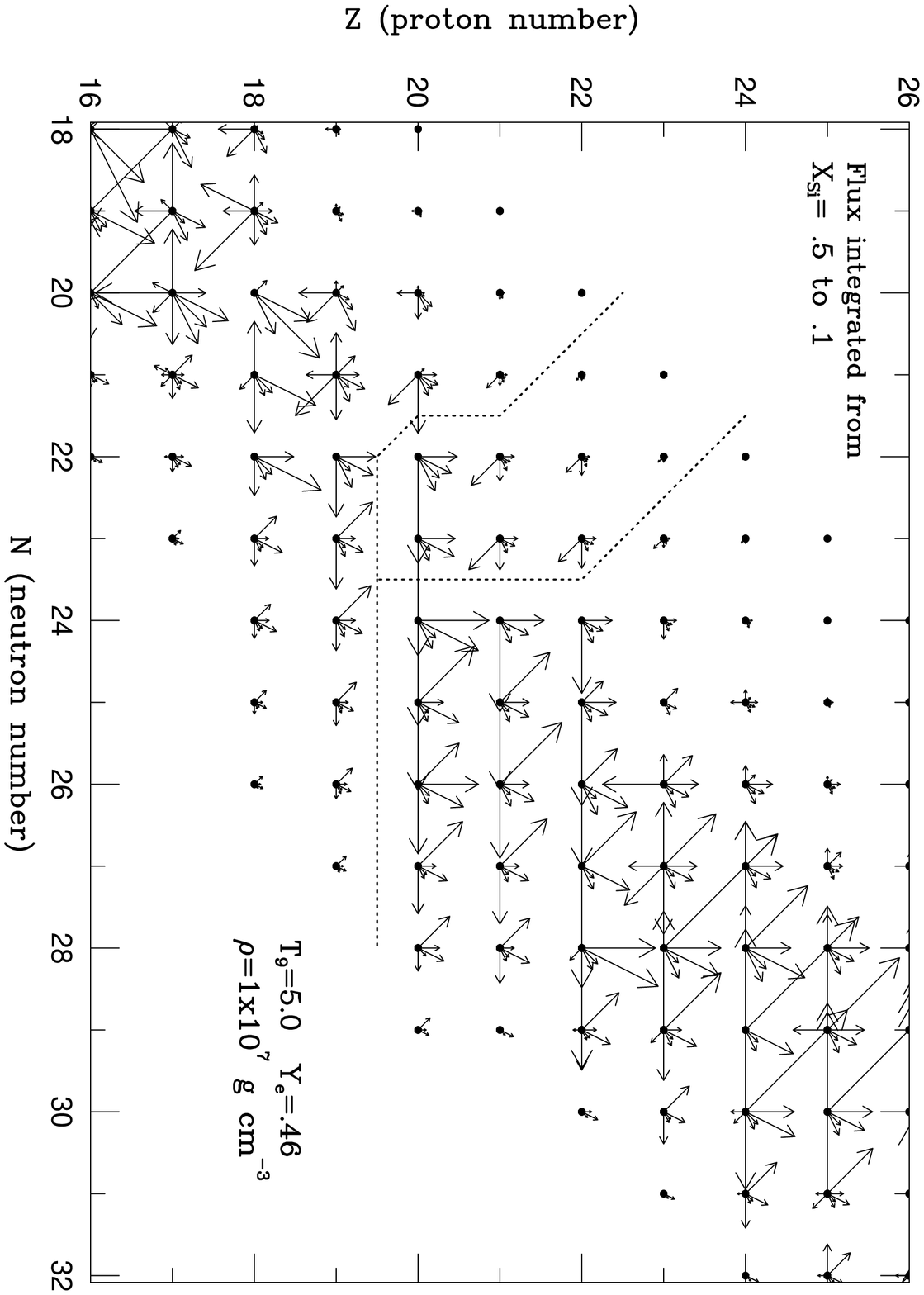}{Figure 6e}{Vector field representing the integrated
reaction fluxes from X(Si group) =.5 to .1, along the boundary between
the silicon group and the iron peak group, for $\t9=5.0$, $\rho=10^7 \gcc$,
and $Y_e=.46$.  The vector magnitudes have a logarithmic dependence
on the size of the integrated flux.}

For $\t9=5.0$, $\rho=10^7 \gcc$, and $Y_e=.498$, we find that at early
times there are two definite quasi-equilibrium groups separated by a thin
boundary region (See Figure 1c).  Although there are some relative changes in
the abundances of these nuclei, variations of temperature, density, and
degree of silicon exhaustion do not sharply alter the membership of this
boundary region.  In Figure 6a this boundary region is enclosed by the
dotted line, with the lower edge of the iron peak group to the right and
above, and the upper edge of the silicon group to the left and below.  As
revealed in Table 2, for $\t9=5.0$, $\rho=10^7 \gcc$, and $Y_e=.498$, by
X(Si group) =.5 there is a single well established QSE group.  In fact,
for these conditions, the species which comprise the iron peak group have
abundances within 10\% of their silicon quasi-equilibrium abundances by the
time X(Si group) $\sim.85$.
As a result, the vectors in Figure 6a represent the integrated reaction
fluxes from X(Si group) =.95, by which time the two QSE groups are well
established, until X(Si group) =.85.  Thus these are the fluxes which bring
the groups into quasi-equilibrium.  It is therefore not surprising that
there are no significant fluxes downward from the iron peak group into the
boundary region, and similarly no fluxes into the silicon group which
originate above.  The dominant flux into the iron peak group is $\rm ^{45}Sc
(p,\gamma) ^{46}Ti$.  A number of other fluxes contribute including $\rm
^{45}Ti (n,\gamma) ^{46}Ti$, $\rm ^{42}Ca (\alpha,\gamma) ^{46}Ti$, $\rm
^{43}Sc (\alpha,p)  ^{46}Ti$, $\rm ^{46}V (n,p) ^{46}Ti$, and $\rm ^{44}Ti
(\alpha,p) ^{47}V$ with fluxes relative to $\rm ^{45}Sc (p,\gamma) ^{46}Ti$
of .12, .09, .04, .05, and .08, respectively.  Thus $\rm ^{45}Sc (p,\gamma)
^{46}Ti$ accounts for approximately 70\% of the entire flux into the iron
peak group.  This agrees well with the findings of WAC for explosive
burning at comparable densities.  This agreement occurs in spite of
differences over group membership between this work and WAC.  In particular,
WAC place \nuc{Sc}{46} as a member of the silicon group (those authors do
not include a boundary region), while our results place it within the iron
peak group for comparable $Y_e$.  The existence of the boundary region
causes a further complication, in that there is not a one to one
correspondence between the flows out of the silicon group and those into
the iron peak group.  In addition to reactions like $\rm ^{44}Ti (\alpha,p)
^{47}V$, $\rm ^{43}Sc (\alpha,p)  ^{46}Ti$, and $\rm ^{42}Ca (\alpha,
\gamma) ^{46}Ti$, which directly connect the silicon group to the iron peak
group, there are a number of reactions which transfer mass from the
silicon group into the boundary region.  Principal among these are $\rm
^{42}Ca (\alpha,n) ^{45}Ti$, $\rm ^{42}Ca (\alpha,p) ^{45}Sc$, $\rm ^{44}Ti
(n, \gamma) ^{45}Ti$, $\rm ^{44}Sc (p,\gamma) ^{45}Ti$, and $\rm ^{44}Sc
(n,p) ^{44}Ca$, with fluxes relative to $\rm ^{45}Sc (p,\gamma) ^{46}Ti$ of
.28, .35, .21, .20, and .12, respectively.  With this many reactions
contributing significantly to the flow of mass into the boundary region,
the dominance
of $\rm ^{45}Sc (p,\gamma) ^{46}Ti$ as a flow out of the boundary requires
substantive reactions among the nuclei of the boundary region.  Thus the
relative values of \rqe\ revealed in Figure 1c for the nuclei within the
boundary are representative of reaction flows within the region and not
just a coincidental occurrence.  Note in particular in Figure 1c that both
\nuc{Ti}{45} and \nuc{Ca}{44} are relatively overabundant when compared to
\nuc{Sc}{45} and that Figure 6a shows significant flow along $\rm ^{45}Ti
(n,p) ^{45}Sc$ and $\rm ^{44}Ca (p,\gamma) ^{45}Sc$.  Thus the links within
the boundary region are important to the behavior of the boundary.  The
thin fringe of nuclei shown in Figure 1c between the silicon and iron peak
groups are not separate beads on strings of reactions between the QSE
groups, but part of an interconnected region which mediates the transfer of
mass between the silicon and iron peak groups.

Once these two QSE groups are replaced by a single group, knowledge of the
flows between the formerly separate groups is  no longer essential, as these
flows represent the changes required under QSE as the free nucleon abundances
converge toward equilibrium.  In the case of $\t9=5.0$, $\rho=10^7 \gcc$, and
$Y_e=.498$, this means that the fluxes for X(Si group) $< .8$ reflect the
movement of the mass into more proton-rich nuclei, as we discussed in Sect.
5.  Figure 6b shows the integrated fluxes from X(Si group) = .75 to .1.  The
dominant fluxes between the former silicon and iron peak groups, are a set
of $\rm (\alpha,p)$ reactions, principally $\rm ^{44}Ti (\alpha,p) ^{47}V$,
$\rm ^{43}Sc (\alpha,p)  ^{46}Ti$, and $\rm ^{45}Ti (\alpha,p) ^{48}V$.
Thus the dominant mass fluxes at late time proceed upward through more
proton-rich nuclei than they did prior to the merger of the QSE groups.
Another reflection of this tendency toward proton-rich nuclei is the
domination of (n,p) reactions by (p,n) after group merger.  Comparison of
Figures 6a and 6b reveals that, while at early time there was considerable
flux within the iron peak QSE group through reactions like $\rm ^{48}Cr
(n,p) ^{48}V$ and $\rm ^{47}V (n,p) ^{47}Ti$, after merger it is the
reverse reactions, $\rm ^{48}V (p,n) ^{48}Cr$ and $\rm ^{47}Ti (p,n)
^{47}V$ in this case, which dominate.  In another important reversal, at
late times there are downward flows from the former iron peak group, most
importantly $\rm ^{46}Ti (\gamma,p) ^{45}Sc$.  For $\t9=3.5$, $\rho=10^{10}
\gcc $, and $Y_e=.498$, a considerably greater density than any considered
by WAC, the scenario is little changed.  As is shown in Figure 6c, the dominant
fluxes, during the time that two separate QSE groups persist, are still $\rm
^{45}Sc (p,\gamma) ^{46}Ti$, $\rm ^{45}Ti (n, \gamma) ^{46}Ti$, $\rm
^{42}Ca (\alpha,\gamma) ^{46}Ti$, $\rm ^{43}Sc (\alpha,p)  ^{46}Ti$, $\rm
^{46}V (p,n) ^{46}Ti$, and $\rm ^{44}Ti (\alpha,p) ^{47}V$, with relative
fluxes of 1, .05, .23, .007, .007, and .32, respectively.  Clearly there
is much change
in the relative fluxes, with $\rm ^{45}Sc (p,\gamma) ^{46}Ti$ carrying only
60\% of the flux into the iron peak group, but the change is merely
in degree.  The variation of temperature and density does not profoundly
change which reaction fluxes bridge the gap between the iron peak and silicon
groups, although the relative importance can vary.  Thus, for $Y_e=.498$ our
analysis of the dominant reaction flows which transfer mass into the iron
peak group is in excellent agreement with that of WAC.  The dominance of $\rm
^{45}Sc (p,\gamma) ^{46}Ti$ as the primary conduit of mass transfer into
the iron peak group is readily apparent.  At late times the $(\alpha,p)$
reactions on more proton-rich nuclei dominate, especially  $\rm ^{44}Ti
(\alpha,p) ^{47}V$, while there is a downward flow via $\rm ^{46}Ti
(\gamma,p) ^{45}Sc$.  Both of these points support the earlier analysis of WAC.

We noted earlier that the variation of $Y_e$ causes the membership of the
QSE groups to change.  These changes are reflected in the comparison of
Figures 1c and 1e.  As we discussed in \S 4, the boundary region for
smaller $Y_e$ is displaced and noticeably broader in N.  This new boundary
region is enclosed within the dotted lines of Figure 6d, which shows
the integrated fluxes for $\t9=5.0$, $\rho=10^{7} \gcc$, and $Y_e=.46$,
from X(Si group) =.95 to .5.  Under these conditions the silicon and iron
peak groups reach $\pm5\%$ of merger for X(Si group) $\sim .1$.  There is
some ambiguity as to the group membership of the neutron-rich isotopes of
K (\nuc{K}{43}, \nuc{K}{44}, \nuc{K}{45}, and \nuc{K}{46}), as they seem
to be members of the silicon group for X(Si group) $\sim .9$ with abundances
approximately 60\% of their silicon QSE abundance, while the members of the
iron peak group are more than an order of magnitude underabundant (Figure
1e).  For X(Si group) $\sim .5$ (Figure 1f), these K isotopes are still
only approximately 80\% of their silicon QSE abundance, while the iron peak
group has converged to within a factor of 2.  Examination of Figure 6d reveals
that the reaction fluxes from these K isotopes into the neutron-rich
isotopes of Ca are much smaller  than the corresponding fluxes from Ca
into Sc and are also much smaller than the flows into the boundary
region from less neutron-rich isotopes of K and Ar.  Thus their inclusion
within the silicon group seems reliable, although imperfect.  The dominant
fluxes out of the silicon group are $\rm ^{38}Ar (\alpha,\gamma) ^{42}Ca$,
$\rm ^{39}Ar (\alpha,n) ^{42}Ca$, $\rm ^{40}Ar (\alpha,n) ^{43}Ca$, $\rm
^{42}K (p,n) ^{42}Ca$, and $\rm ^{43}K (p,n) ^{43}Ca$, with relative fluxes
of .47, .34, 1, .1, and .30, respectively. There is also a small group of
reactions, most
importantly $\rm ^{41}Ar (\alpha,n) ^{44}Ca$, $\rm ^{42}Ar (\alpha,n)
^{45}Ca$, $\rm ^{43}K (\alpha,n) ^{46}Sc$, $\rm ^{44}K (p,n) ^{44}Ca$, and
$\rm ^{45}K (p,n) ^{45}Ca$, which directly link the silicon group to the
iron peak group.  However, the sum of these fluxes is less than 15\% of the
flux carried by $\rm ^{40}Ar (\alpha,n) ^{43}Ca$.  These small fluxes are
also only a minor contribution to the flux into the iron peak group, which
is overwhelmingly dominated by $\rm ^{43}Ca (n,\gamma) ^{44}Ca$.  The flux
through this reaction is 2.3 times larger than the flux through $\rm
^{40}Ar (\alpha,n) ^{43}Ca$ and thus represents over 90\% of the flux into
the iron peak group.  The flow within the iron peak group is dominated by
a series of $\rm (n,\gamma)$ reactions between isotopes of Ca, out to $\rm
^{48}Ca$.  The flow upward from Ca is largely by (p,n) and $\rm (\alpha,
n)$ reactions.   This transfer of mass into the iron peak nuclei via the
neutron-rich isotopes of Ca is reminiscent of the exploratory results TA
found for lower mass stars.  In Figure 6c we examined the integrated
fluxes for low temperature and high density, conditions one expects for
core silicon burning in lower mass stars, but with $Y_e=.498$, and found
little difference between the low temperature/high density case and the
high temperature/low density case, except for an increase in the relative
importance of flows through the more proton-rich nuclei.  From Figure 6d,
we conclude that the flows through neutron-rich Ca which TA described are
due to the greater neutronization found in the core of less massive stars.

Examination of Figure 6e reveals that the trend noted earlier for less
neutron-rich nuclei to increase in importance at later times is also true
for $Y_e= .46$.  With merger of the QSE groups occurring for X(Si group)
$\sim .1$ under these conditions, the integrated flux from X(Si group) =.5
to .1 is still reflective of the flows which bring the groups to
equilibrium.  At these later times, the reaction $\rm ^{38}Ar (\alpha,
\gamma) ^{42}Ca$ carries a larger flux than the reactions $\rm ^{39}Ar
(\alpha,n) ^{42}Ca$ and $\rm ^{40}Ar (\alpha,n) ^{43}Ca$ combined.  Although
the dominant flow into the iron peak group is still $\rm ^{43}Ca (n,\gamma)
^{44}Ca$, comparison of Figure 6e to Figure 6d shows that the series of large
fluxes through $\rm (n,\gamma)$ reactions on Ca isotopes is greatly
diminished for \nuc{Ca}{46} and \nuc{Ca}{47}.  Instead the flow proceeds
upward at lower N.  This process continues after group merger as the average
Z/A within the iron peak group approaches $Y_e$.  Clearly the convergence
of the average Z/A within the iron group toward $Y_e$, as we discussed in
\S 5, plays a role in making proton-rich nuclei more important as silicon
is exhausted.  With the iron group dominated at early times by nuclei with
Z/A $< Y_e$, the reactions through the neutron-rich isotopes are naturally
enhanced by the increased abundance.  As the abundance of free neutrons
declines from the initial overabundance, the relative importance of the
less neutron-rich nuclei, reflected by the average Z/A, grows.  This is
particularly true at low $Y_e$, since the silicon group is dominated by
nuclei with Z/A $\sim .5$.

Thus the reactions which bridge the gap between the silicon group and the
iron peak group reflect the underlying changes in abundance.  The
interconnection of these groups is also reflective of the group membership
and of changes in the relative importance of these members.  For high
$Y_e$ the dominant flow into the iron peak group prior to the merger of the
groups is $\rm ^{45}Sc (p,\gamma) ^{46}Ti$.  The flow which feeds
\nuc{Sc}{45} exits the silicon group via a series of reactions involving
\nuc{Ca}{40}, \nuc{Ca}{42}, \nuc{Sc}{43}, \nuc{Sc}{44}, \nuc{Ti}{44} and
\nuc{Ti}{45}.  For somewhat smaller $Y_e$ the flow into the iron peak group
via $\rm ^{45}Sc (p,\gamma) ^{46}Ti$ is still important, but it is now fed
through reactions involving \nuc{Ca}{42}, \nuc{Ca}{44}, and \nuc{Sc}{46}.
As the degree of neutronization increases, the dominant flow between the
groups naturally winds its way through more neutron-rich nuclei as the
abundances of these nuclei become more important.  For $Y_e=.46$ this
means that the merger is achieved via a flow that proceeds through a long
series of $\rm (n,\gamma)$ captures on Ca.  As time proceeds, as we showed
in \S 5, the average Z/A within the iron peak group converges toward the
global value of $Y_e$.  The network of reactions responds to/causes this
greater abundance of less neutron-rich nuclei by increasing the importance
of bridging reactions involving more proton-rich nuclei and by increasing
the flux through (p,n) reactions within the iron peak group.  It is the
neutronization
of material which dominates the behavior of the QSE groups, and therefore
the reactions that merge the groups.  Although the variation of $Y_e$ has
little effect on the cross sections and rate coefficients, the reaction flows
across the boundary change in response to the changing abundances within
the groups.  Since these reactions, particularly those from the bottom of
the silicon quasi-equilibrium group, govern the rate of conversion of
silicon into iron peak nuclei, these differences in reaction flows will
clearly be reflected in the rate of energy production.

\bigskip

\sect{7. Energetics of Silicon Burning}

\medskip

\topinsert
\centerline{Table 4 The Energy Reservoir}
\nobreak
\medskip
\moveright .5in
\vbox{ \baselineskip=16 true pt
\tabskip 1 pc
\halign{\hfil$#$\hfil&\hfil$#$\hfil&\hfil$#$\hfil &\hfil$#$\hfil&
\hfil$#$\hfil&\hfil$#$\hfil \cr
\noalign{\smallskip \thinrule \medskip}
{\t9}&\rho (\gcc)&Y_e&{\rm Init.\ BE\ (erg/g)}&
{\rm NSE\ BE\ (erg/g)}&{\rm Diff.\ (erg/g)} \cr
\noalign{\medskip \thinrule \medskip}
3.5&10^7&.498&8.155\times10^{18}&8.346\times10^{18}&1.91\times10^{17} \cr
\ &\ &.48&8.193\times10^{18}&8.431\times10^{18}&2.38\times10^{17} \cr
\ &\ &.46&8.172\times10^{18}&8.477\times10^{18}&3.05\times10^{17} \cr
\noalign{\smallskip \thinrule \medskip}
3.5&10^{10}&.498&8.155\times10^{18}&8.348\times10^{18}&1.93\times10^{17} \cr
\ &\ &.49&8.172\times10^{18}&8.386\times10^{18}&2.14\times10^{17} \cr
\ &\ &.48&8.193\times10^{18}&8.431\times10^{18}&2.38\times10^{17} \cr
\ &\ &.47&8.214\times10^{18}&8.461\times10^{18}&2.47\times10^{17} \cr
\ &\ &.46&8.172\times10^{18}&8.479\times10^{18}&3.07\times10^{17} \cr
\noalign{\smallskip \thinrule \medskip}
5.0&10^7&.498&8.155\times10^{18}&8.198\times10^{18}&0.43\times10^{17} \cr
\ &\ &.49&8.172\times10^{18}&8.283\times10^{18}&1.11\times10^{17} \cr
\ &\ &.48&8.193\times10^{18}&8.369\times10^{18}&1.76\times10^{17} \cr
\ &\ &.47&8.214\times10^{18}&8.420\times10^{18}&2.06\times10^{17} \cr
\ &\ &.46&8.172\times10^{18}&8.452\times10^{18}&2.80\times10^{17} \cr
\noalign{\smallskip \thinrule \medskip}
5.0&10^{10}&.498&8.155\times10^{18}&8.345\times10^{18}&1.90\times10^{17} \cr
\ &\ &.48&8.193\times10^{18}&8.424\times10^{18}&2.31\times10^{17} \cr
\ &\ &.46&8.172\times10^{18}&8.472\times10^{18}&3.00\times10^{17} \cr
\noalign{\smallskip \thinrule \medskip}
}}
\endinsert

\topfig{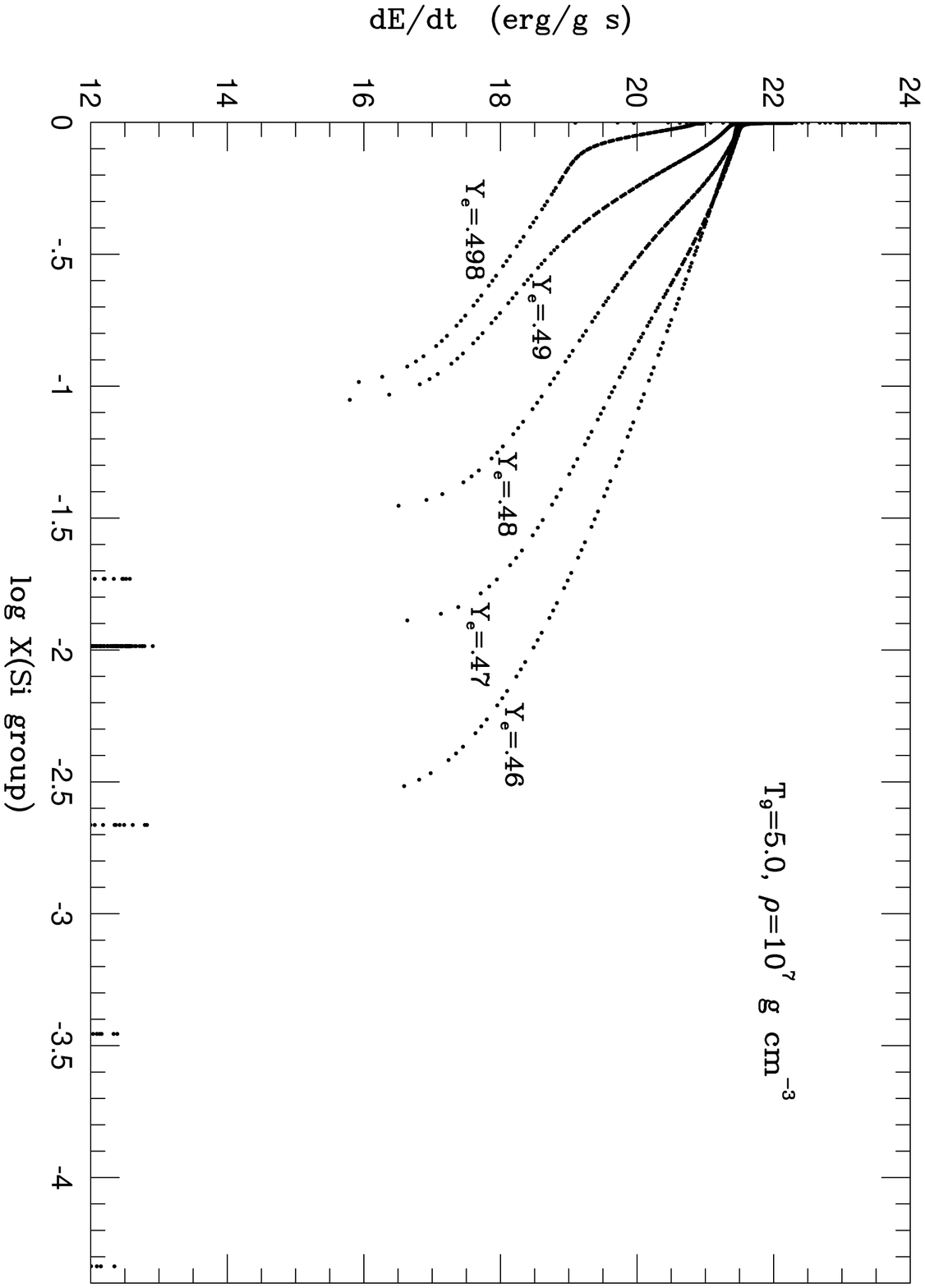}{Figure 7a}{Rate of energy generation as a function
of degree of silicon exhaustion, for $\t9=5.0$, and $\rm \rho=10^7 \gcc $,
with $Y_e$ varying from .498 to .46.}

\topfig{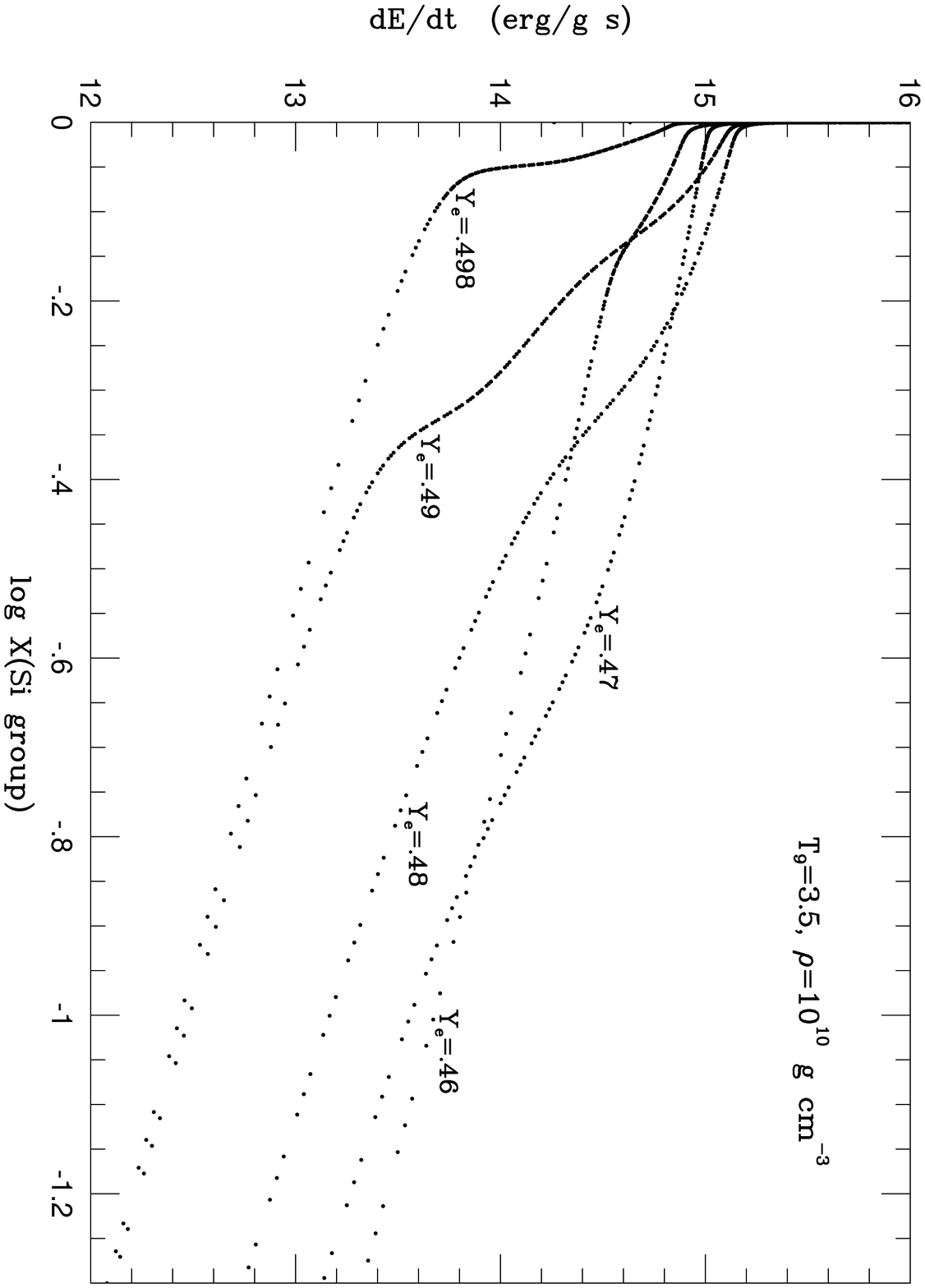}{Figure 7b}{Rate of energy generation as a function
of degree of silicon exhaustion, for $\t9=3.5$, and $\rm \rho=10^{10} \gcc $,
with $Y_e$ varying from .498 to .46.}

Before we examine the energy production from the actual network
calculations, it is instructive to make some simple energetic arguments
about silicon burning.  If we take as the final distribution of each of our
constant temperature, density, and $Y_e$ cases the corresponding screened
NSE distribution and compare this with the initial distribution, we can
calculate the amount of energy available from silicon burning.  Table 4
shows the binding energies of the initial and NSE distributions and their
difference for a few cases, all in ergs/g.  Comparison of cases which differ
only in $Y_e$ reveals that much more energy is available at lower $Y_e$.
This reflects the larger binding energy per nucleon of \nuc{Fe}{56},
\nuc{Fe}{54}, and \nuc{Ni}{58} in comparison to \nuc{Ni}{56}, a fact which,
for example,  accounts for much of the light curve of supernovae.  The
comparison of cases which differ in temperature and density shows a greater
availability  of energy at lower temperatures and higher densities.  The
lower the temperature, the more the $\exp (B(^AZ)/kT)$ term in the nuclear
Saha  equation dominates, causing a more prominent peak in the abundance
distribution composed of the most bound nuclei.
Similarly, a higher density also favors more massive nuclei which (up to
Ni/Fe) have a larger binding energy.  For example, for $Y_e = .498$ and
$\rho = 10^{10} \gcc$, the change from $\t9=3.5$ to $\t9=5.0$ causes a
change in the mass fraction concentrated in isotopes of Mn, Fe, Co, and Ni,
from .9996 to .996.  For $Y_e = .498$ and $\rho = 10^7 \gcc$, the same
change in temperature results in changes in this mass fraction from .9991
to .918.  With as much as 6 times more energy available per silicon nucleus
at low $Y_e$, even if the rate of silicon depletion is the same in all cases,
there can be tremendous differences in the energy generation and therefore
in the hydrodynamic conditions which cause this process.  This is particularly
important as electron captures cause $Y_e$ to drop.

The lack of a single dominating reaction in silicon burning makes
prediction of the energy production complicated.  In general, the rate of
energy production per gram by a nuclear process, $\epsilon$ is given in
terms of the reaction rates, $r_{ij}$ (reactions $\rm cm^{-3} s^{-1}$), by
$$
\epsilon = {\sum_{ij} r_{ij} Q_{ij} \over \rho} \ ,
\eqno(15)
$$
where $Q_{ij}$ is the Q-value for the reaction between $i$ and $j$.  In
cases where a single reaction dominates the process the sum is unnecessary.
Although it is possible, as was shown above, to calculate the energy store
available for silicon burning, the plethora of possible governing reactions
revealed in \S 6 makes estimation of the temporal behavior of this
energy release difficult.  For each reaction $ij$ in silicon burning, there
is a reverse reaction $k\ell$ which is approaching equilibrium with it.
Thus Equation (15) becomes
$$
\epsilon = {\sum_{ij} \left(r_{ij} -r_{k\ell(ij)} \right) Q_{ij} \over
\rho} \ ,
\eqno(16)
$$
where $Q_{k\ell} = -Q_{ij}$.  As the abundances approach equilibrium,
$\epsilon$ approaches zero, not necessarily because of the exhaustion of
fuel but because
each pair of reactions is balancing.  While Table 4 shows the global value
of ${\cal E}=\int_{0}^{NSE} \epsilon \ dt$, it says nothing about the
interplay of the various $r_{ij}$'s and $r_{k\ell}$'s.  These, however,
determine the temporal behavior of $\epsilon$.

In \S 6 we concluded that there are significant differences in the
relative importance of the many reaction flows which typify silicon burning.
In particular, the variation of $Y_e$ has a profound effect on which
reactions are the most important links between the quasi-equilibrium
groups and which reactions are primarily responsible for flow downward
from the silicon group.  Examination of Figure 7a shows that these
differences in reactions result in tremendous variation in $\epsilon$ as
a function of X(Si group) and $Y_e$.  These particular curves are for
$\t9=5$ and $\rm \rho=10^7 \gcc$.  While the behavior is similar in each
case as equilibrium is approached, the initial
stages, which are the most important from the point of view of energy
production, vary greatly.  Clearly the different reactions which dominate
as $Y_e$ varies result in very different behavior.  As we noted in \S 4,
the lower $Y_e$ cases take less time to reach a similar degree of silicon
exhaustion.  Table 5 shows the elapsed time for several degrees of silicon
exhaustion.  Silicon burning at $Y_e =.498$ is much slower than silicon
burning at $Y_e=.46$.  With the elapsed time to X(Si group) $\sim .9$
almost an order of magnitude larger for high $Y_e$ and the energy available,
the global ${\cal E}$, much smaller, it is not surprising that $\epsilon$
for $Y_e=.498$ is more than an order of magnitude smaller than $\epsilon$
for $Y_e=.46$ at early times.  Further, with the elapsed time between
X(Si group) $\sim .5$ and .1 being more than 100 times larger for $Y_e=
.498$ than for $Y_e=.46$, it is not surprising that $\epsilon$ falls off
more precipitously for higher $Y_e$.  These very different temporal
evolutions in energy generation are then directly attributable to the
very different fluxes seen in \S 6.  Clearly simple approximations which
multiply a temperature- and density-dependent formula for $\epsilon$ by
a correction dependent on $Y_e$
are ruled out.  The behavior with changing $Y_e$ is dependent on the complex
interaction of the different reactions out of the silicon group.  Figure 7b
shows that at high density/low temperature there is a similar tangle of
curves as a function of $Y_e$.  Any successful approximation of the energy
generation for silicon burning needs to take into account the complex
behavior of the reactions out of the quasi-equilibrium groups.  As we
established in \S\S 4 and 5, the reactions within the groups can be well
understood by the use of quasi-equilibrium.  The many reactions, shown
in \S 6 to be  important for the understanding of the flows among the
groups, are also the governing reactions here, particularly those reactions
which transfer mass downward from the silicon group.  The sheer number of
important reactions, and their response to changes in $Y_e$ and degree of
silicon exhaustion, preclude a simple analytic model.  Instead, in Paper
II, we will discuss a promising approximation to the network calculation
of silicon burning employing quasi-equilibrium.  For now, however, we will
discuss a few additional general characteristics of the energy generation
during silicon burning.

\topinsert
\centerline{Table 5 Elapsed time verses Degree of Silicon Exhaustion}
\nobreak
\medskip
\moveright .5in
\vbox{ \offinterlineskip
\tabskip .6 pc
\halign{#\strut&\hfil$#$\hfil&\hfil$#$\hfil&\hfil$#$\hfil &#\vrule&
\hfil$#$\hfil&\hfil$#$\hfil&\hfil$#$\hfil \cr
\noalign{\smallskip \thinrule \smallskip \thinrule}
&\multispan 3 \hfil {\rm Conditions} \hfil&&\multispan 3 \hfil \rm
{Elapsed time in seconds for } \hfil \cr
&\t9&\rho (\gcc)&Y_e&&{\rm X(Si group)=.9}&
{\rm X(Si group)=.5}&{\rm X(Si group)=.1}\cr
\noalign{\thinrule}
&5.0&10^7&.498&&5.6\times10^{-5}&1.9\times10^{-3}&3.6\times10^{-2} \cr
&5.0&10^7&.48&&8.0\times10^{-6}&7.2\times10^{-5}&1.7\times10^{-3} \cr
&5.0&10^7&.46&&1.1\times10^{-5}&5.8\times10^{-5}&2.6\times10^{-4} \cr
&5.0&5\times10^7&.46&&1.1\times10^{-5}&5.8\times10^{-5}&2.5\times10^{-4} \cr
&5.0&2\times10^8&.46&&1.0\times10^{-5}&5.5\times10^{-5}&2.2\times10^{-4} \cr
&5.0&10^9&.46&&9.4\times10^{-6}&5.0\times10^{-5}&2.0\times10^{-4} \cr
&5.0&10^{10}&.46&&7.9\times10^{-6}&4.1\times10^{-5}&1.5\times10^{-4} \cr
&4.5&10^{10}&.46&&4.7\times10^{-4}&2.6\times10^{-3}&8.8\times10^{-3} \cr
&4.0&10^{10}&.46&&6.6\times10^{-2}&4.4\times10^{-1}&1.6\times10^{0} \cr
&3.5&10^{10}&.46&&3.3\times10^{1}&2.8\times10^{2}&1.1\times10^{3} \cr
\noalign{\thinrule}
}}
\endinsert

One particularly noteworthy feature, common to all of the cases pictured in
Figure 7a, is the tail off, with the energy generation actually becoming an
energy depletion before the silicon group reaches its equilibrium value.
This occurs once the iron peak nuclei dominate the mass fraction and thus the
energy generation from the destruction of silicon group elements has dropped.
The convergence of the free proton and free neutron fractions to their
equilibrium values results in an increasing portion of the mass within the
iron peak group being found in elements with higher $Z$ and $A$.  These
nuclei are not as bound as the Fe and Ni isotopes which dominate the NSE
distribution, hence their production is endoergic.  Furthermore, the
production of the free nucleons is also endoergic.  As $Y_e$ decreases,
the equilibrium mass fraction of these nuclei increases, requiring more
energy to produce them and therefore causing the tail-off to begin further
from equilibrium, i.e., while the energy generated from the conversion of
silicon into iron peak nuclei is larger. This results in the increasing
gap between the last positive value of the energy generation and the
equilibrium silicon group mass fraction as $Y_e$ decreases.  For $Y_e \sim
.5$, this effect is exacerbated by the transition as silicon exhaustion
increases in which nucleus is most abundant.  It is this transition,
demonstrated in Figure 3a, from \nuc{Fe}{56} and \nuc{Fe}{54} to the less
bound \nuc{Ni}{56} which is responsible for the sharp decline of $\epsilon$
seen for larger values of $Y_e$.

\topfig{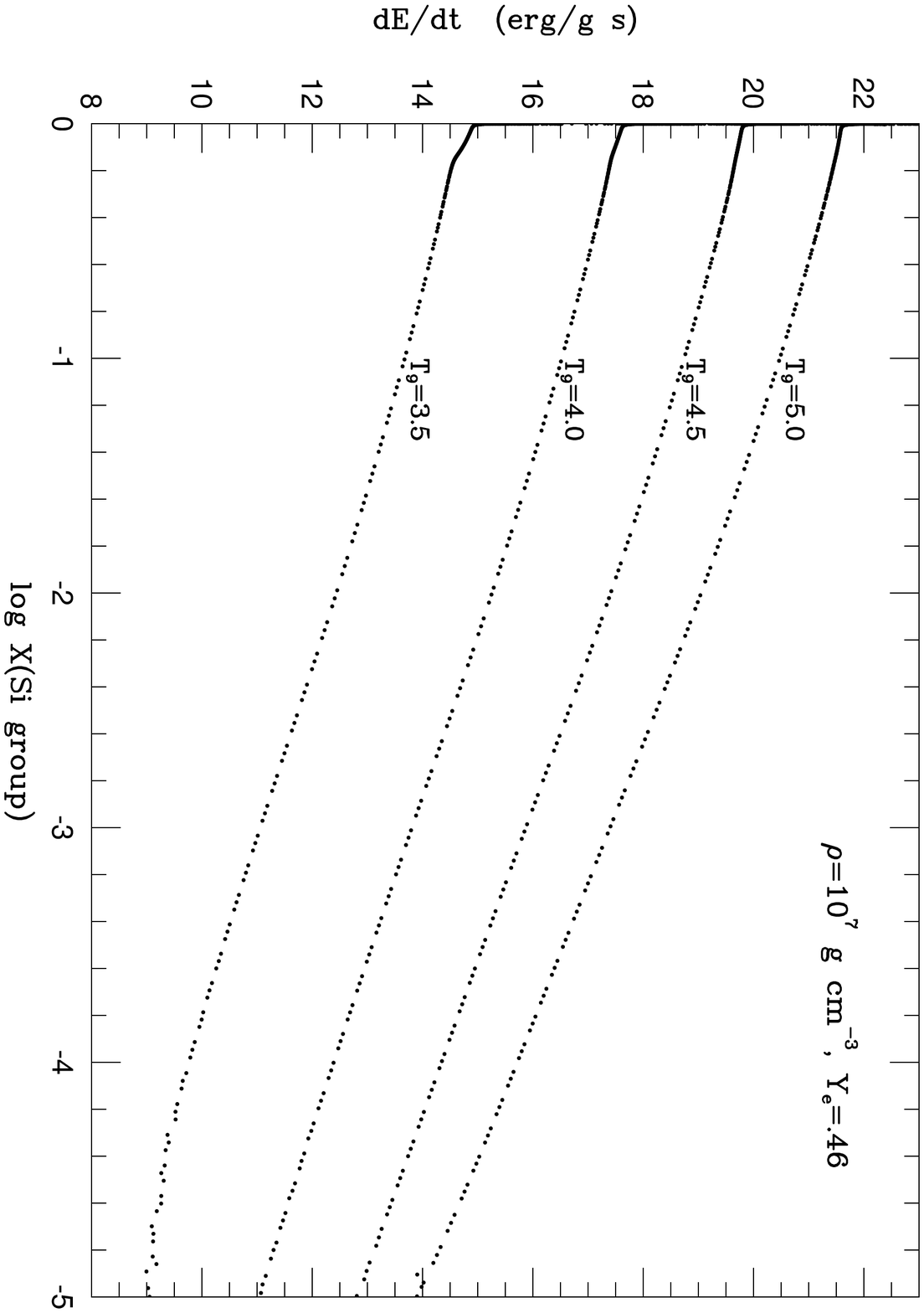}{Figure 8a}{Rate of energy generation as a function
of degree of silicon exhaustion, for $\rm \rho=10^7 \gcc$ and $Y_e=.46$,
with $\t9$ varying from 3.5 to 5.0.}

\topfig{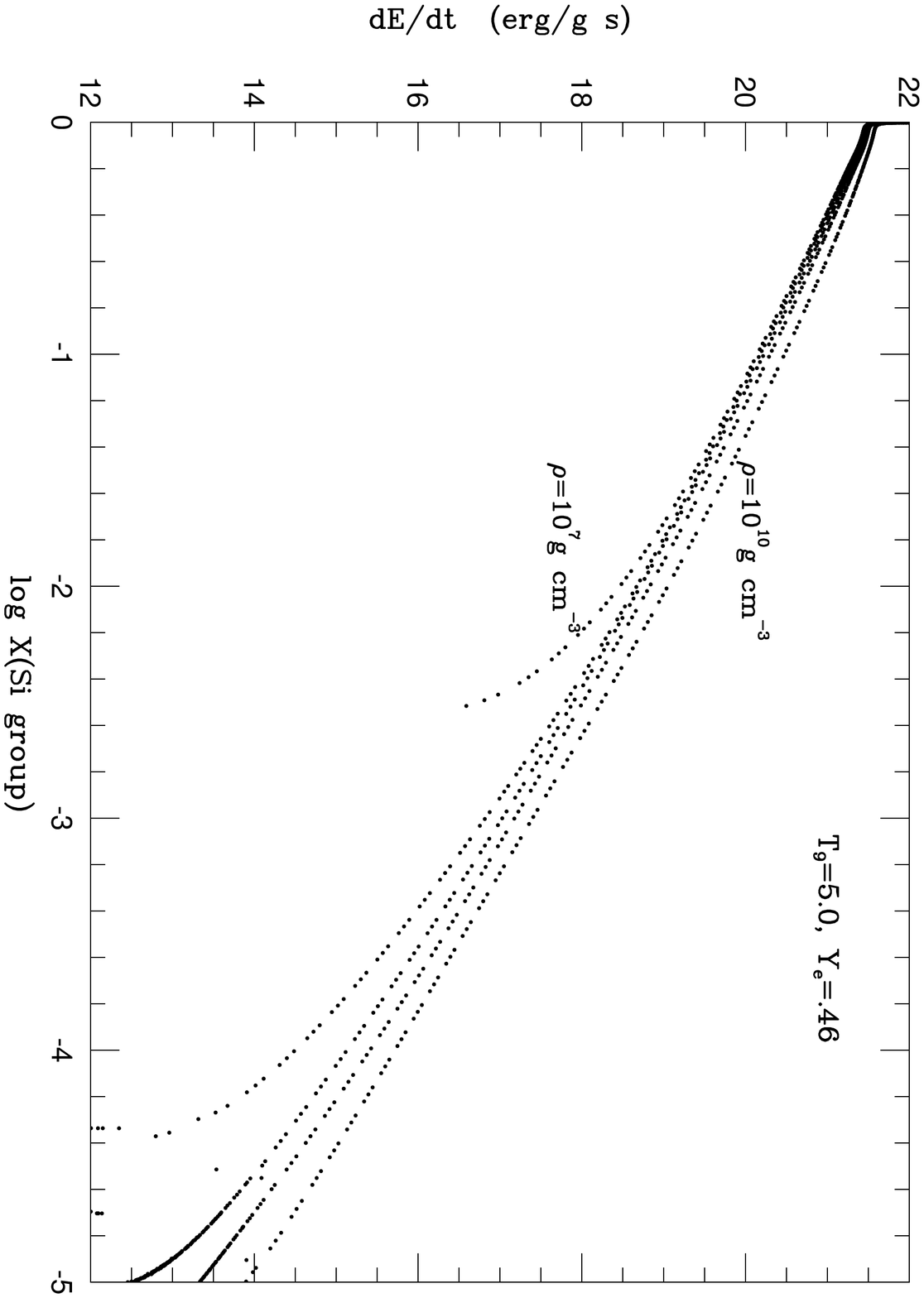}{Figure 8b}{Rate of energy generation as a function
of degree of silicon exhaustion, for $\t9=5.0$ and $Y_e=.46$, with $\rho$
varying from $\rm 10^7 \gcc$ to $\rm 10^{10} \gcc$.}

Figure 8a shows the effect of temperature variations on $\epsilon$ for
$\rho=10^{10} \gcc $ and $Y_e=.46$.  The resemblance these curves share
is striking.  Each is a power law which tails off as each case approaches
equilibrium.  The lack of convergence between these curves indicates that
changes due to variations in temperature are variations in rates, not
changes in the relative importance of different reactions, which agrees
with our analysis of the reaction fluxes in \S 6.  From Table 4 it is
apparent that the available energy supply is essentially unchanged by
temperature variations for this density.  Thus the differences in $\epsilon$
are due to differing timescales for the burning.  Comparison of the times
necessary for these conditions to result in 10\% exhaustion of the silicon
group attest to this assertion.  From the elapsed times for X(Si group)
$\sim .9$ shown in Table 5, we find that the differences in $\epsilon$ are
very similar to the differences in elapsed time.  Clearly the increase in
temperature increases the photodissociation rates, thus providing for
quicker destruction of the silicon group.  However, the variation in
timescale with temperature is not consistent with the variations in the
photodisintegration rate of \nuc{Si}{28}.  This is in keeping with the
argument by BCF, supported by our analysis, that it is not the
photodissociation rate of silicon which governs the rate of silicon
destruction; rather, it is the downward flow from the silicon group
which determines the burning timescale.

Variations in density have effects similar to those of temperature.  Figure
8b portrays the effects of the variation of density, for $\t9=5.0$, and
$Y_e=.46$.  Once again we see power laws that tail off as each case
approaches its equilibrium.  Note that the variation of $\epsilon$ with
density, particularly at early times, is small.  The variations in elapsed
time to reach various degrees of silicon exhaustion as a function of
density are also shown in Table 5.  The ratio of $\epsilon$ for $\rho =
10^{10} \gcc$ to that for $\rho = 10^{7} \gcc$ is only 1.5 for X(Si group)
$\sim .9$, and 1.4 for
X(Si group) $\sim .5$.  Within the silicon QSE group, the increase in density
results in an increase in the relative importance of Mg.  The ratio of the
mass fractions in Mg for $\rho = 10^{10} \gcc$  compared to that for $\rho
= 10^{7} \gcc$ is 1.3 for both X(Si group) $\sim .9$ and X(Si group) $\sim
.5$.  This suggests that the difference in the energy generation rate as a
function of density is largely due to differences in quasi-equilibrium
abundances and in the reaction flows which result from them.

In this section we showed that increases in temperature or density have
large effects on the rate of energy generation, greatly enhancing the rate
at which silicon is destroyed.  Variations in neutronization also have a
large effect on the rate of energy generation, changing the reaction
pathways by which silicon is converted into iron peak elements.  Thus, the
evidence presented in this section points to the need to approximate the
complete behavior of silicon burning in order to approximate the energy
generation.  This requires keeping track of many of the 300 nuclei and
3000 reactions used in this nuclear network.  Fortunately, as we will show
in a subsequent paper, quasi-equilibrium provides the means to approximate
energy generation accurately at considerable savings in computation.
Quasi-equilibrium greatly reduces the number of reactions which need to be
considered, since only those which leave their group are important to the
evolution of the abundances within the group.  Further, the prediction of
abundances based on the quasi-equilibrium abundances greatly reduces the
amount of nuclear accounting which must be done.

\bigskip

\sect{8. Conclusions}

\medskip

We have performed a detailed reexamination of silicon burning as a function
of temperature, density, and neutronization,  using a large nuclear network.
Central to understanding this process is the concept of quasi-equilibrium.
A natural extension of nuclear statistical equilibrium, which it approaches
in the asymptotic limit, quasi-equilibrium reflects the near balance of the
rapid formation and destruction of nuclei during silicon burning.  The net
gain in abundance of a species is much smaller than either its formation
or destruction rates.  We rederived the equations of quasi-equilibrium,
including for the first time the effects of Coulomb screening on the
equilibrium distribution.  This has been found by Hix \etal\ (1996) to be
important for Nuclear Statistical Equilibrium, and we showed that Coulomb
screening is also important in reconciling the network abundance calculations
with quasi-equilibrium.  We further demonstrated the real usefulness of
quasi-equilibrium in describing the network evolution, with many of the
most abundant nuclei forming two quasi-equilibrium groups, one focused on
silicon and the other on the iron peak nuclei.  While previous authors
have seen such separate groups at early times, we discovered that, with
increasing neutronization, the separation of these groups persists through
a much more significant portion of silicon burning.  Furthermore, we showed
not only that the duration of this two group phase is a function of $Y_e$,
but also that the membership in the groups is dependent on $Y_e$.  Since
previous work ignored extensive variation of neutronization and its effect
on quasi-equilibrium, this behavior has not been seen before.

We demonstrated the effects of these interacting quasi-equilibrium groups
on the abundances as the distribution evolves.  One noticeable effect is
the excessive production of neutron-rich iron peak nuclei during incomplete
silicon burning, resulting in an iron peak group noticeably more neutronized
than the global neutronization would imply.  This effect is missed by many
approximations to silicon burning, particularly those using narrow nuclear
networks.  With quasi-equilibrium governing the relative abundances within
the groups, the principal evolution of the distribution is due to the
reactions linking the groups.  We further demonstrated that the important
reactions for the destruction of silicon and the formation of iron peak
nuclei are highly dependent on the degree of neutronization.  While
variations in temperature and density affect the relative rates of reactions
and the overall speed of silicon destruction, they do not radically alter
the reaction paths into and out of the quasi-equilibrium groups in the way
that the variation of $Y_e$ does.  For relatively unneutronized material,
the principal reaction flow linking the silicon group with the iron peak
group proceeds through isotopes of Sc and Ti on the proton-rich side
of stability, in agreement with WAC.  For larger neutronization, the
increasing availability of free neutrons, and the corresponding changes in
quasi-equilibrium group membership, result in the dominant flow proceeding
through neutron-rich isotopes of Ca, as TA suggested.  Furthermore, the
rate at which matter flows along these reaction paths is also strongly
enhanced under these conditions.  With the entire process and especially
the rate of silicon destruction strongly dependent on $Y_e$, it is not
surprising to find that the energy generation by silicon burning has a
strong and complex dependence on $Y_e$.  As we also showed, the variations
in energy generation due to changes in temperature and density are not as
convoluted as those due to changes in $Y_e$.

Taken together, these effects of neutronization on the process of silicon
burning imply that successful modeling of silicon burning in its
hydrodynamic context needs to account for the variety of reaction paths
and full range of important nuclei.  Because large networks like the one
used here are a cumbersome addition to hydrodynamic calculations, much
previous work has tried to approximate silicon burning using narrow networks
and other simplified schemes.  The present results warn of the danger of
applying such schemes, developed for slight neutronization ($Y_e \sim .5$),
beyond the context in which they were developed.  In the context of reduced
nuclear reaction networks, this implies that the networks not only need
to stretch from H to the iron peak but also must have considerable
breadth.  This seemingly returns us to the cumbersome large networks.
However, in a forthcoming paper we will demonstrate an alternative.
Instead of reducing the number of nuclei by restricting the width, we find
it possible to use the physics of quasi-equilibrium to intelligently reduce
the number of abundances which must be evolved to a more manageable size.

\medskip

The list of people who have contributed to this work is naturally larger
than the author list.  In particular, we would like to thank Dave Arnett,
Ken Nomoto, and Masaki Hashimoto for motivating this work, and Al Cameron
for his interest and aid.  W.R.H. was supported in part by a NASA Graduate
Student Researcher Fellowship.  F.-K.T. was supported in part by NSF grant
89-13799 and the Swiss Nationalfonds.

\vfill\supereject

\parindent=30truept
\parskip=\smallskipamount
\def\pgf{\noindent\hangindent=0.5in\hangafter=1}
\def\pubfont{\it}
\def\voln#1{{\bf #1}}
\def\astap{{\pubfont Astron. Astrophys.} }
\def\apj{{\pubfont Ap. J.} }

\def\apjsup{{\pubfont Ap. J. Suppl.} }
\def\atnucdata{{\pubfont At. Nucl. Data Tables} }
\def\canjphys{{\pubfont Can. J. Phys.} }
\def\physrep{{\pubfont Phys. Rep.} }

\def\procastsocaust{{\pubfont Proc.  Astron. Soc. Australia} }
\def\progthphys{{\pubfont Prog. Theor. Phys.} }
\def\revmodphys{{\pubfont Rev. Mod. Phys.} }

\sect{References}

\def\aufdnion{\pgf
Aufderheide, M., Baron, E., \& Thielemann, F.-K. 1991, \apj {\bf370}, 630}
\aufdnion

\def\bakaeise{\pgf
Bao, Z.Y., \& K\"appeler, F. 1987, \atnucdata \voln{36}, 411}
\bakaeise

\def\bocfsiei{\pgf
    Bodansky, D., Clayton, D.D., \& Fowler, W.A. 1968, \apjsup \voln{16} 299}
\bocfsiei

\def\bbfhfise{\pgf
    Burbidge, E.M., Burbidge, G.R., Fowler, A.A., \& Hoyle, F. 1957,
    \revmodphys \voln{29}, 547}
\bbfhfise

\def\cafoeiei{\pgf
    Caughlan, G.R., \& Fowler, W.A. 1988, \atnucdata \voln{40}, 283}
\cafoeiei

\def\cottnion{\pgf
    Cowan, J.J., Thielemann, F.-K., Truran, J. W. 1991, \physrep \voln{208},
    267}
\cottnion

\def\fohosifo{\pgf
    Fowler, W.A., \& Hoyle, F. 1964, \apjsup \voln{9}, 201}
\fohosifo

\def\grabseth{\pgf
    Graboske, H.C., DeWitt, H.E., Grossman, A.S., \& Cooper, M.S. 1973,
    \apj \voln{181}, 457}
\grabseth

\def\haweeifi{\pgf
    Hartmann, D., Woosley, S.E., \& El Eid, M.F. 1985, \apj \voln{297}, 837}
\haweeifi

\def\hanofini{\pgf
    Hayashi, C., Nishida, M., Ohyama, N., \& Tsuda, H. 1959,
    \progthphys \voln{22}, 101}
\hanofini

\def\hithnisi{\pgf
    Hix, W.R., \& Thielemann F.-K. 1996, \apj, submitted}
\hithnisi

\def\htftnisi{\pgf
    Hix, W.R., Thielemann, F.-K., Fushiki, I., \& Truran, J.W. 1996,
    \apj, submitted}
\htftnisi

\def\itkmnize{\pgf
    Itoh, N., Kuwashima, F., \& Munakatu, H. 1990, \apj \voln{362}, 620}
\itkmnize

\def\mifosetw{\pgf
    Michaud, G., \& Fowler, W.A. 1972, \apj \voln{173}, 157}
\mifosetw

\def\nohaeiei{\pgf
    Nomoto, K., \& Hashimoto, M. 1988, \physrep \voln{163}, 13}
\nohaeiei

\def\thareifi{\pgf
    Thielemann, F.-K., \& Arnett, W.D. 1985, \apj \voln{295}, 604}
\thareifi

\def\thateise{\pgf
    Thielemann, F.-K., Arnould, M., \& Truran, J.W. 1987,
    in {\pubfont Advances in Nuclear Astrophysics}, ed. E. Vangioni-Flam
    et al. (Gif sur Yvette:Editions Fronti\`eres), 525}
\thateise

\def\thhnnize{\pgf
    Thielemann, F.-K., Hashimoto, M., \& Nomoto, K. 1990,  \apj \voln{349},
222}
\thhnnize

\def\theanith{\pgf
    Thielemann, F.-K., Bitouzet, J.-P., Kratz, K.-L., M\"oller, P.,
    Cowan, J.J., \& Truran, J.W. 1993,  \physrep \voln{227}, 269}
\theanith

\def\thnhnifo{\pgf
    Thielemann, F.-K., Nomoto, K., \& Hashimoto, M. 1994, in {\pubfont
    Supernovae, Les Houches, Session LIV}, ed. S. Bludman, R. Mochkovitch, \&
    J. Zinn-Justin (Amsterdam:Elsevier), 629}
\thnhnifo

\def\trcgsifi{\pgf
    Truran, J.W., Cameron, A.G.W., \& Gilbert, A. 1965, \canjphys \voln{44},
    563 }
\trcgsifi

\def\wagosini{\pgf
    Wagoner, R.V. 1969, \apjsup \voln{18}, 247}
\wagosini

\def\wafhsise{\pgf
    Wagoner, R.V., Fowler, W.A., \& Hoyle, F. 1967, \apj \voln{148}, 3}
\wafhsise

\def\wewfeifi{\pgf
    Weaver, T.A., Woosley, S.E., \& Fuller, G.M. 1985, in {\pubfont Numerical
    Astrophysics}, ed. J. Centrella, J. Leblanc, \& R. Bowers (Boston:Jones
    \& Bartlet), 374}
\wewfeifi

\def\wgtreisi{\pgf
    Wiescher, M., G\"orres, J., Thielemann, F.-K., \& Ritter, H. 1986,
    \astap \voln{160}, 56}
\wgtreisi

\def\whgteise{\pgf
    Wiescher, M., Harms, V., G\"orres, J., Thielemann, F.-K., \& Rybarcyk, L.J.
   1987, \apj \voln{316}, 162}
\whgteise

\def\wggbeini{\pgf
    Wiescher, M., G\"orres, J., Graaf, S, Buchmann, L., \& Thielemann, F.-K.
    1989, \apj \voln{343}, 352}
\wggbeini

\def\wigtnize{\pgf
    Wiescher, M., G\"orres, J., \& Thielemann, F.-K. 1990, \apj \voln{363},
340}
\wigtnize

\def\wooseisi{\pgf
    Woosley, S.E. 1986, in {\pubfont Nucleosynthesis and Chemical Evolution},
    16th Advanced Course of the Swiss Society of Astrophysics and
    Astronomy, ed. B. Hauck, A. Maeder, \& G. Meynet (Geneva:Geneva Obs.), 1}
\wooseisi

\def\woacseth{\pgf
    Woosley, S.E., Arnett, W.D., \& Clayton, D.D. 1973, \apjsup \voln{26}, 231}
\woacseth

\def\wohonitw{\pgf
    Woosley, S.E., \& Hoffman, R. 1992, \apj \voln{395}, 202}
\wohonitw

\def\wopweiei{\pgf
    Woosley, S.E., Pinto, P.A., \& Weaver, T.A. 1988, \procastsocaust
    \voln{7}, 355}
\wopweiei

\vfill\eject\end